\documentclass[english]{aa}
\usepackage[T1]{fontenc}
\usepackage[latin1]{inputenc}
\usepackage{graphicx}
\usepackage{epstopdf}

\usepackage{amssymb}
\usepackage[authoryear]{natbib}

\usepackage{textcomp}

 \usepackage{esint}

\makeatletter

\usepackage{aas_macros}

\usepackage{babel}
\makeatother
\begin{document}

\title{Simulations of the cosmic infrared and submillimeter background for future large
surveys: II. Removing the low-redshift contribution to the anisotropies using stacking}

\author{N. Fernandez-Conde, G. Lagache, J.-L. Puget, H. Dole}

\institute{Institut d'Astrophysique Spatiale (IAS), b\^atiment 121, Universit\'e
Paris-Sud 11 and CNRS (UMR 8617), 91405 Orsay, France e-mails: nestorconde@gmail.com, {[}guilaine.lagache;jean-loup.puget;herve.dole]@ias.u-psud.fr }

\offprints{G. Lagache}


\titlerunning{Stacking analysis and study of CIB anisotropies}

\authorrunning{N. Fernandez-Conde}

\abstract{Herschel and Planck are surveying the sky at unprecedented angular scales and sensitivities over large areas. But both experiments are limited by source confusion in the submillimeter. The high confusion noise in particular restricts the study of the clustering properties of the sources that
dominate the cosmic infrared background. At these wavelengths, it is more appropriate to consider the statistics of the unresolved component. In particular, high clustering will contribute in excess of Poisson noise in the power spectra of CIB anisotropies.}{These power spectra contain contributions from sources at all redshift. We show how the stacking technique can be used to separate the different redshift contributions to the power spectra.}{We use simulations of CIB representative of realistic Spitzer, Herschel, Planck, and SCUBA-2 observations. We stack the 24~$\mu$m sources in longer wavelengths maps to measure mean colors per redshift and flux bins. The information retrieved on the mean spectral energy distribution obtained with the stacking technique is then used to clean the maps, in particular to remove the contribution of low-redshift undetected sources to
the anisotropies.}{Using the stacking, we measure the 
mean flux of populations 4 to 6 times fainter than the total noise at 350~$\mu$m at redshifts $z=1$ and $z=2$, respectively,
and as faint as 6 to 10 times fainter than the total noise at 850~$\mu$m at the same redshifts. In the deep Spitzer fields, the detected 24~$\mu$m sources up to z$\sim$2 contribute significantly to the submillimeter anisotropies. We show that the method provides excellent (using COSMOS 24~$\mu$m data) to good (using SWIRE 24~$\mu$m data) removal of the $z<2$ (COSMOS) and $z<1$ (SWIRE) anisotropies.}{Using this cleaning method, we then hope to have a set of large maps dominated by high redshift galaxies for galaxy evolution study (e.g., clustering, luminosity density). }

\maketitle

\section{Introduction}

The first observational evidence of the cosmic infrared background (CIB) was
reported by \citet{1996A&A...308L...5P} and confirmed by \citet{1998ApJ...508..123F} and \citet{1998ApJ...508...25H}.
The CIB is composed of the relic emission at infrared wavelengths
of the formation and evolution of galaxies and consists of contributions
from infrared starburst galaxies and to a lesser degree from active galactic
nuclei. Deep cosmological surveys of this background
have been carried out with ISO  \citep [see] [for reviews] {2000ARA&A..38..761G,2005SSRv..119...93E}
mainly at 15~$\mu$m
with ISOCAM \citep [e.g.,] [] {2002A&A...384..848E}; at 90 and 170~$\mu$m
with ISOPHOT \citep [e.g.,] [] {2001A&A...372..364D}; with Spitzer at 24,
70, and 160~$\mu$m  \citep [e.g.,] []
{2004ApJS..154...70P,2004ApJS..154...87D} 
and with ground-based instruments SCUBA \citep [e.g.,] [] {2002PhR...369..111B}, LABOCA
 \citep [e.g.,][]{2008A&A...485..645B} ,and MAMBO \citep [e.g.,] []{2000astro.ph.10553B} at 850, 870, and 1300~$\mu$m
respectively. The balloon-borne experiment BLAST performed the first deep extragalactic 
surveys at wavelengths 250-500$\mu$m capable of measuring large numbers of star-forming 
galaxies, and their contributions to the CIB \cite[][]{2009Natur.458..737D}.
These surveys allowed us to obtain a far clearer understanding
of the CIB and its sources \citep [see][for a general review]
{2005ARA&A..43..727L}  but many questions remain unanswered such as 
the evolution of their spatial distribution with redshift. \\

The spatial distribution of infrared galaxies as a function
of redshift is a key component of the scenario of galaxy formation and evolution.
However, its study
has been hampered by high confusion and instrumental noise
and/or by the small size of the fields of observation. 
Tentative studies, with a small number of sources at 850~$\mu$m
\citep{2004ApJ...611..725B}, found evidence of a relationship
between submillimeter galaxies and the formation of massive galaxies
in dense environments. 
Works by \citet{2006ApJ...641L..17F} and \citet{2008MNRAS.383.1131M}
measured a strong clustering of ultra luminous infrared galaxies
(ULIRG) detected with Spitzer at high redshifts. 
Alternatively, the infrared background anisotropies could also 
provide information about the correlation between the sources of the
CIB and dark matter \citep{2000ApJ...530..124H,2001ApJ...550....7K,2007ApJ...670..903A},
and its redshift evolution.
\citet{2007ApJ...665L..89L} and \citet{2009arXiv0904.1200V} reported the detection of a correlated component
in the background anisotropies using Spitzer/MIPS (160~$\mu$m) and BLAST (250, 350,
and 500~$\mu$m) data. These authors found that star formation is highly biased at z$>$0.8. The strong evolution
of the bias parameter with redshift, caused by the shifting
of star formation to more massive halos with increasing
redshift, infers that environmental effects influence
the vigorous star formation.\\

To improve our understanding of the formation and evolution of galaxies using 
CIB anisotropies, we need more information about the redshift of the sources contributing to the
CIB. We also need a method that allows to go deeper than the confusion noise level. In this context,
 an invaluable tool is the stacking technique, which allows a statistical 
study of groups of sources that cannot be detected individually at a given wavelength. Its requires
the knowledge of the positions of the sources being ``stacked''
as inferred from their individual detection at another wavelength. This knowledge
is then used to stack the signal of the sources at the wavelength
at which they cannot be detected individually. Since the signal of the
sources increases with the number of sources N and the noise (if
Gaussian) increases with $\sqrt{N}$, the signal-to-noise ratio
will increase with $\sqrt{N}$. For an additional description of the basics of stacking techniques we
refer to for example \citet[][]{2006A&A...451..417D} and \citet[][]{2009arXiv0904.1205M}.\\ 

Stacking was used to measure the contribution
of 24~$\mu$m galaxies to the background at 70 and 160~$\mu$m using
MIPS data \citep{2006A&A...451..417D}. Contribution from galaxies
down to 60 $\mu Jy$ at 24~$\mu$m is at least 79\% of the 24~$\mu$m,
and 80\% of the 70 and 160~$\mu$m backgrounds, respectively. At longer wavelengths
studies used this technique to determine the contribution
of populations selected in the near- and mid-infrared to the FIRB (far-infrared background) background: 
3.6~$\mu m$ selected sources to the 850~$\mu$m background \citep{2006ApJ...647...74W}
and 8~$\mu$m and 24~$\mu$m selected sources to the 850~$\mu$m
and 450~$\mu$m backgrounds \citep{2006ApJ...644..769D, 2008MNRAS.386.1907S}. 
Finally, \citet[][]{2009arXiv0904.1205M} measured total submillimeter intensities 
associated with all 24~$\mu$m sources
that are consistent with 24 micron-selected galaxies generating the full intensity of the FIRB.
Similar studies with Planck and Herschel will provide even more evidence about
the nature of the FIRB sources.\\

Theoretically, a stacking technique also could be used to study
the mean SED (spectral energy distribution) of the stacked
sources \citep[e.g.,][]{2007ApJ...670..301Z}. 
The main potential limitations would be caused by
 the errors in the redshifts of the sources and an
insufficiently large number of sources to stack per redshift bin. 
The observation
of sufficiently large fields to which the technique can be applied is now
assured by the
to Spitzer legacy surveys FIDEL, COSMOS,  and
SWIRE\footnote{http://ssc.spitzer.caltech.edu/legacy/} 
and Planck and Herschel surveys. 
Advances in the measurement of the redshift have also been
accomplished, although for very small fields for 
sources up to $z\sim2$ \citep [e.g.][] {2006ApJ...637..727C}, and for the
larger COSMOS fields up to $z\sim1.3$ with very high accuracy \citep{2009apj...690.1236I}. Future
surveys are planned to measure the redshifts
in larger fields such as the dark energy survey (DES\footnote{http://www.darkenergysurvey.org/}) or the GAMA
spectroscopic survey \citep [e.g.][] {2008IAUS..245...83B}.\\

The difficulties in separating the contribution to the signal coming from different redshifts
have handicapped the study of CIB anisotropies. 
However, once the mean SEDs of infrared galaxies per redshift bin are obtained we can use this
information to analyze CIB anisotropies.
The SEDs obtained with the stacking technique
can be used to ``clean'' the low-redshift anisotropies (or at
least a significant part of them) from the CIB maps. 
This can be performed by subtracting the undetected low-redshift ($z<1-2$) populations from
the maps using their mean colors and thus build maps dominated by sources at higher redshifts.
This also facilitates the study of the evolution of large-scale structures at high redshift by removing
the noise coming from low redshifts. \\

In this paper, we use the simulations and catalogs presented in
\citet{2008A&A...481..885F}\footnote{The simulations are publicly available at http://www.ias.u-psud.fr/irgalaxies}
to study the limitations of stacking techniques in CIB anisotropy analysis. 
We stack 24~$\mu$m sources detected with MIPS in Planck, Herschel, and SCUBA-2 simulated observations. 
The catalogs and maps were created for
different levels of bias between the fluctuations of infrared galaxy emissivities and the dark matter density field. 
We use a bias $b=1.5$, which is very close to that measured by \citet{2007ApJ...665L..89L}.\\

The paper is organized as follows. In Sect. \ref{sec:Methods}, we 
explain the method used to study the capabilities of the stacking once the redshift of the
sources is known. Section \ref{sec:Limits} details the elements that limit
the accuracy of the stacking technique. In Sect. \ref{sec:Test}, we test the
technique for studying the mean SEDs of galaxies. In Sect. \ref{sec:Cleaning}, the
feasibility of using information about the SEDs to clean the
observations of low-redshift anisotropies is studied. The results are summarized
in Sect. \ref{sec:Summary}. Throughout this paper, the cosmological
parameters are assumed to be $h=0.71,\Omega_{\Lambda}=0.73,\Omega_{m}=0.27$.
For the dark-matter linear clustering, we set the normalization to be
$\sigma_{8}=0.8$.

\section{Description of the method\label{sec:Methods}}

\citet{2006A&A...451..417D} considered every MIPS 24~$\mu$m source in selected fields
with fluxes  $>$60~$\mu$Jy and then sorted the 24~$\mu$m
sources by decreasing flux at 24~$\mu$m (hereafter $S_{24}$). 
The sources were placed in 20 bins of increasing flux density. These
bins were of equal logarithmic width $\bigtriangleup S_{24}/S_{24}\sim0.15$,
except for the bin corresponding to the brightest flux, to
take all the bright sources. They then corrected the average flux obtained
by stacking each $S_{24}$ bin for incompleteness using the correction
of \citet{2004ApJS..154...70P}. This allowed them to determine lower limits
to the CIB at 70~$\mu$m and 160~$\mu$m, and to find the contribution from galaxies
down to 60~$\mu$Jy at 24~$\mu$m to be at least 79\% of the 24~$\mu$m,
and 80\% of the 70 and 160~$\mu$m backgrounds. \\

While these measurements of the total flux are useful for estimating
the overall energy emitted by these populations \citep [see also][] {2009arXiv0904.1205M},
it does little to improve our knowledge of individual
sources. To use the average flux efficiently we have to decrease the
dispersion in the individual fluxes (at the long wavelength) around the average flux of the
population. We can do this by  separating large populations of sources 
into smaller and more homogeneous SED populations. \\

One of the main sources of flux dispersion is the measurement 
of the mean flux using galaxies with very different redshifts.
The lack of accurate redshifts (up to z$\sim$2) across large fields has so far limited 
the use of detailed redshift information in stacking analysis.
Because of this, the fluxes of sources with different SEDs are averaged together
and the mean flux is a poor estimator of the fluxes of individual
sources. However advances in the measurement of the redshifts are expected in the coming years
with the  new generation of spectroscopic and photometric redshift surveys such as 
GAMA \citep [e.g.][] {2008IAUS..245...83B}, (Big-)BOSS\footnote{http://www.sdss3.org/cosmology.php}, 
DES\footnote{http://www.darkenergysurvey.org/}.
We developed a method that assumes that redshifts are known and investigated
the limitations of stacking techniques caused by the uncertainties in the redshifts.
We assessed the dispersion in the fluxes
of individual sources with different redshift errors and the influence of this dispersion on the
quality of the results using our
simulations since this information will not be available in the real
observations.

\subsection{Stacking technique \label{sub:Stacking-technique}}

We used our simulations to study the limitations of the stacking technique using 24~$\mu$m MIPS
sources in Planck, Herschel, and SCUBA-2 observations. The choice of
this wavelength (24~$\mu$m) is motivated by several reasons. Firstly,
24~$\mu$m is a good tracer of infrared galaxies (unlike e.g., near-infrared
detections). Secondly, 24~$\mu$m-selected galaxies emit the bulk of the CIB up
to at least 500~$\mu$m \citep{2006A&A...451..417D, 2009arXiv0904.1205M}. Thirdly, 24~$\mu$m Spitzer
observations provide large and deep surveys, with redshift distribution
of its sources extending up to redshift $z\sim2.5$.
The schematic description of our stacking process follows. The only
requirements are knowledge of both the redshifts of the
sources and their fluxes at 24~$\mu$m.\\

The detected sources at 24~$\mu$m will be characterized by two parameters
$S_{24}$ and z. 
We first remove from the long wavelength map (hereafter $\lambda$ map)
the sources detected individually, using the criteria described in
\citet{2008A&A...481..885F}. These sources are no longer considered in
the discussion, so whenever we refer to sources 
we refer to those detected at 24~$\mu$m with
$S_{24}$ greater than the detection threshold and not those detected
individually in the $\lambda$ map.
The sources are then distributed into redshift bins.
The width of the redshift bins have to be optimized for each observation.
These bins cover the redshift interval between $z=0$ and $z=z_{max}$,
where $z_{max}$ is chosen depending on the goals of the work\footnote{We 
analyze the stacking up to $z_{max}=2$ since
reliable estimates of the redshift up to that
redshift are available (although over quite
small areas).}. We stack independently the sources in each redshift slice. For the
sources in a given redshift slice $i$ ($z_{Slice}^{i}$), the process
of detection is as follows:

\begin{enumerate}
\item Firstly, we order the sources by decreasing $S_{24}$.
We start by stacking in the $\lambda$ map the sub-images of the two
sources with higher $S_{24}$ (that have not been detected individually).
Then we measure the signal-to-noise ratio of the resulting image. A detection
is achieved when the signal-to-noise ratio is higher than a certain
detection threshold. This detection threshold is optimized for different
observations. For the cases discussed in this paper, we use a
detection threshold of three.
If we do not achieve a detection we stack more sources
 (always selecting the next brighter sources at 24 $\mu
m$)\footnote{To decrease the computation time, we increase the number of sources
to be stacked using a logarithmic step of $dN/N=~1.5$.}. 
This is done until we attain the required signal-to-noise ratio. 
\item Once a detection is achieved, we assign to all sources stacked
together a flux equal to the total flux measured in the stacked image
divided by the number of sources.
\item After detection, we restart the process starting from the brightest
sources that we have not yet stacked.
\item Sometimes the last (and therefore faintest) group of sources in the
redshift slice is not successfully stacked by this algorithm because
an insufficient number of faint sources remains to
be stacked in this last iteration. To correct for this, we simply carry
out the algorithm starting this time from the faintest sources and
stacking progressively brighter sources until we achieve a detection.
Although in this
procedure the last two mean flux bins are not independent, the
consequences in terms of systematic errors are negligible (since the
sources affected are few, faint, and the relative error in the stacking
is small).
\end{enumerate}
Once this process is complete we assign a mean
"stacked'' flux to every source of the redshift slice. 
The errors in the fluxes of the sources measured by 
stacking are computed to be the total noise measured in the map 
(following the method described in \citet{2008A&A...481..885F}) 
multiplied by $\sqrt{N}/N$, where N is the number of stacked sources.
We repeat
this process for all the redshift slices until we have a measurement
of the flux at $\lambda$ for all the sources in the catalog. In the
3 dimensional space of $S_{24}-z-S_{\lambda}$, we then have a
set of points $S_{24}^{St}-z^{St}-S_{\lambda}^{St}$ corresponding
to different successful stackings. For each successful stacking, the
coordinates in each of the three axes are the following:
\begin{itemize}
\item $S_{\alpha}^{St^{i}}$: The mean $S_{\alpha}$ of the sources
of the $i^{th}$ stacked population, where $\alpha$ is the reference
wavelength (here 24~$\mu$m). 
\item $z^{St^{i}}$: The mean redshift of the sources of the $i^{th}$
stacked population. 
\item $S_{\lambda}^{St^{i}}$: The mean $S_{\lambda}$ found for the sources
using the stacking technique for the $i^{th}$ stacked population.
\end{itemize}

\paragraph{Redshift slice optimization:}
Our algorithm assumes that sources at similar
$z$ and of similar $S_{24}$ have similar characteristics
at other wavelengths. Our
best option to avoid substantial variance in $S_{\lambda}$ between the
stacked sources is to try to avoid stacking together
sources of very different $S_{24}$ or $z$. 
In this context, the size of
the redshift bins were empirically optimized to ensure that (1)
our detections are of high signal-to-noise ratio, (2)
we achieve successful detections in each redshift slices
without having to stack together sources of very different $S_{\alpha}$
(by more than a factor of three), and (3) the redshift slices are
as thin as possible while complying with the first conditions.
The redshift slices are chosen differently for each observation
to comply with these criteria.

\subsection{Color smoothing\label{sec:Smoothing}}

\begin{figure}
\begin{centering}
\includegraphics[width=6.cm]{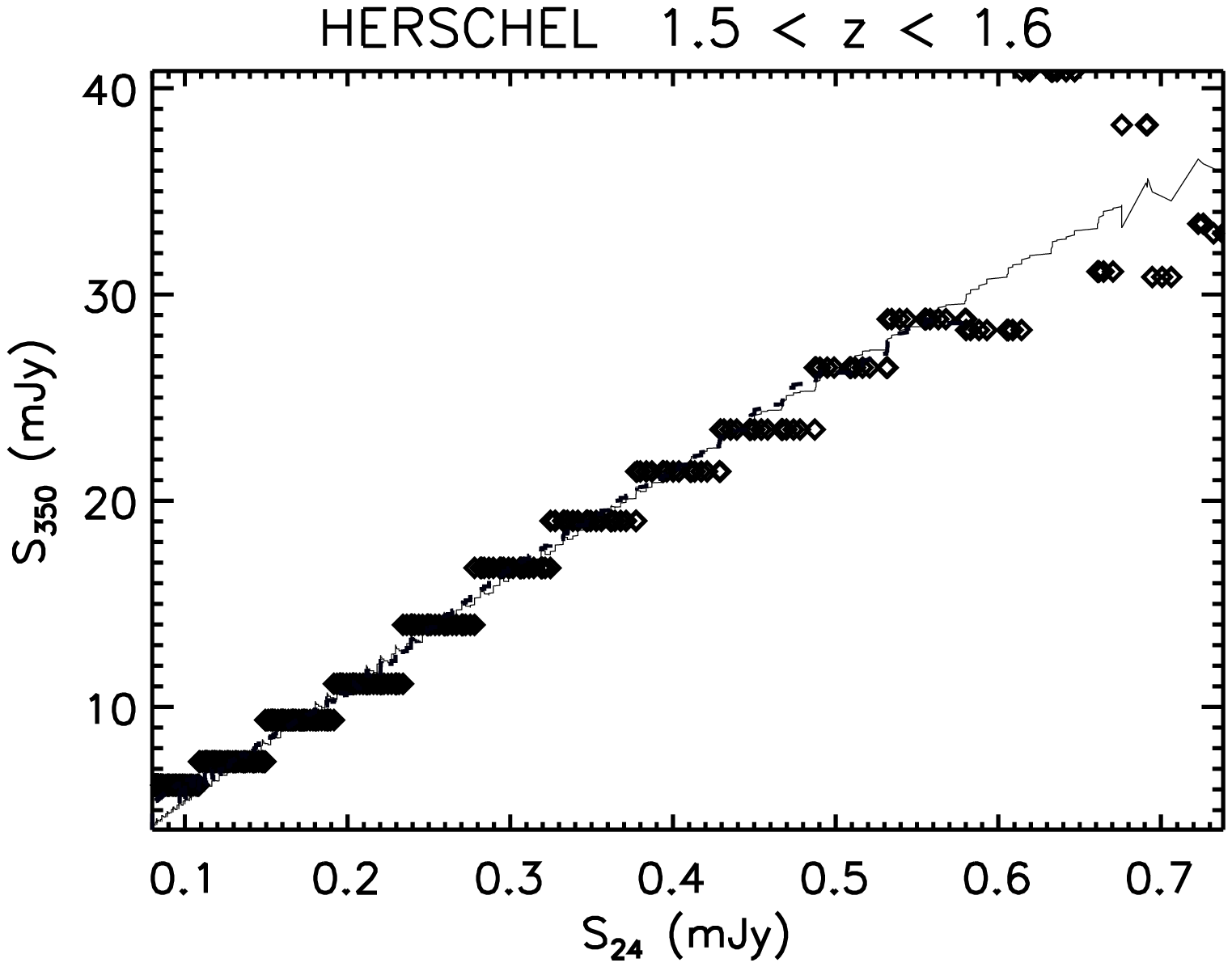}
\includegraphics[width=6.cm]{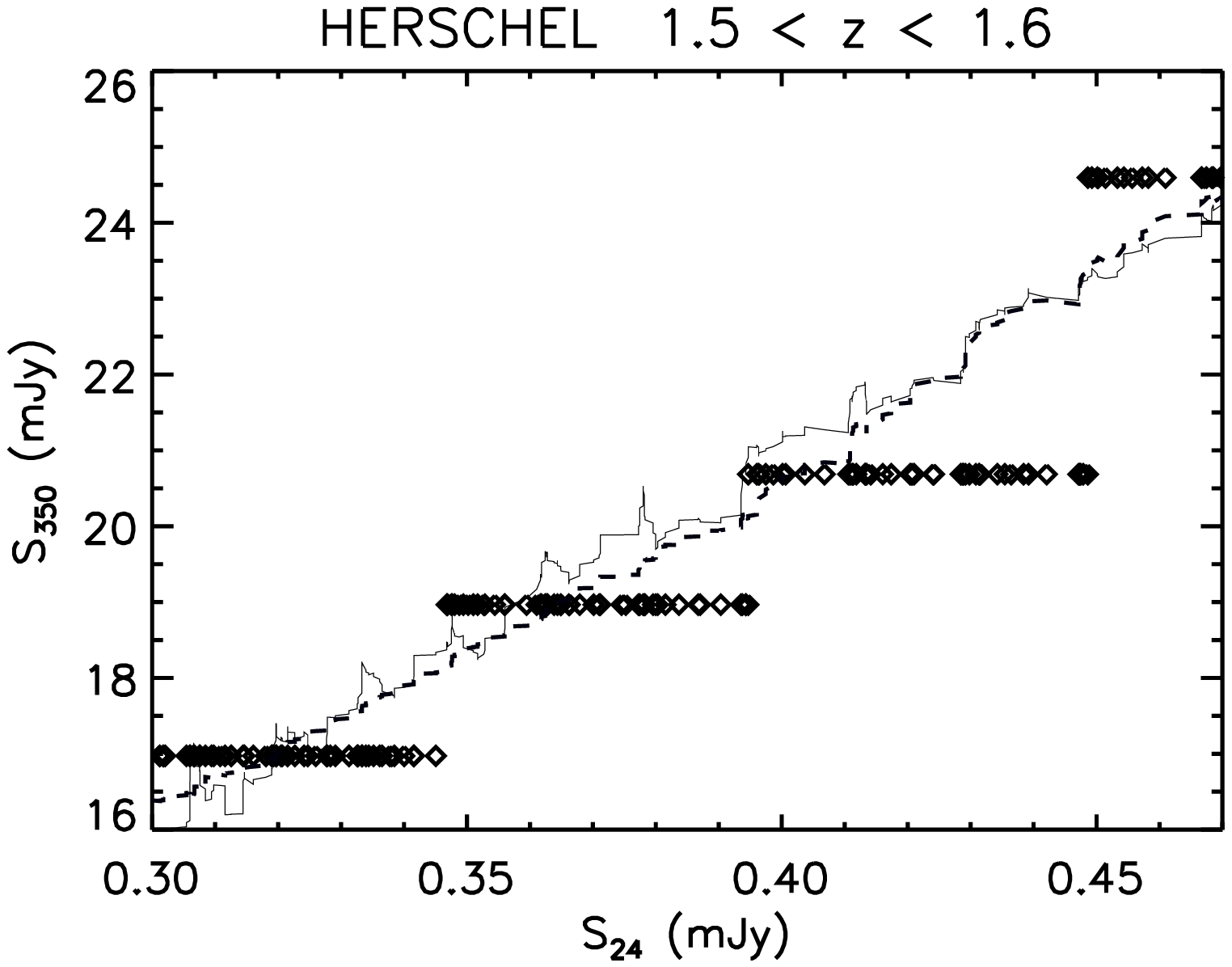}
\par\end{centering}
\caption{Top: Input fluxes of the sources in the redshift slice $1.5<z<1.6$ (solid line) together with
estimates of the fluxes of the sources using the smoothing technique
(dashed line) and estimates of the fluxes of the sources without
smoothing (diamond). Bottom: The same but zoomed for $0.3\, mJy<S_{24}<0.47\, mJy$.\label{fig:Smoothing Cut}}
\end{figure}

The algorithm discussed above is quite simplified because it
assumes that all sources detected in the same redshift bin
have the same color $S_{\lambda}/S_{\alpha}$. In contrast we would expect there to be a continuous
variation of  $S_{\lambda}/S_{\alpha}$ with both $S_{\alpha}$ and $z$. 
Following this assumption allows us to interpolate
values between detections at different $S_{\alpha}$ for each
redshift slice. 
A more complicated means of correction is to
smooth our predictions by interpolating $S_{\lambda}$ through the
grid formed by the set of points $S_{\alpha}^{St}-z^{St}-S_{\lambda}^{St}$
found with the stacking algorithm described above for the whole $S_{\alpha}-z$
plane. We do this with the IDL function TRIGRID, which given data points
defined by the parameters $S_{\alpha}^{St}-z^{St}-S_{\lambda}^{St}$
and a triangulation of the planar set of points determined by $S_{24}^{St}$
and $z^{St}$ returns a regular grid of interpolated $S_{\lambda}$
values. We tried both approaches and found that the differences between 
the results for the two different smoothings is very small so from now on we use only the  ``$S_{\lambda}$ smoothing''.
Figure \ref{fig:Smoothing Cut}.a shows the fluxes at 350~$\mu$m (with $1.5<z<1.6$) before and after the
two dimensional smoothing. It shows the real fluxes of the sources (known from the simulations),
the recovered fluxes using the smoothing 
technique, and the recovered fluxes without
smoothing. We can see that the smoothing greatly improves
the accuracy of the fluxes. After this
correction, the results are in very good agreement with the input fluxes.

\section{Limitations of the method\label{sec:Limits}}

We now test the limitations of the method related to 
the difficulties we expect to
face when real data are analyzed (e.g., intrinsic dispersion in the colors of the sources,
errors in the measurement of the fluxes and in redshifts, clustering)\footnote{The
problems associated with errors in the measurement of $S_{24}$
are considered negligible \citep [see][] {2007ApJS..172...86S}.}.
To illustrate the limitations, in this section we use the simulations
at 350~$\mu$m. We reached the same conclusions using other
far-infrared and submillimeter wavelengths. The size of the redshift slices that divide the
$S_{24}-z$ space was chosen to be $dz=0.1$;  
wider redshift slices would stack together sources with very different fluxes; smaller
redshift slices led to too low signal-to-noise ratios.\\

Two different Spitzer surveys are used, COSMOS and SWIRE. COSMOS
is a deep observation with a completeness of $\sim$100\%
up to $S_{24}=80\,\mu$Jy \citep{2007ApJS..172...86S}. It allows
us to test the stacking of faint sources. COSMOS covers a smaller field than SWIRE (2 sq. deg. 
versus ~50 sq. deg.) hence its stacking measurements are less accurate for bright sources.
Thus we also use the much larger SWIRE survey \citep {2004ApJS..154...54L}, which is less deep ($S_{24}>270\,\mu$Jy)
but covers $\sim$25 times more area\footnote{And therefore should have $\sqrt{N}=5$ times more signal-to-noise
ratio for similar populations of sources.}. 
We analyze the stacking of 24~$\mu$m sources 
for two study cases: observations in the far-infrared with Herschel
at 350$\,\mu$m and (in the next section) observations in the submillimeter with Planck
and SCUBA-2 at 850$\,\mu$m. 
The characteristics of the Herschel/SPIRE, Planck/HFI, and SCUBA-2 observations are the following:

\paragraph{Stacking in the COSMOS field:}
\begin{itemize}
\item Detection limit: $S_{24}^{D}>80\,\mu$Jy at $24\,\mu$m.
\item Size of the field: $2$ sq. deg. 
\item Linear bias: $b=1.5$.
\item Type of observation with Herschel: 350$\mu$m ``Deep'' (with  1$\sigma$=12.3~mJy).
\item Type of observation with SCUBA-2: 850$\mu$m (with  1$\sigma$=1mJy).
\end{itemize}

\paragraph{Stacking in the SWIRE fields:}
\begin{itemize}
\item Detection limit: $S_{24}^{D}>270\,\mu$Jy at 24$\,\mu$m.
\item Size of the field: $50$ sq. deg. 
\item Linear bias: $b=1.5$.
\item Type of observation with Herschel: 350$\:\mu$m ``deep'' (with 1$\sigma$=12.3~mJy).
\item Type of observation with SCUBA-2 and Planck: 850$\,\mu$m (with 1$\sigma$=1~mJy and 
1$\sigma$=46.7~mJy -- see Table 4 from \citet{2008A&A...481..885F}-- respectively).
\end{itemize}

\subsection{Cold and starburst populations}
\begin{figure}
\begin{centering}
\includegraphics[width=6.cm]{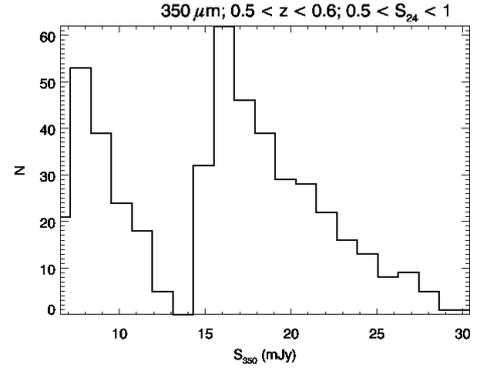}
\par\end{centering}
\caption{Histogram of the fluxes at 350~$\mu$m for a stacking box with $0.5<z<0.6$ and
$0.5<S_{24}<1$~mJy. The mean value of the sources 
is $S_{350}\sim17$~mJy. The two different populations
are the normal cold sources ({\it left} population) and the starburst sources
({\it right} population). It is clear that the main cause of error in our flux measurement  
comes from us stacking together two different
populations. As expected, we checked that reducing the redshift slice does not reduce
the dispersion.}
\label{fig:HistogramsStackBox2populations}
\end{figure}

\begin{figure}
\begin{centering}
\includegraphics[width=0.4\columnwidth]{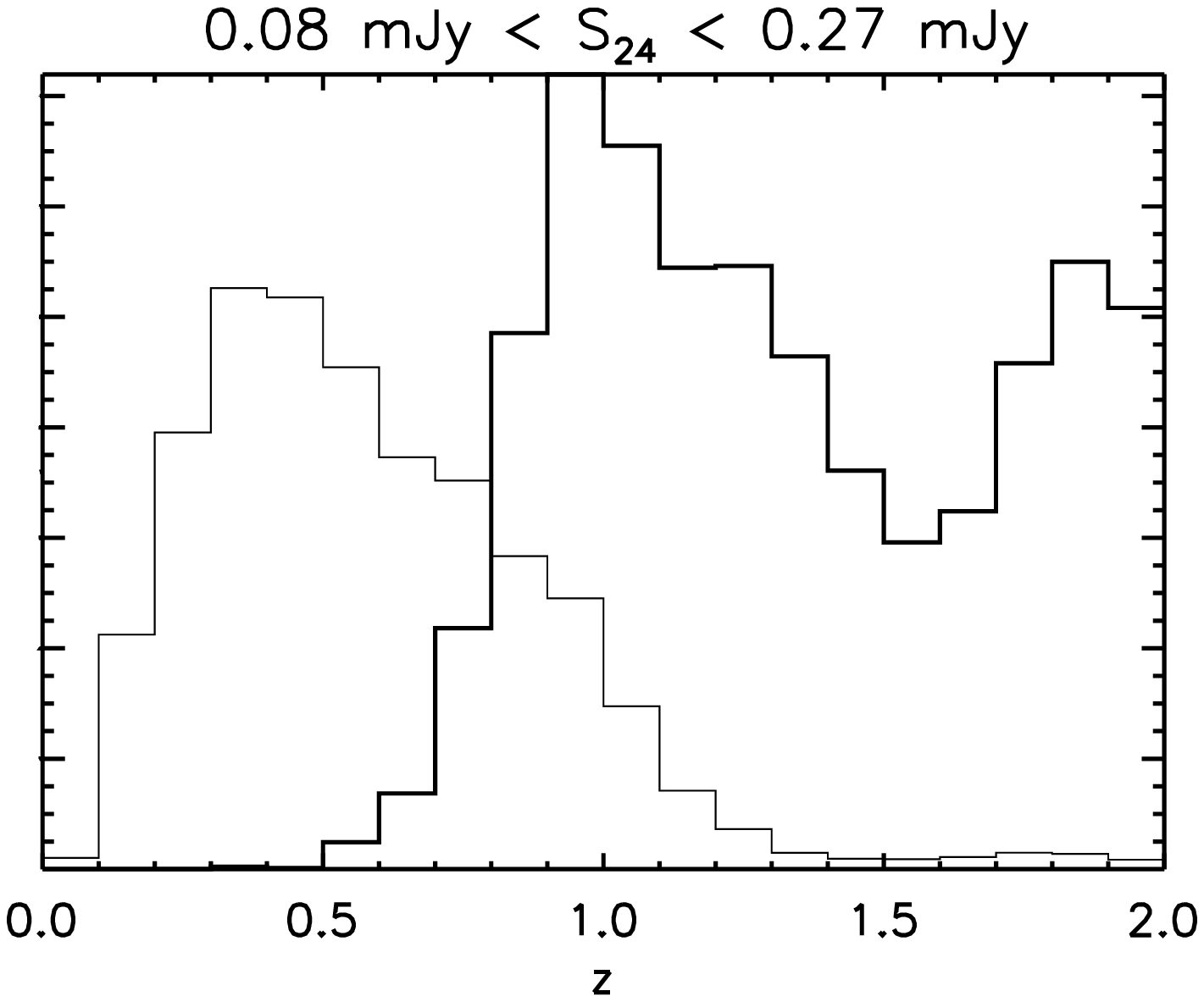}\includegraphics[width=0.4\columnwidth]{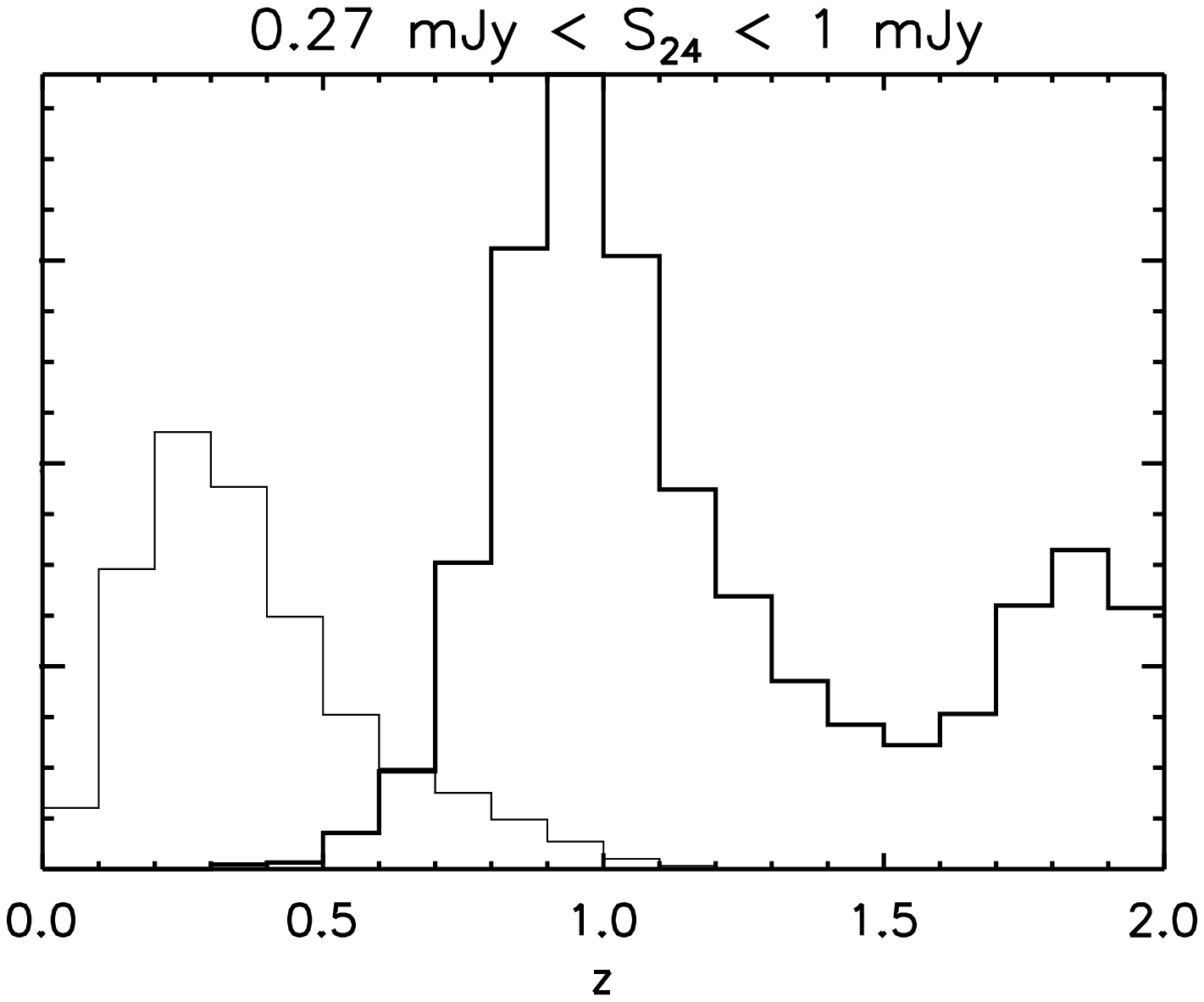}
\includegraphics[width=0.4\columnwidth]{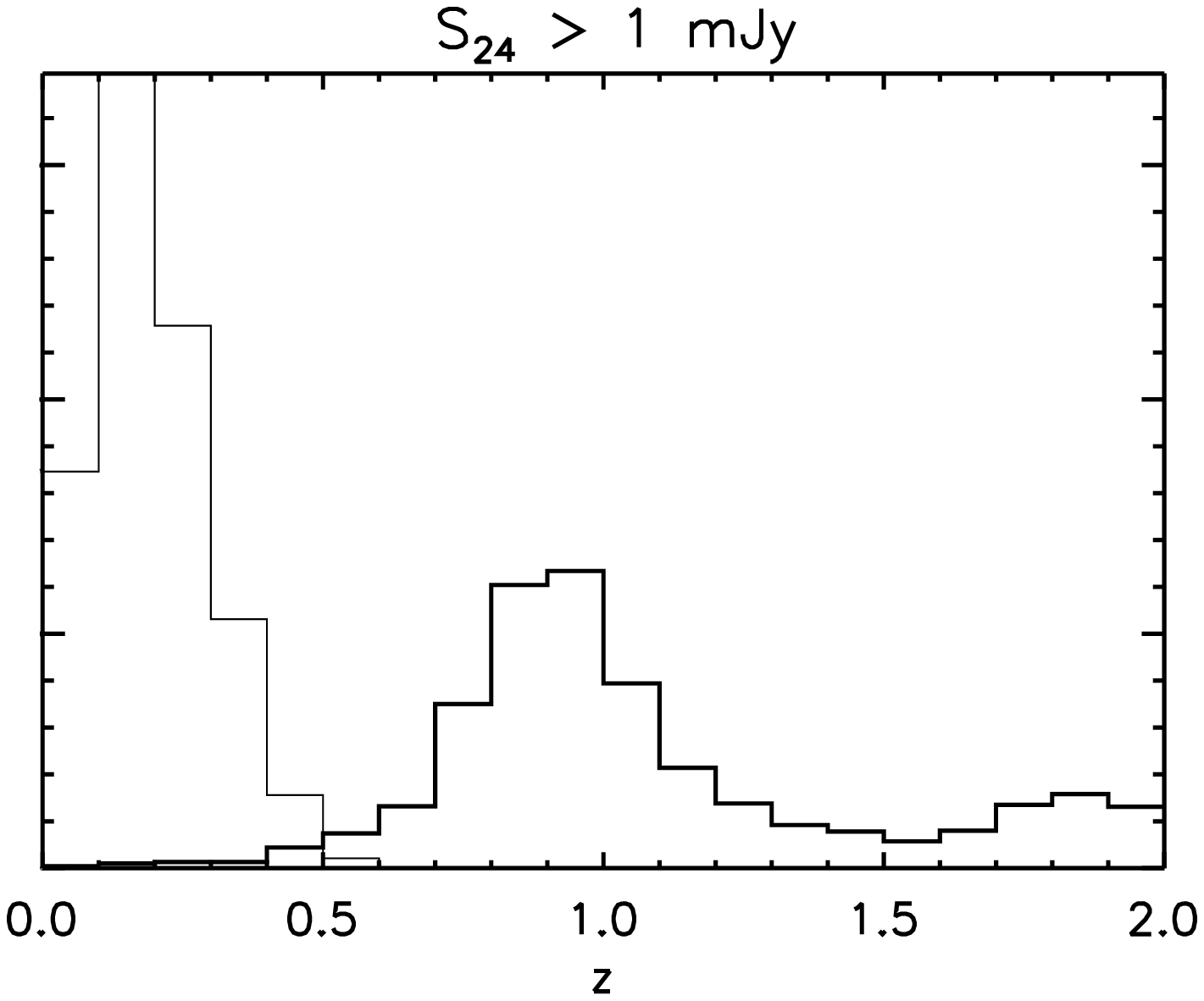}
\par\end{centering}
\caption{Histograms of the number of cold (thin line) and starburst (thick line)
sources per 24~$\mu$m flux bin (histograms are normalized to the higher number of sources per histogram). 
The problematic regions are those where both populations
have similar number of galaxies. This is especially important for
$0.6<z<0.7$ and faint sources. }
\label{fig:SBvsCOLD}
\end{figure}

Figure \ref{fig:HistogramsStackBox2populations} shows the 
histograms of the fluxes at 350~$\mu$m for a stacking box with $0.5<z<0.6$ and
$0.5<S_{24}<1$ mJy.
The main source of error in the estimate of the fluxes for this case would not be the dispersion in either $S_{24}$
or $z$ but the presence of two different populations, which are indistinguishable
using observations at shorter wavelengths. These two populations are
the starburst and the normal (cold) populations described in \citet{2003MNRAS.338..555L}.
Figure \ref{fig:SBvsCOLD} shows the number of starburst and normal
sources as a function of $z$ for sources with $80<S_{24}<270\,\mu$Jy,
$0.27<S_{24}<1$~mJy, and $S_{24}>1$~mJy. For the three afore mentioned cases, the cold sources
are the dominant population for $z<0.8,\, z<0.6$, and $z<0.5$ respectively.
There are no effective ways of separating these two populations, and this will cause poor estimates of
the mean colors of each population. This is particularly
important when the number of sources of each type is approximately equal. This
is because we add together two populations of very different $S_{350}$
(cold sources
are in general brighter in the submillimeter than starburst sources at the same redshifts
and with similar $S_{24}$).
When one of the populations dominates, this problem becomes negligible.

\subsection{Errors caused by intrinsic dispersion in colors \label{sec:dispersion}}
\begin{figure}
\begin{centering}
\includegraphics[width=0.7\columnwidth]{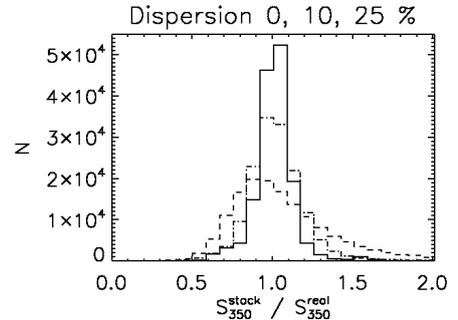}
\par\end{centering}
\caption{Ratio of recovered fluxes from stacking ($S_{350}^{stack}$) to input fluxes in the simulation ($S_{350}^{real}$)
for sources with $S_{24}>$270$\,\mu$Jy, with no additional  dispersion in the fluxes at 350~$\mu$m 
(thick solid line), and with $10\%$ (dotted-dashed line) and $25\%$ (dashed line) additional dispersion.  \label{fig:Histograms-of-the}}
\end{figure}

Because of the lack of constraints on SEDs at long wavelengths and their
evolution with redshift, 
the \citet{2004ApJS..154..112L} model does not take into account that galaxies of the same 
luminosity and redshift could have different
values of $S_{\lambda}$ (apart from the distinction between
  normal and starburst sources). To assess the effect of this
dispersion, we introduce a random Gaussian error into the flux estimated
with the stacking for each of the stacked sources. The errors that we 
make using this procedure are equivalent to those that we would make
if we were to use a model with an intrinsic Gaussian dispersion in the
$S_{\lambda}$ of the sources.  This type of error does not affect the results
for the mean of the sources but the average difference between this
mean and the fluxes of the individual sources. We test the effect on our results 
for different levels of dispersion (measured
in terms of the standard deviation in the dispersion compared to the mean
flux of the sources). In Fig. \ref{fig:Histograms-of-the}, we can
see the histograms of the errors for a dispersion of $0\%$, $10\%$, $25\%$ for
all sources with $S_{24}>270\,\mu Jy$. As expected, the figure illustrates how
the histograms broaden with dispersion. For a standard deviation in
the errors of the fluxes associated with the stacking $\sigma_{St}$ and
a standard deviation associated with the fluxes  $\sigma_{Disp}$ ,the
final standard deviation in our errors $\sigma_{Tot}$ would be  $\sigma_{Tot}=\sqrt{\sigma_{St}^2+\sigma_{Disp}^2}$.
We do not analyze other statistical representations of this effect
(i.e., non-Gaussian intrinsic dispersion) since we do not have any
strong observational constraints.

\subsection{Redshift uncertainty} 
The effect of redshift errors are difficult to evaluate. This is because they combine with 
the non-linear k-correction, making the variation in $S_{\lambda}$ with $z$ complex.
In Sect. \ref{sec:Test}, we study the effect of redshift errors 
for two different relative errors $\frac{\triangle z}{z}=3\%$ and $\frac{\triangle z}{z}=10\%$.

\subsection{Problematic areas of the S24-z space\label{sec:Problematic}} 

\begin{figure}
\includegraphics[width=4.3cm]{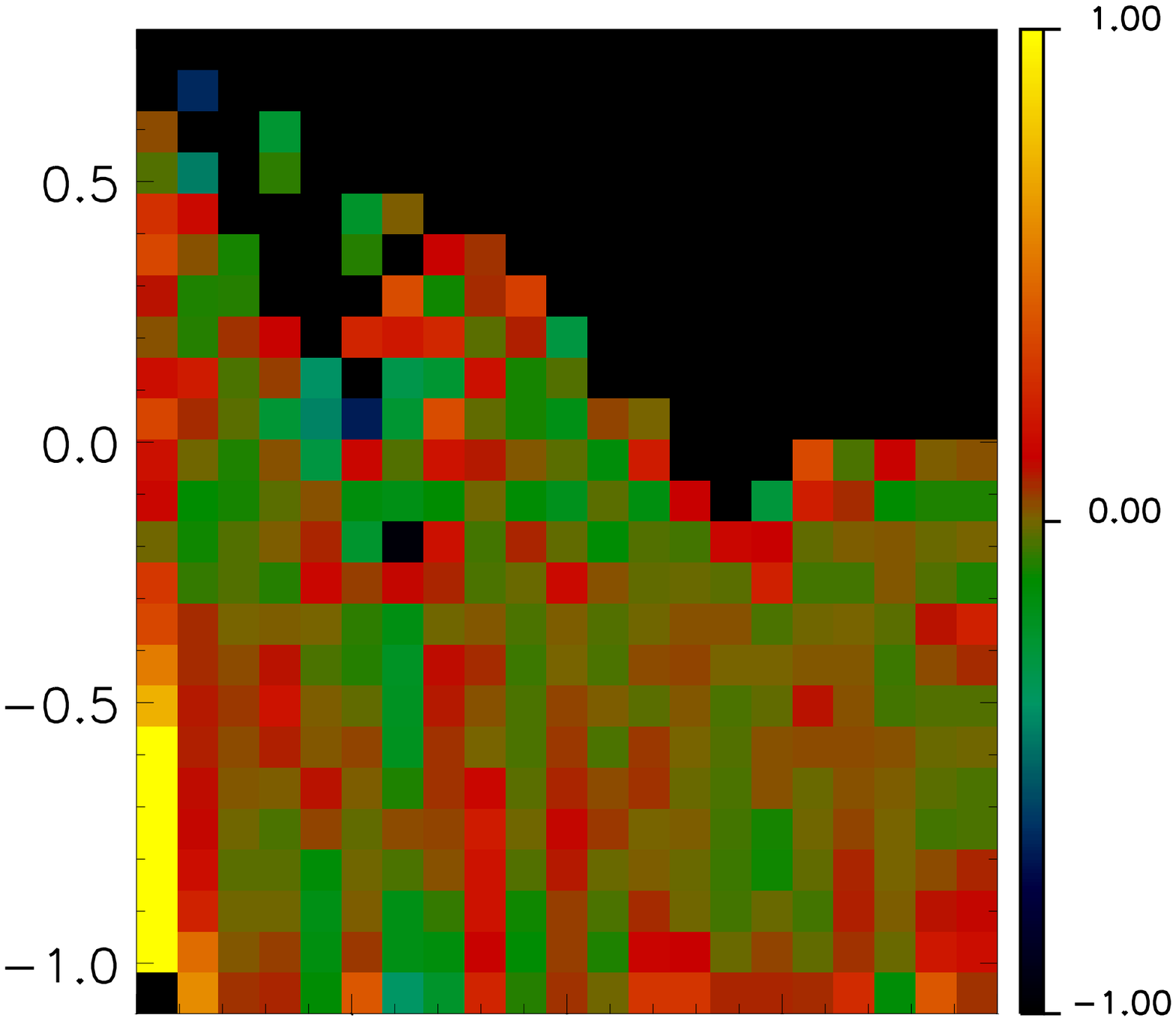}\includegraphics[width=4.3cm]{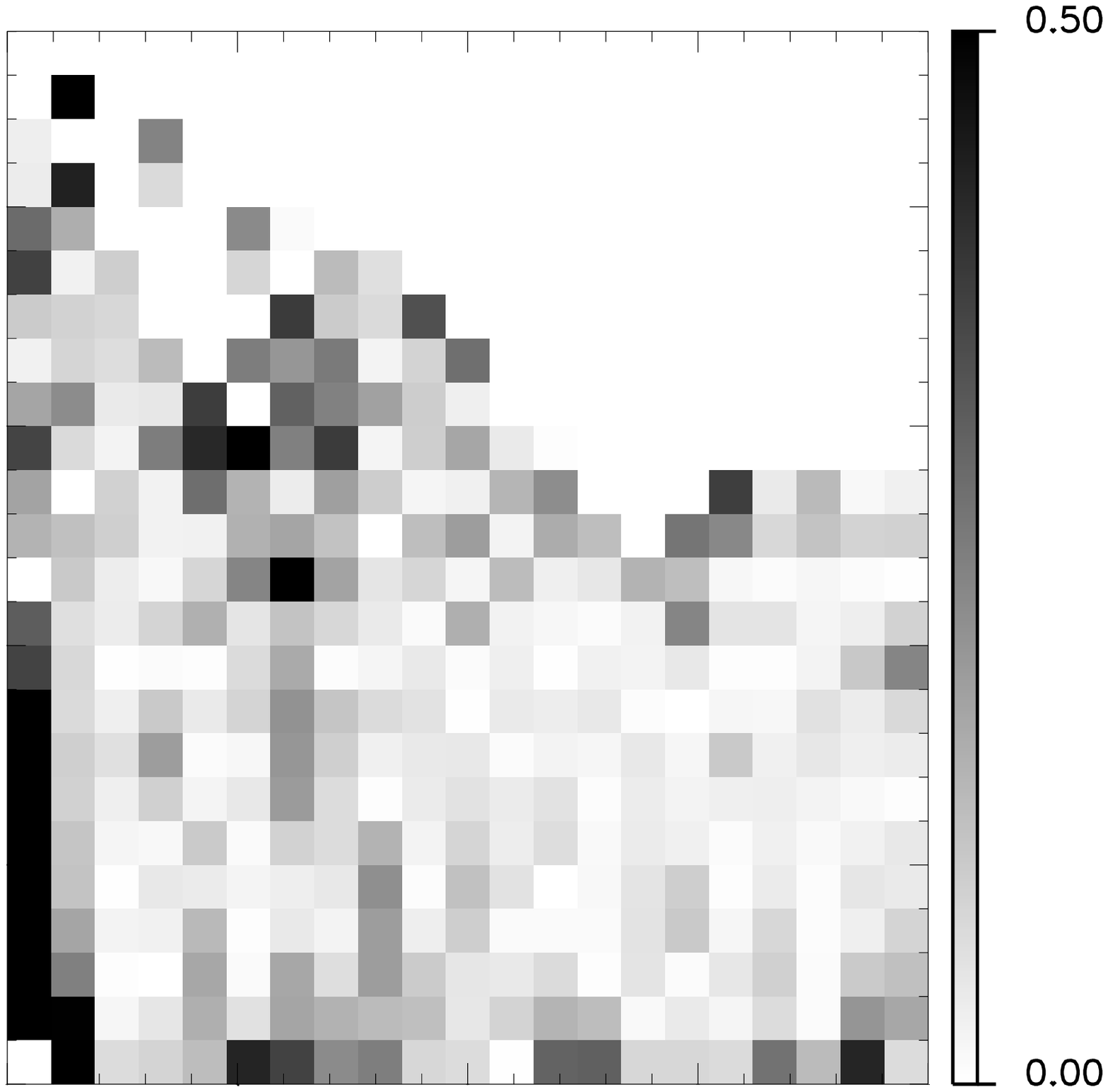}
\includegraphics[width=4.3cm]{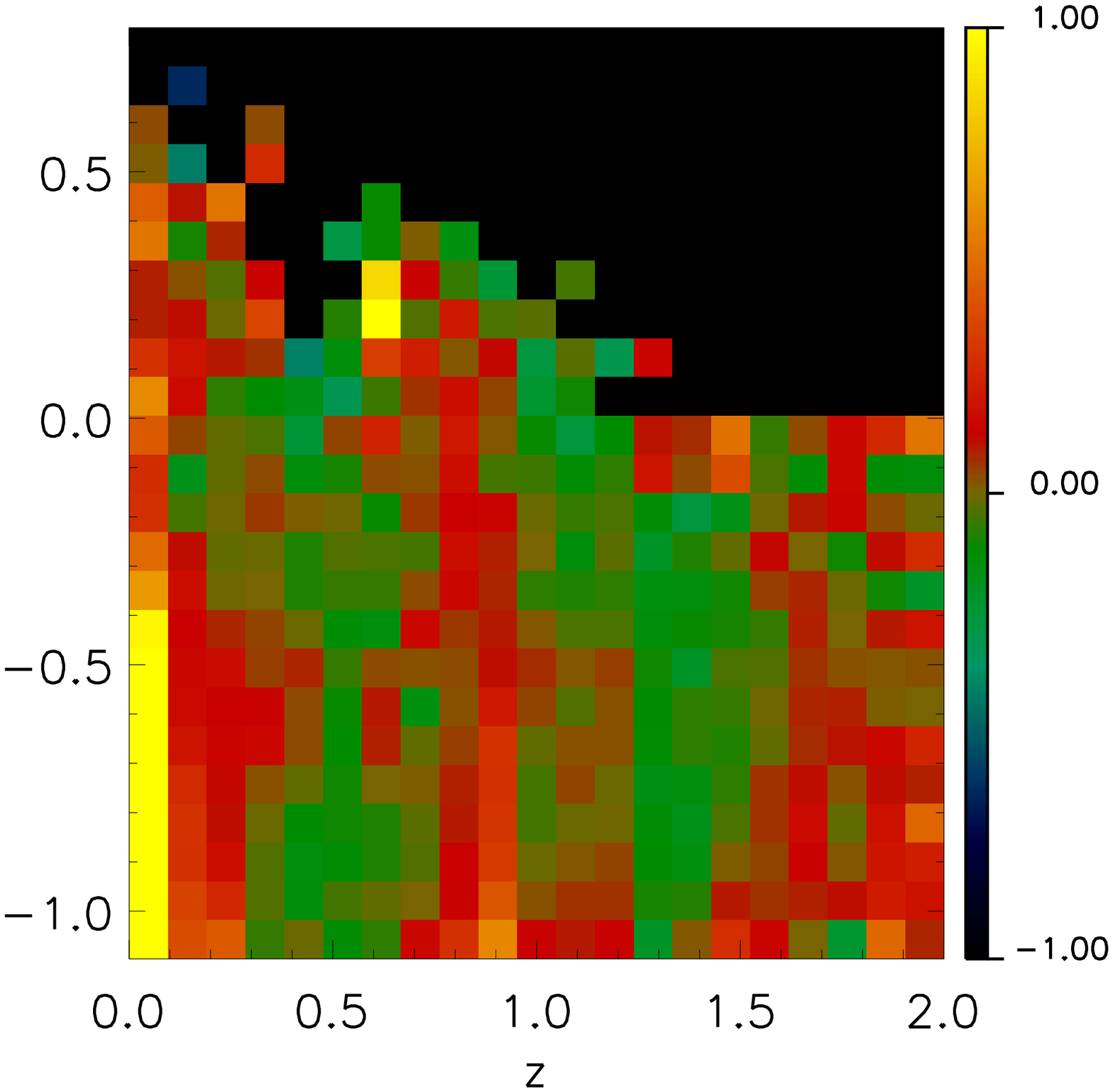} \hspace{0.2cm} \includegraphics[width=4.cm]{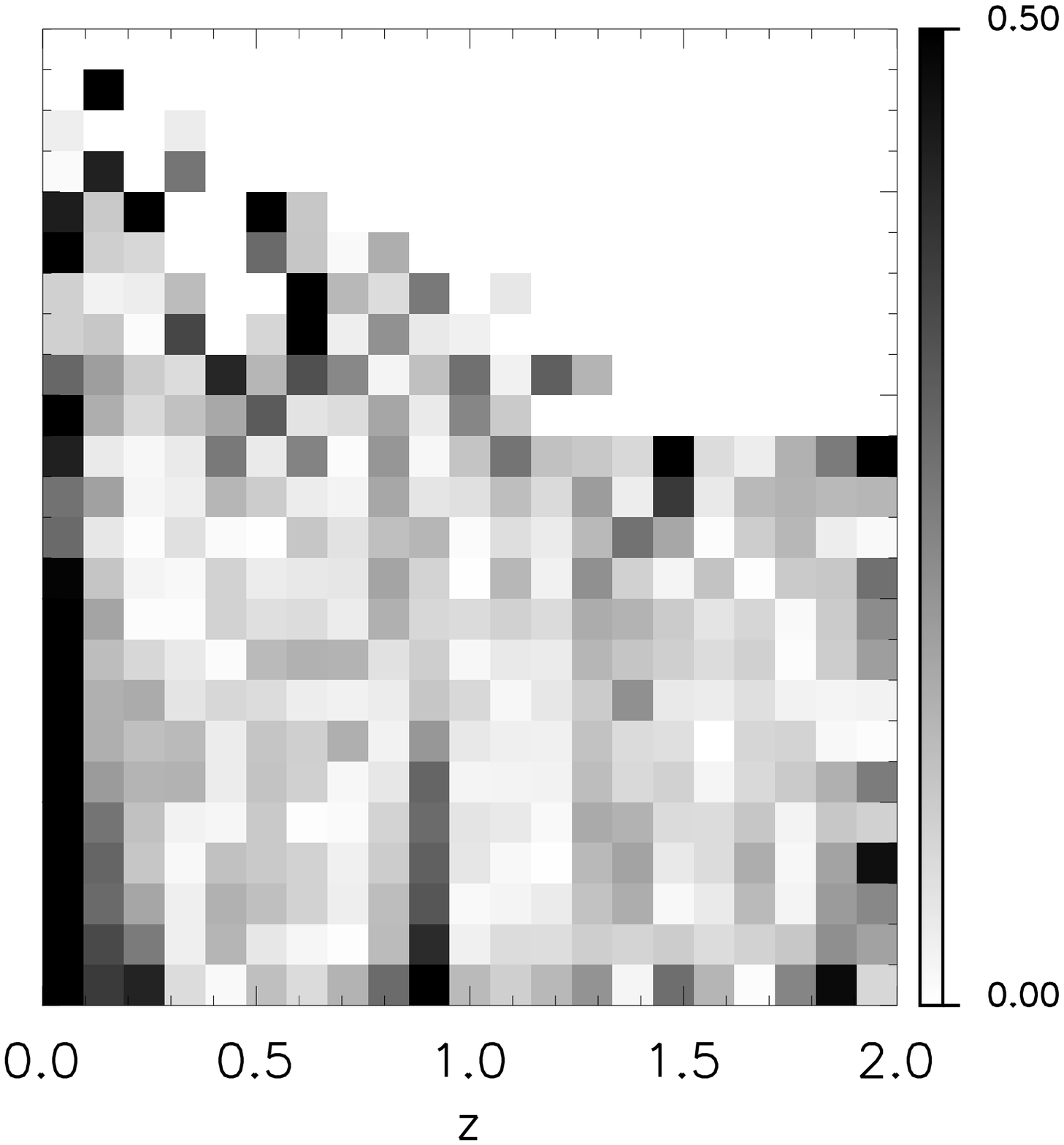}
\caption{Accuracy in the mean recovered fluxes at 350~$\mu$m in a COSMOS-like observation
when considering redshift errors of $\frac{\triangle z}{z}=3\%$ (top) and $\frac{\triangle z}{z}=10\%$ (bottom). 
The colors (shading) correspond to different
values of the accuracy, while the vertical axis is $S_{24}$ and the horizontal axis is the redshift bin.
Left: relative errors in the mean recovered
fluxes ($\nabla\bar{S}_{350}^{Stack}=(\bar{S}_{350}^{Stack}-\bar{S}_{350}^{Real})/\bar{S}_{350}^{Real}$)
for all the $S_{24}-z$ space. Right: the same
but in absolute values $\left|\nabla\bar{S}_{350}^{Stack}\right|=\left|(\bar{S}_{350}^{Stack}-\bar{S}_{350}^{Real})/\bar{S}_{350}^{Real}\right|$
and decreasing the dynamic range of the plot (0-0.5) to illustrate the
errors more clearly. The $S_{24}-z$ space is divided linearly in $z$ and logarithmically
in $S_{24}$. Redshifts are given on the bottom-right figure, $Log(S_{24})$ (in mJy) 
on the left figures.
\label{fig:Cosmos-observation-stacked-HArray}}
\end{figure}

Figure \ref{fig:Cosmos-observation-stacked-HArray} shows the errors
in the estimate of the mean fluxes in the $S_{24}-z$
space for a 350~$\mu$m Herschel observation of the COSMOS field
with redshift errors $\frac{\triangle z}{z}=3\%$ and $\frac{\triangle z}{z}=10\%$. 
For the estimates of the fluxes,
we can easily identify several problematic areas in the $S_{24}-z$ space.
These are: data points at very low redshifts ($z<0.1$), 
the brightest sources because of small number statistics and the faintest sources because
of flux errors\footnote{Note that the top right area with no data plotted corresponds to a region
where they are no sources at 24 microns; note also that in color representations as in Fig. \ref{fig:Cosmos-observation-stacked-HArray}, small differences in estimated value can have a great visual impact due to the variation in colors. A mere 20\% change in the 
estimate can change the color from green to red. The general variation is consistent with our detection threshold 
of 3$\sigma$.}.

\paragraph{Low z:}
There are very few sources at $z<0.1$. This prevents the stacking
from achieving high signal-to-noise ratio levels. This translates into large
errors in the measurement of the mean fluxes for sources with $z<0.1$.

\paragraph{Bright sources:}
These sources are rare and we are therefore unable to reach
signal-to-noise ratios as good as for fainter sources.  We expect
the results for bright sources to be better when the stacking technique
is applied to larger fields (for example using the WISE survey \citep{2005SPIE.5899..262M}).
We should keep this in mind when analyzing the results in our study cases.

\paragraph{Faint sources:}
Another shortcoming of the method is that the smoothing
techniques cannot be applied to sources fainter than the stacked
flux of the faintest bin. 
The best solution is to assume for the last point given by the stacking that all the sources have the same
color, which is equivalent to assuming that their color is the same
as that of the sources that are slightly brighter than them.

\section{Application of the method\label{sec:Test}}

We now verify the accuracy of the method with realistic
simulations of observations including redshift errors and by using existing observations
at 24~$\mu$m with Spitzer.

\subsection{Stacking Herschel data in the far-infrared: 350~$\mu$m}

We comment on the main issues and sources of error encountered
when stacking 24~$\mu$m sources in Herschel/Spire observations
at 350~$\mu$m and considering a detection threshold of
$3\,\sigma$. We note that the difficulties faced by the stacking technique 
at 250 and 500~$\mu$m are similar. 
We use a division in the z axis with redshift slices of $dz=0.1$. 
We analyze the results for two redshift errors, an 
optimistic one of $\frac{\triangle z}{z}=3\%$ and a pessimistic one of $\frac{\triangle z}{z}=10\%$.
This illustrates the degradation in the quality of the results with redshift error.

\begin{figure*}
\includegraphics[width=6.3cm]{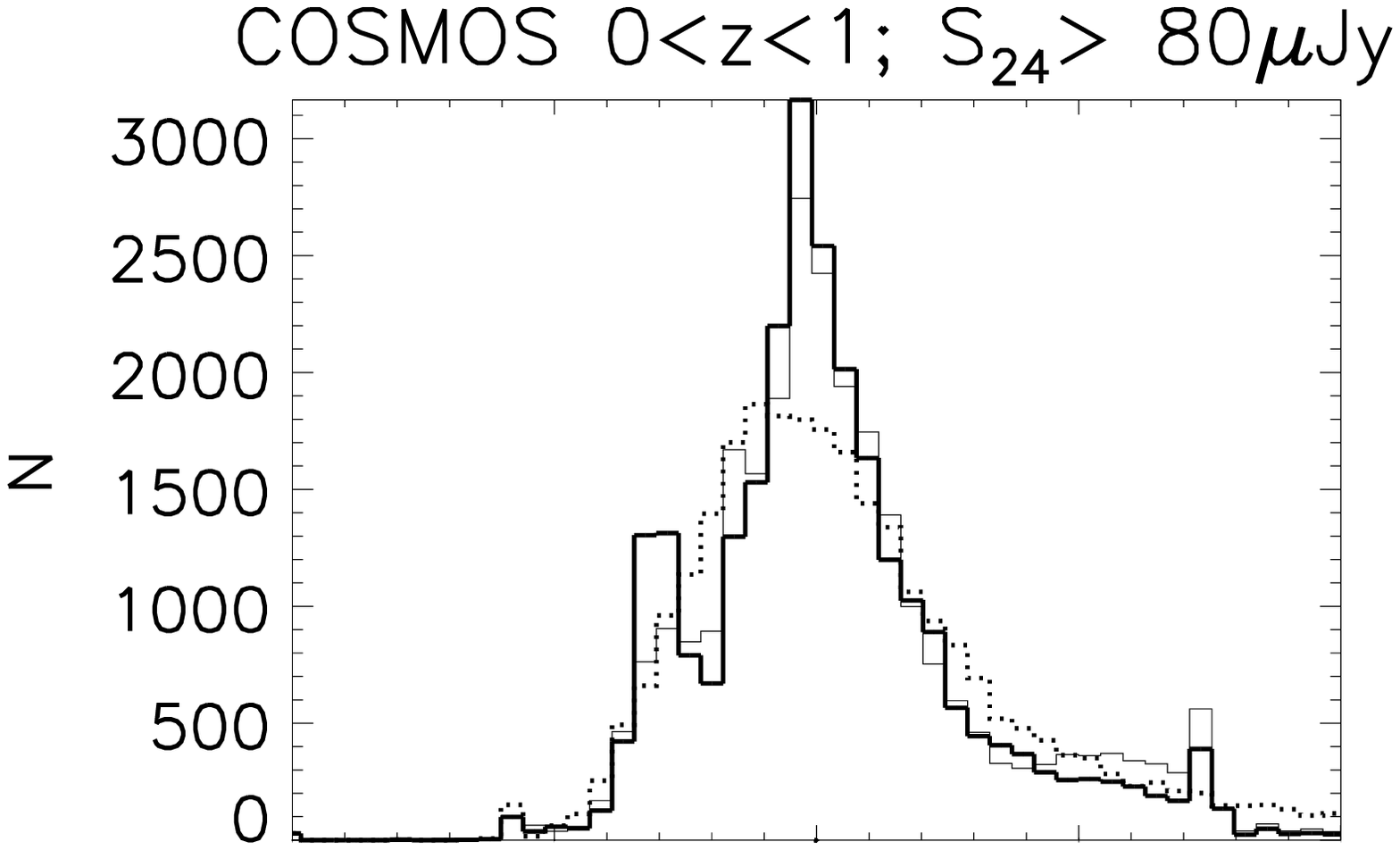}
\includegraphics[width=6.3cm]{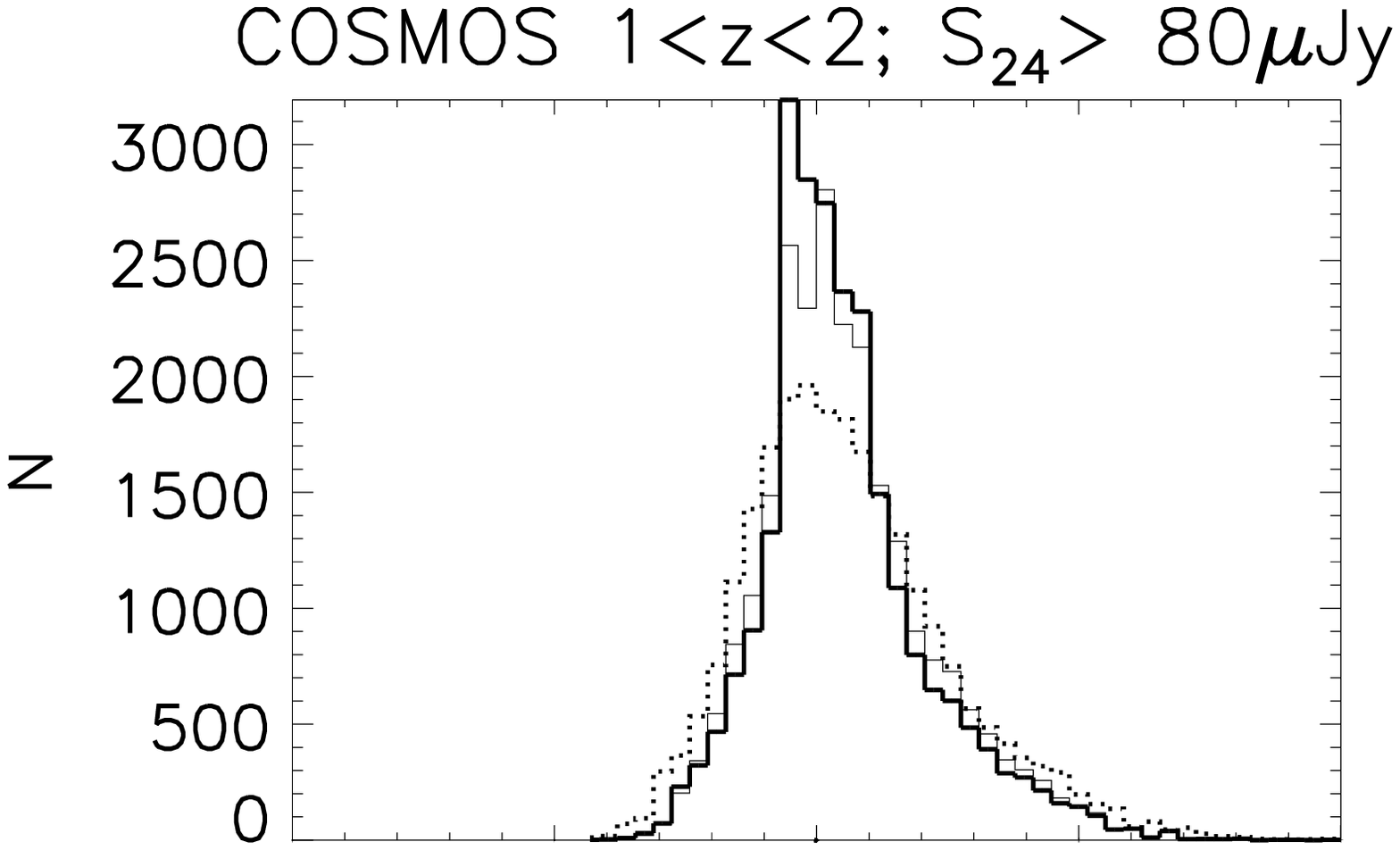}
\par\smallskip\noindent
\includegraphics[width=6.3cm]{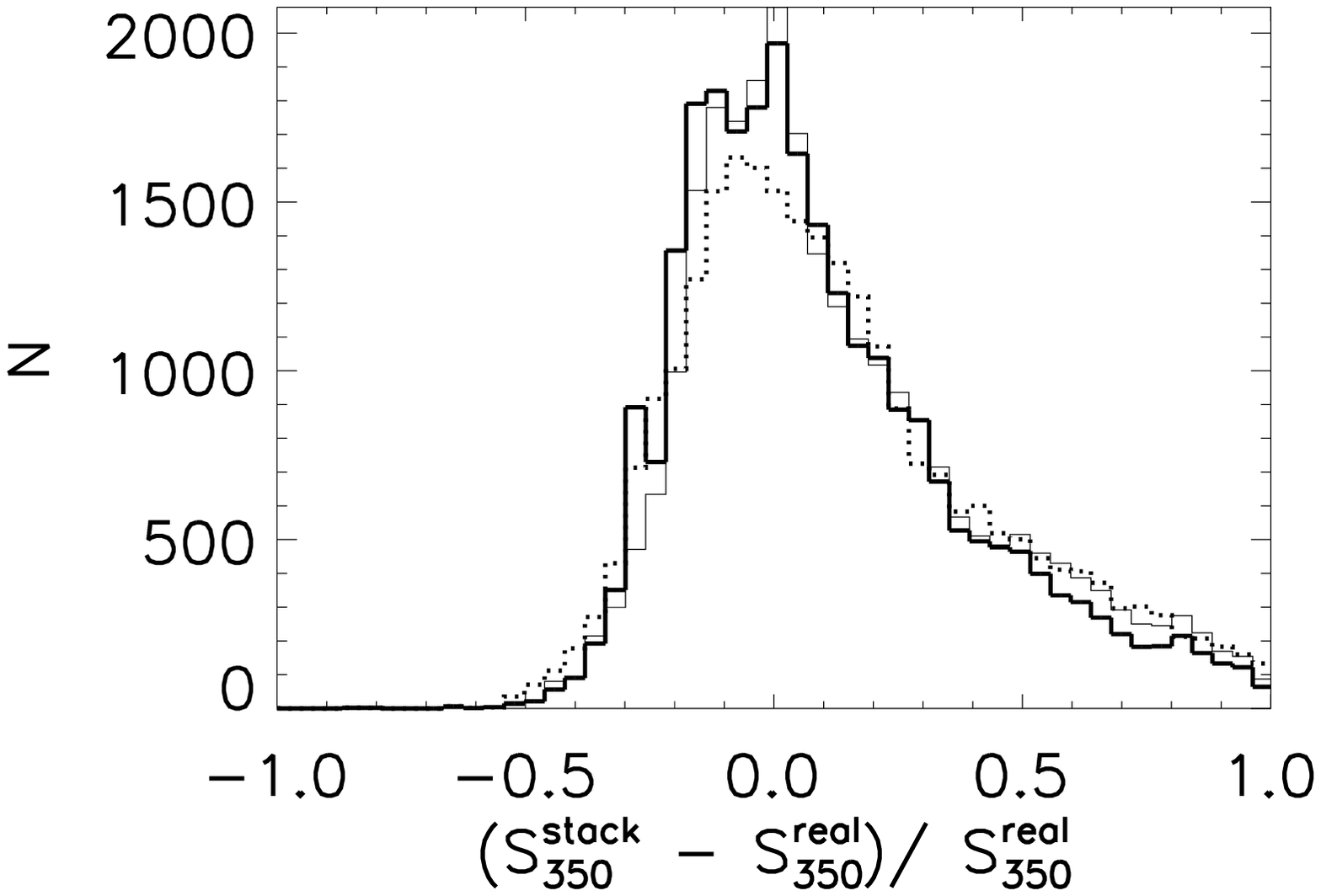}
\includegraphics[width=7.2cm]{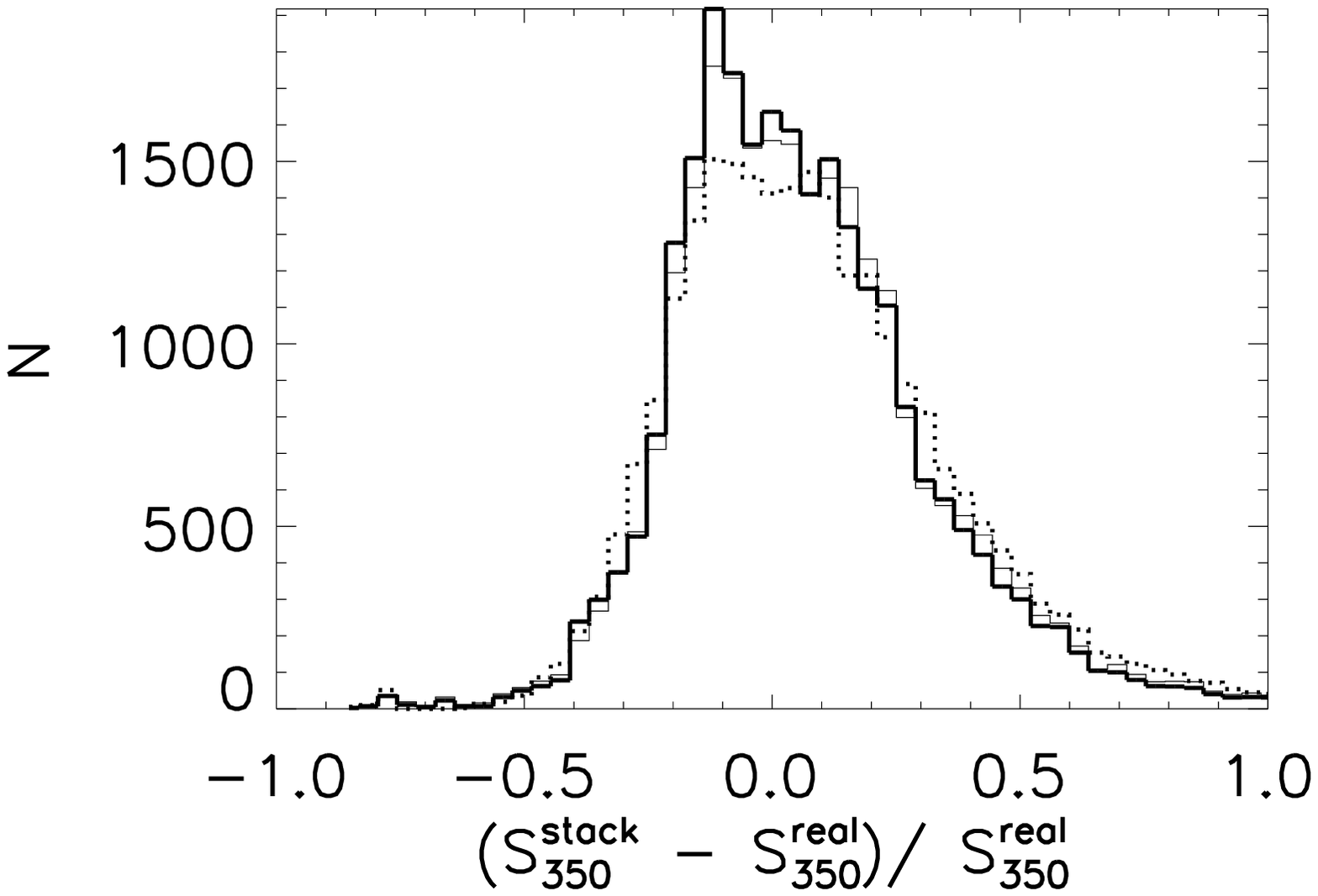}
\caption{Relative errors in recovered fluxes for individual sources at 350~$\mu$m in a COSMOS-like observation
for redshift
errors $\frac{\triangle z}{z}=3\%$ (top figures) and $\frac{\triangle z}{z}=10\%$ (bottom
figures). Left: for $S_{24}>80\,\mu$Jy and redshifts
$0<z<1$. Right:  for $S_{24}>80\,\mu$Jy and $1<z<2$. The zeros
represents a perfect estimate. Three estimates are shown: direct
values obtained with the stacking (dotted line); values obtained with the
stacking and smoothed in $z$ (thin solid line), and smoothed both in $z$ and in $S_{24}$ (thick
solid line). \label{fig:Cosmos-observation-stackedH3}}
\end{figure*}

\begin{figure*}
\includegraphics[width=6.5cm]{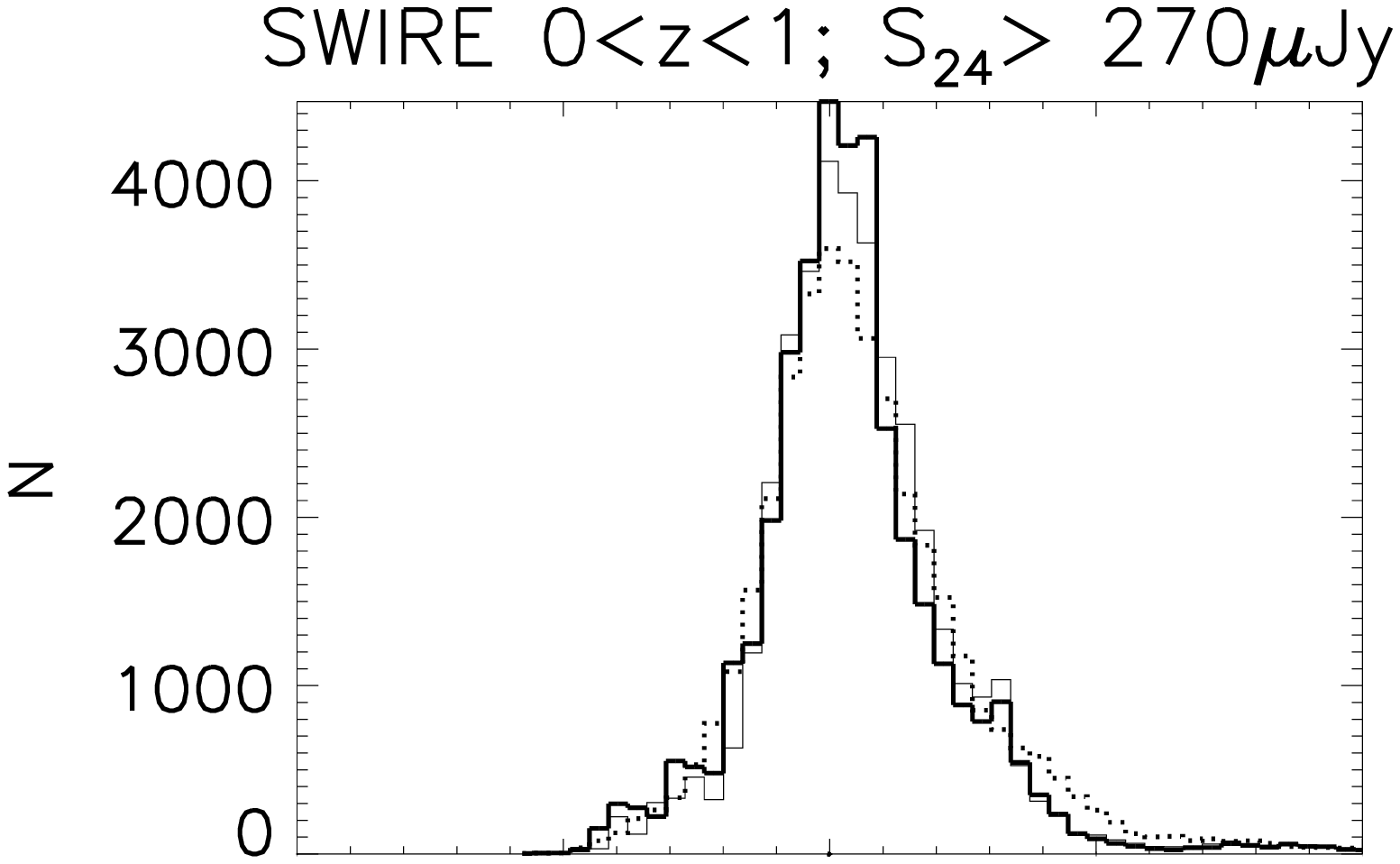}
\includegraphics[width=6.5cm]{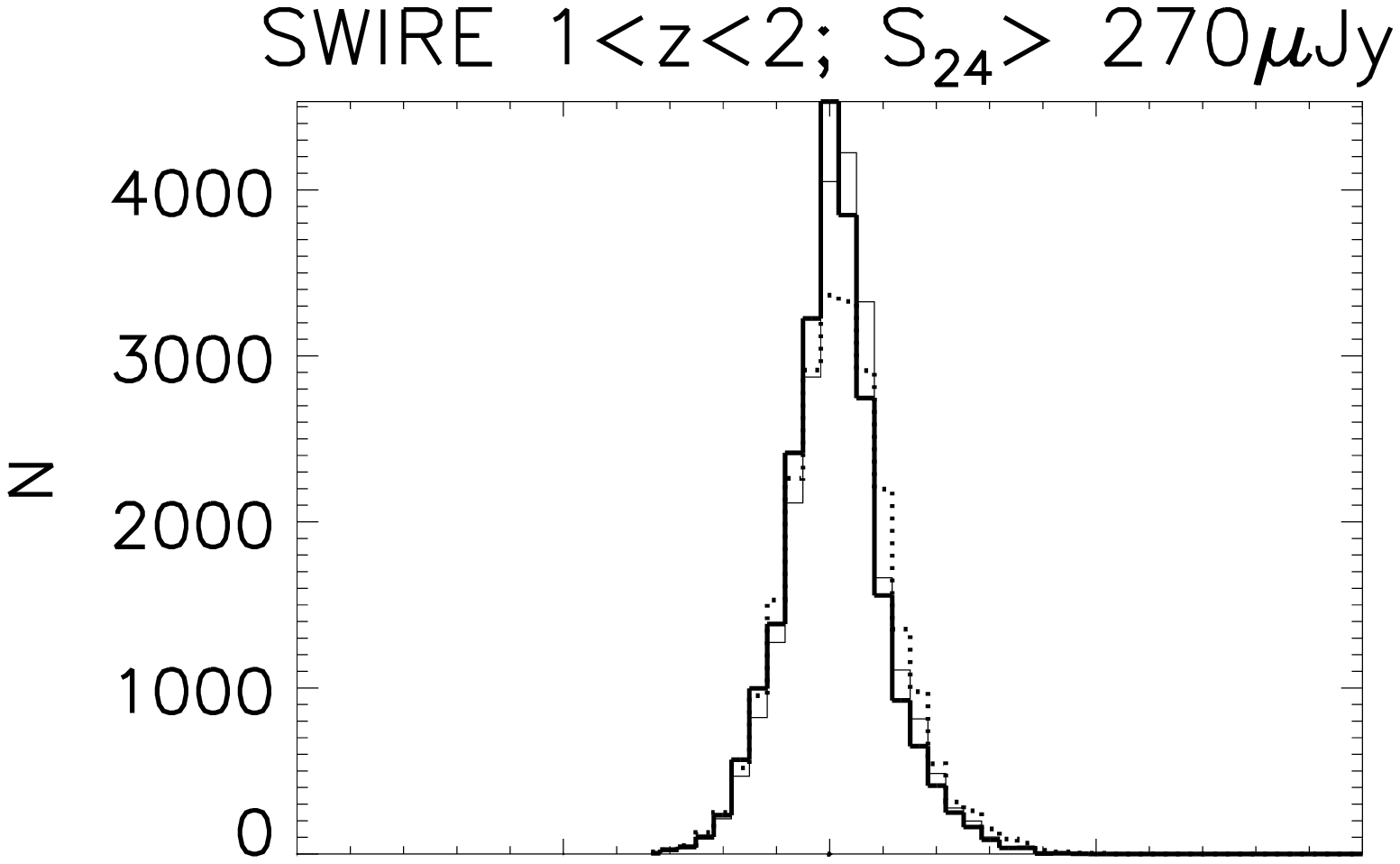}
\par\medskip\noindent
\includegraphics[width=6.5cm]{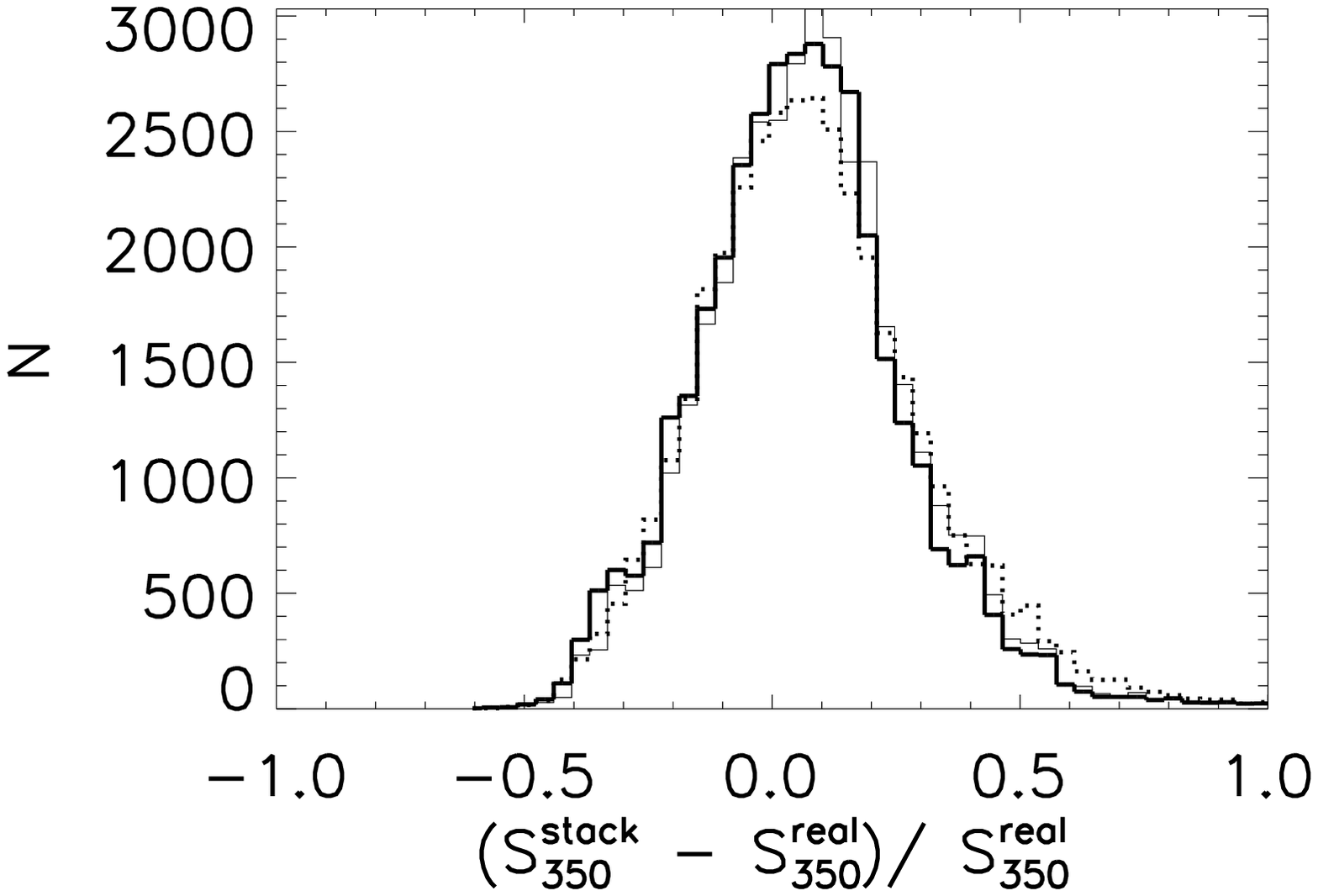}
\includegraphics[width=6.5cm]{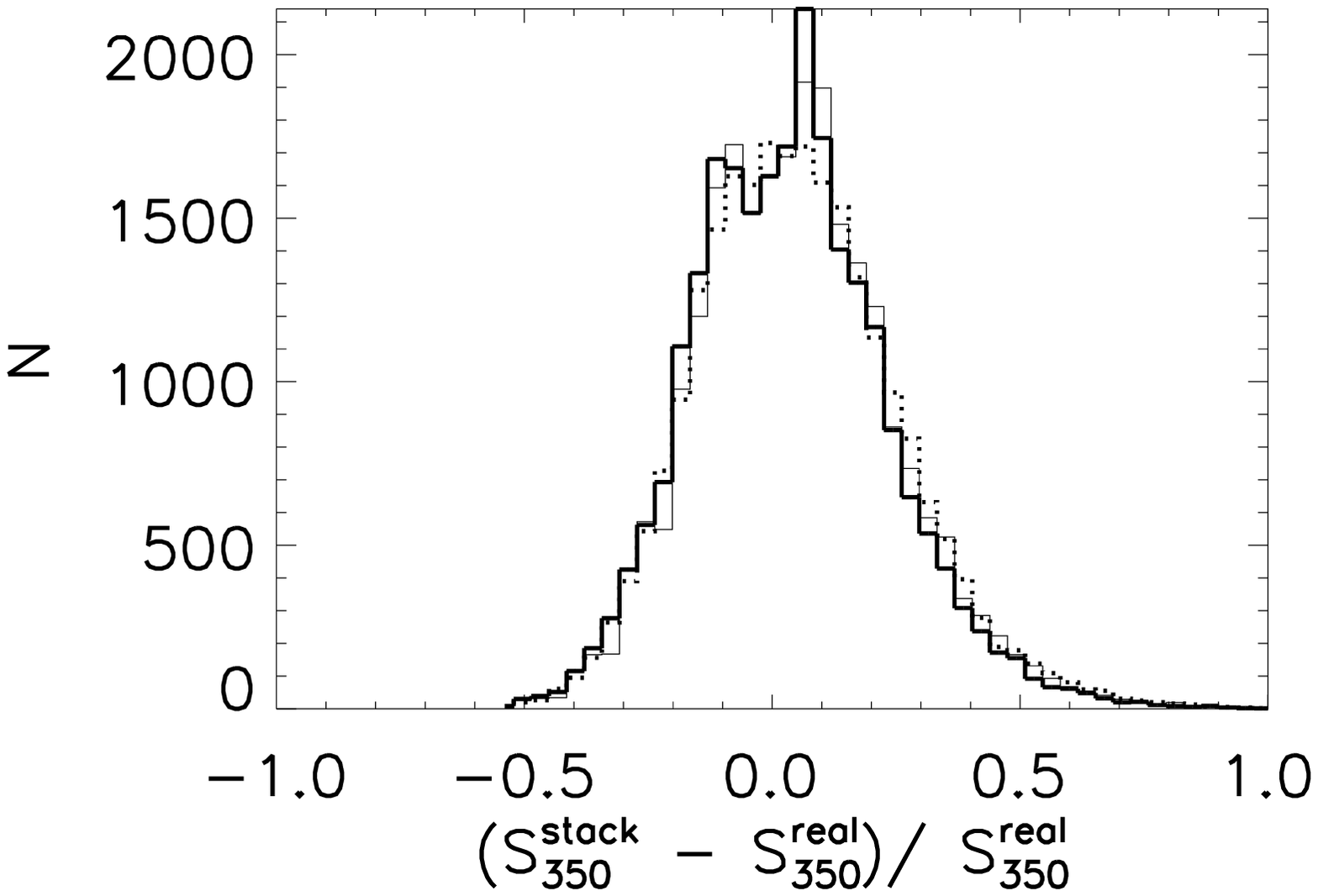}
\caption{Same as Fig. \ref{fig:Cosmos-observation-stackedH3} but for an
observation at 350~$\mu$m if the SWIRE fields.\label{fig:Swire-observation-stackedH3-10}}
\end{figure*}

\paragraph{Errors in individual recovered fluxes:}
Figure \ref{fig:Cosmos-observation-stackedH3}
shows the errors in the estimate of the fluxes of
the sources with
the stacking technique for redshifts $0<z<1$ and $1<z<2$ for an observation
of the COSMOS field at 350~$\mu$m.  Three
different estimates are shown: one compiled using stacking without
{}``smoothing'' and two others created using two different smoothing techniques (in either z or both $z$ and 
$S_{24}$, cf. Sect. \ref{sec:Smoothing}). The differences between the
estimates obtained using the two smoothing techniques are quite small for most sources. 
The figures show rather good
agreement between the input values of the fluxes and those found by
the stacking technique. The results improve for $z>1$ compared
to those at $z<1$. This is because of the low signal-to-noise ratio at low z
and the two-population problem. As expected, the results degrade with the redshift error.
The results also improve when either of the two
smoothing algorithms are used. 
Figure \ref{fig:Swire-observation-stackedH3-10} shows the results
for a SWIRE observation. The recovered fluxes are more accurate because the larger number of
sources allows us to obtain higher signal-to-noise ratios for the
stacking (but it is limited to $S_{24}>270\,\mu$Jy).

\paragraph{Limit for faint sources:}
Stacking in the COSMOS field allows the detection of sources as faint
as $S_{350}=2.1\pm0.7$ mJy at z$\sim1$, which is 6 times lower
than the noise (1 $\sigma$). At z$\sim2$, we achieve detections for sources with
$S_{350}=3\pm1$ mJy or 4 times lower than the noise. This is
equivalent to a gain in the signal-to-noise ratio of a factor of 18
and 12, respectively, with respect to the 3$\sigma$ detection. If the Spitzer data were complete 
down to lower fluxes, we should be able to successfully detect those sources too. The stacking
method at 350~$\mu$m is limited by the Spitzer
detection limits.

\begin{figure}
\includegraphics[width=4.cm, height=5.35cm]{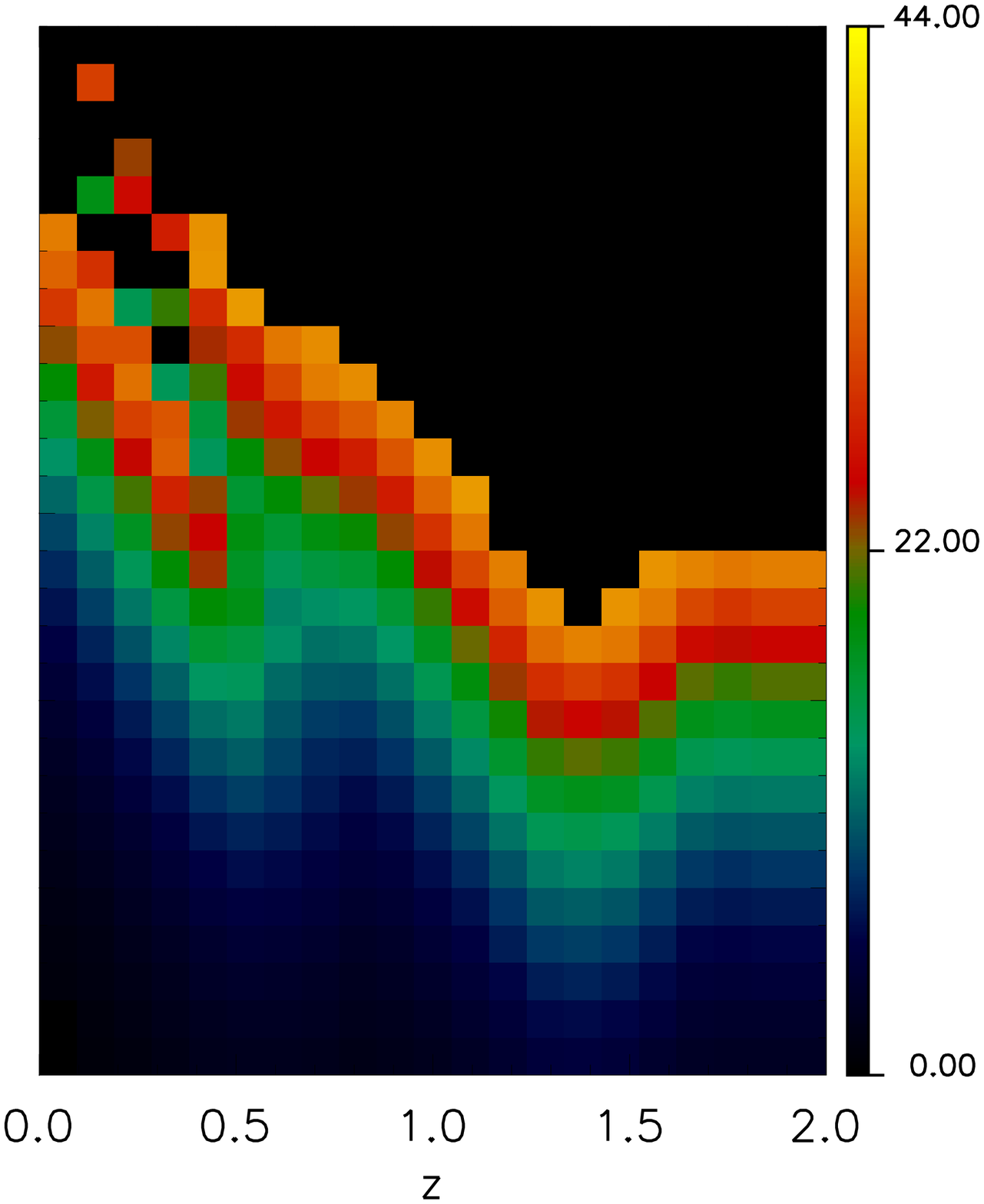}
\includegraphics[width=4.cm, height=5.35cm]{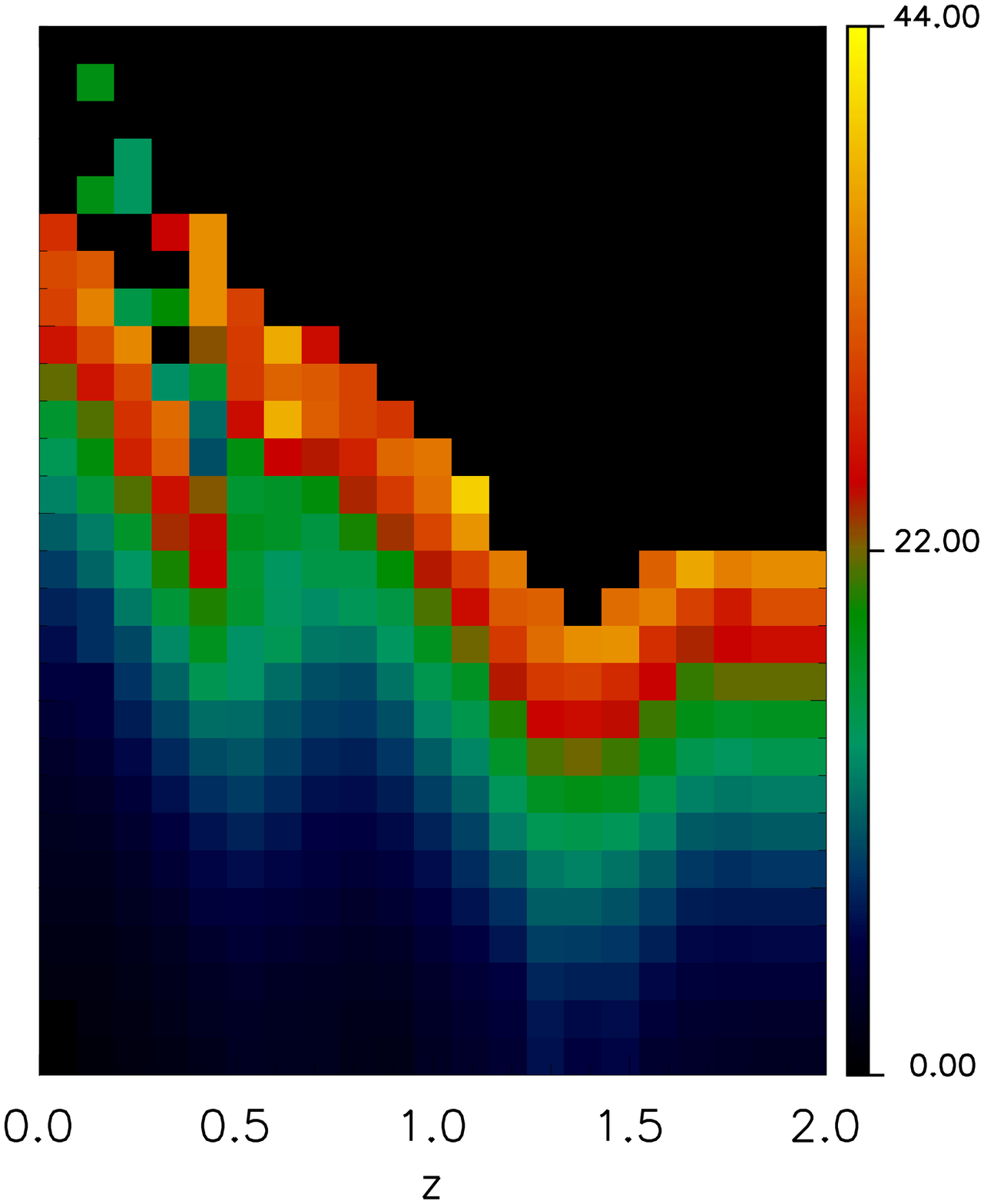}
\caption{Left: Real fluxes of the sources at 350~$\mu$m (in mJy) in the space
$S_{24}-z$. Right: fluxes found by the smoothed stacking technique.
The colors correspond to different
values of the flux, while the vertical axis is $S_{24}$ and the horizontal axis is the redshift bin.The $S_{24}-z$ space is divided linearly in $z$ and logarithmically
in $S_{24}$.\label{fig:Real-fluxes-of}}
\end{figure}

\begin{figure}
\includegraphics[width=4.cm, height=5.35cm]{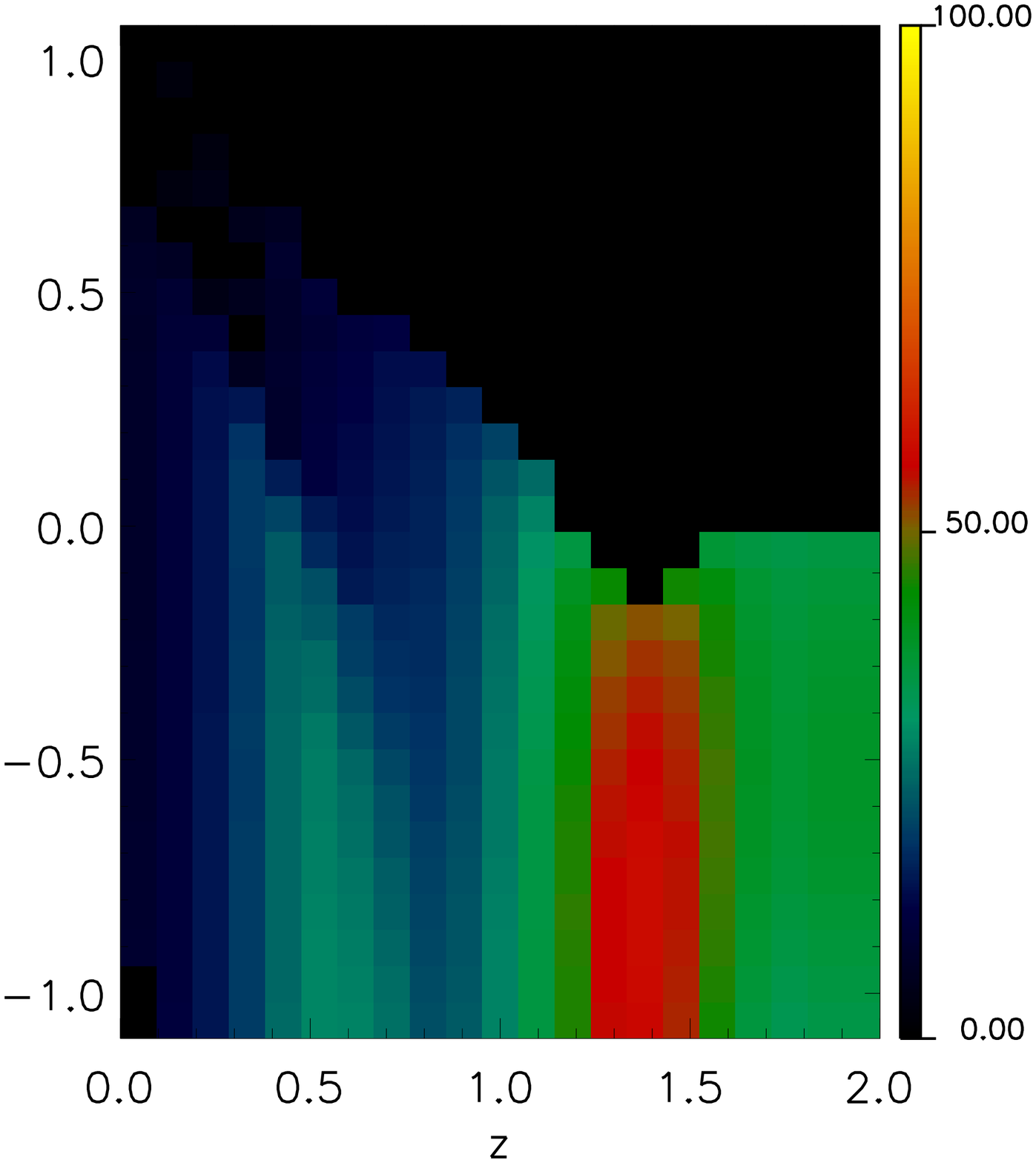}
\includegraphics[width=4.cm]{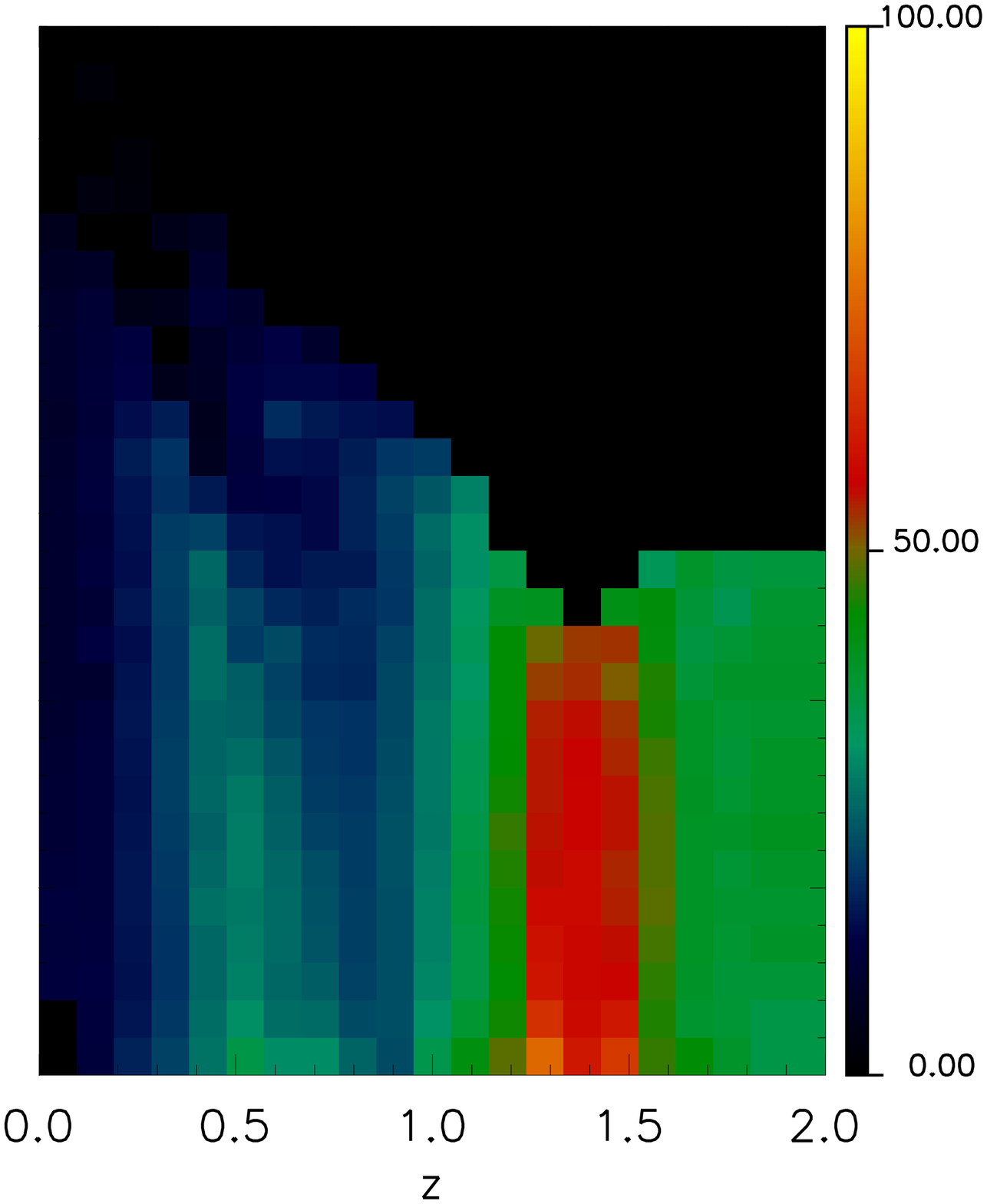}
\caption{Left: Real $S_{350}/S_{24}$ flux ratio of the sources in the space
$S_{24}-z$. Right: $S_{350}/S_{24}$ flux ratio found by the smoothed stacking technique
(right). The colors correspond to different
values of the ratio, while the vertical axis is $S_{24}$ and the horizontal axis is the redshift bin.
The $S_{24}-z$ space is divided linearly in $z$ and logarithmically
in $S_{24}$. 
\label{fig:Real-colours-of}}
\end{figure}

\begin{figure}
\includegraphics[width=4.cm]{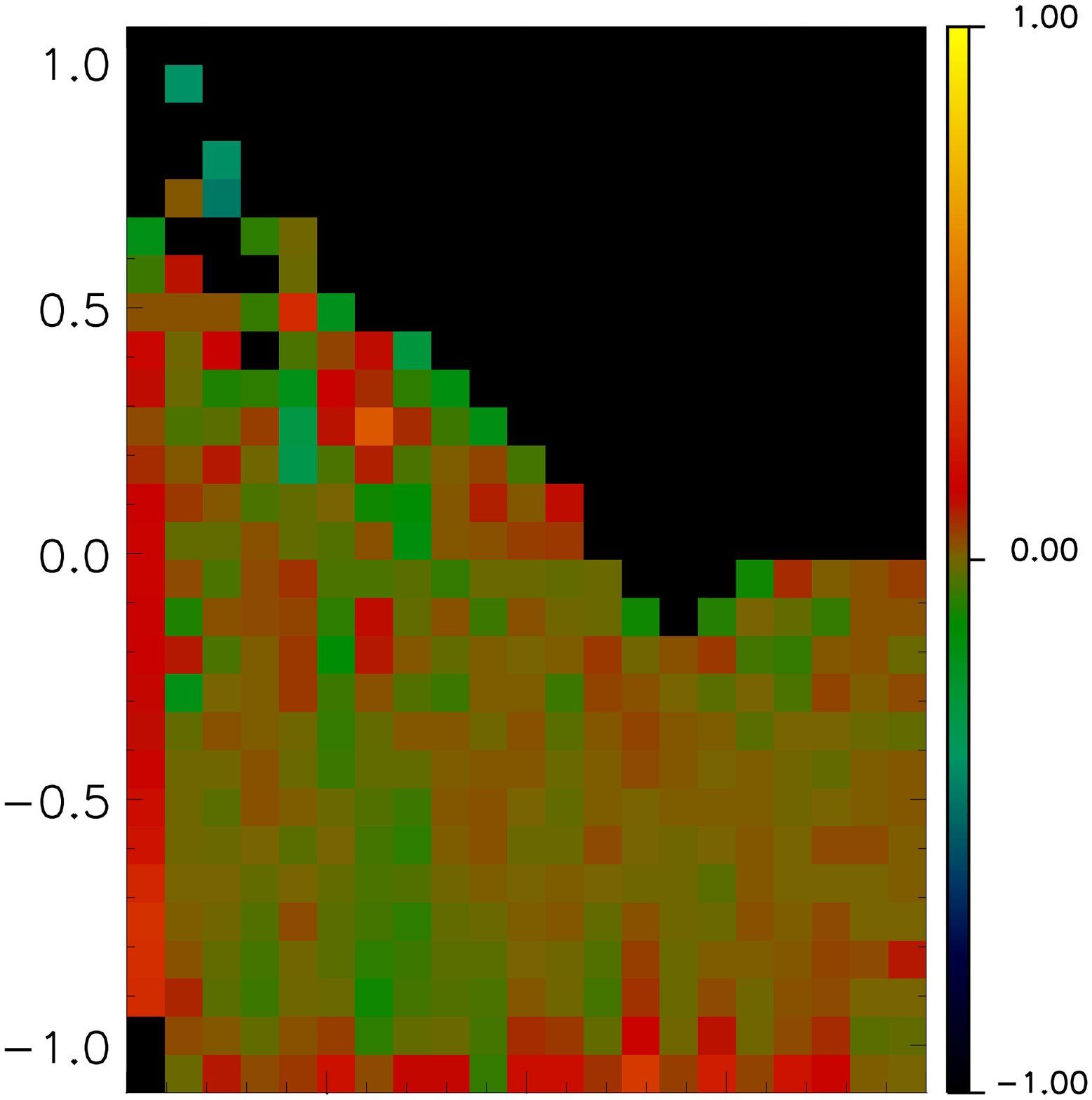}
\includegraphics[width=4.cm]{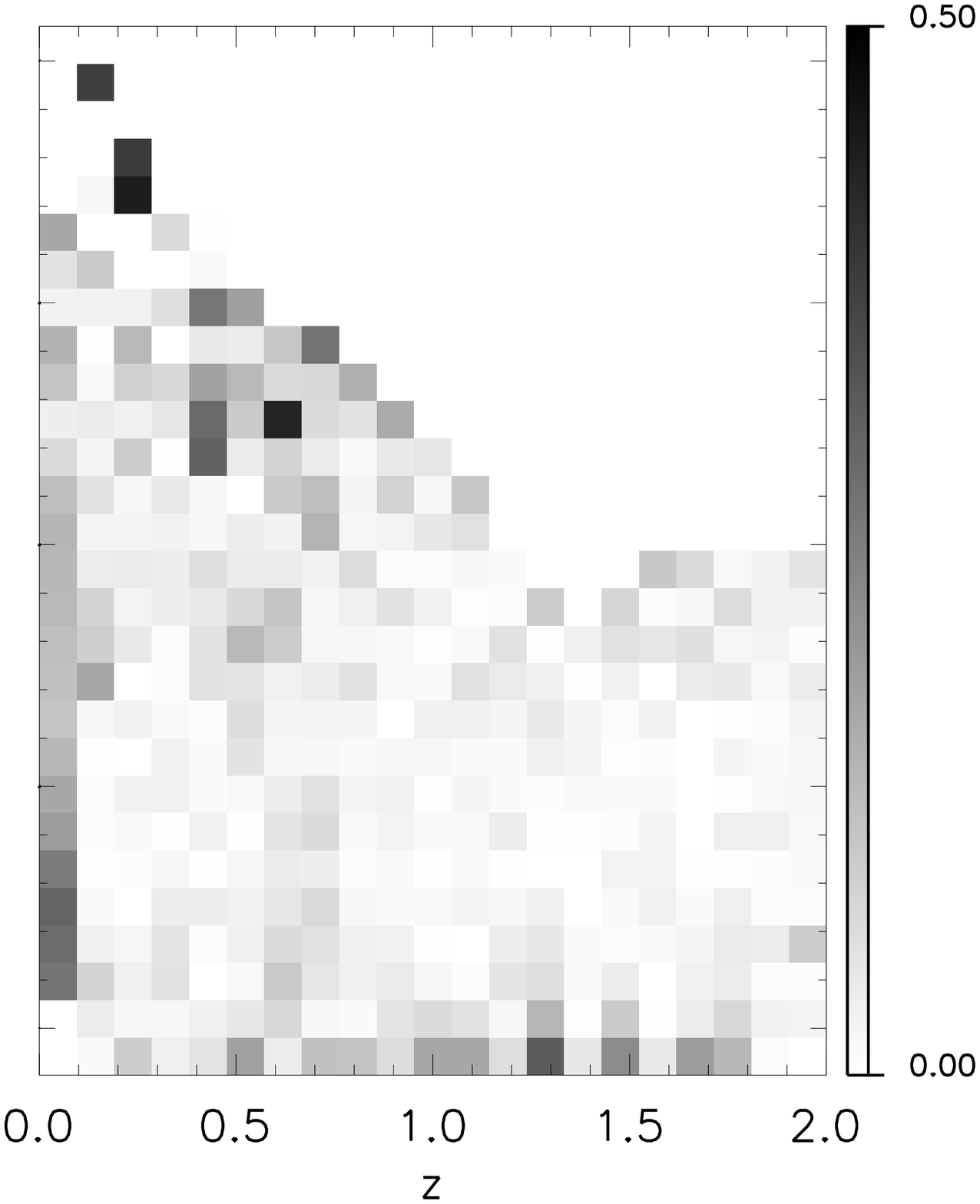}
\caption{Same as Fig. \ref{fig:Cosmos-observation-stacked-HArray} but with no redshift errors. \label{fig:FluxErrStc}}
\end{figure}

\paragraph{Mean errors:}
The final results for the fluxes and colors of the sources obtained using the
stacking technique are compared with the {\it real} (input) values in Figs. \ref{fig:Real-fluxes-of}
and \ref{fig:Real-colours-of}. They are in very good agreement with the input fluxes
(called {\it real} fluxes in the figures)
but to obtain a clearer idea of the errors we show on Fig. \ref{fig:FluxErrStc} two plots
of the mean flux relative error\footnote{Note that the mean flux relative error is equivalent to the mean
color relative error since there is no error in our $S_{24}$ measurements.} 
per box of $S_{24}-z$. The left figure shows the relative differences
between our mean estimated flux (using the stacking technique) and
the flux of the sources introduced in the model. 
Yellowish colors represent overestimates
of the source fluxes compared to their input fluxes. Darker
colors represent underestimates. The right figure shows the same relative
error but this time in absolute value. We can see that the larger
errors, which can be as high as 50\%, are made for sources at $z<0.1$. 
This is because the small number of sources at these
redshifts prevents the stacking from achieving sufficiently high signal-to-noise ratios. 
For the bulk of sources
however, the errors in the mean flux are smaller than 10\%. The errors associated with
the problem of 2 populations cannot be illustrated by these figures because
this problem does not affect the accuracy of the mean value found
for a set of sources but the dispersion in the fluxes of individual
sources around this mean value.

\subsection{Stacking Planck and SCUBA-2 data at 850~$\mu$m \label{sub:Planck-850}}
\begin{figure}
\begin{centering}
\includegraphics[width=0.6\columnwidth]{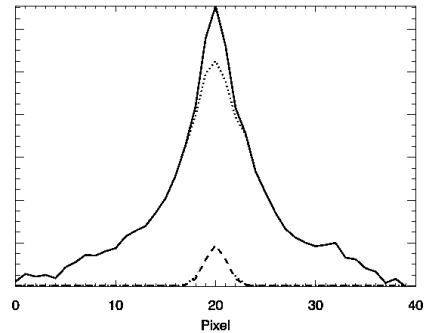}
\par\end{centering}
\caption{Lateral cut of a stacking image (Planck/HFI at 850~$\mu$m) for very faint sources
($S_{24}\sim100\,\mu Jy$). The total signal (normalized to 1 at the peak), the signal from both 
the clustering
and the sources are the solid, dotted, and dashed lines, respectively. One pixel equals 25 arcsec.\label{fig:Correlation-estimation-sources}}
\end{figure}

When applying the same technique to Planck observations 
at 850~$\mu$m, we encounter a fundamental limitation
of the stacking technique. In the stacked image, we can discern two
contributions to the peak, one associated
with the stacked sources, which has the shape of the PSF, and another broader
peak around it which is associated with the sources correlated with the stacked
sources. The method works easily when the PSF width is much smaller
than the width of the correlation peak. However, this condition is
not fulfilled for Planck observations where the width of
the correlation signal around sources is not very different from the
width of the PSF. Furthermore, when stacking faint sources,
$S_{24}\sim100\,\mu Jy$, the signal associated with the correlations
is much stronger than that of the sources: it becomes impossible
to distinguish between the signal from the sources 
being stacked and the signal from the clustering.
Figure \ref{fig:Correlation-estimation-sources} shows a cut of a
stacked image for very faint sources ($S_{24}\sim100\,\mu Jy$).
The figure shows the total signal, the 
signal coming from both the clustering and the 
sources. For these faint sources, we can see that
the signal from the clustering of the sources is more important
than that of the stacked sources and their FWHMs are very similar.
Several attempts were made to correct this problem. By far the
most effective solution is to use additional observations with
a narrower PSF at similar wavelengths to estimate the fraction of the
flux that is associated with the clustering. This method is described
hereafter. Another possible solution
that does not rely on complementary observations is presented in Appendix A.\\

The problem caused by the clustering contribution to the flux measured
with Planck/HFI makes it difficult to use this instrument alone to
estimate the fluxes accurately . It is therefore necessary to
use observations with other instruments with smaller FWHM. In the far-infrared,
we could use Herschel (for the same channel as Planck at 350~$\mu$m). 
For the submillimeter observations, we will have to use ground-based submillimeter instruments
(e.g., future camera SCUBA-2 at 850~$\mu$m or LABOCA at 870~$\mu$m).

\begin{figure}
\begin{centering}
\includegraphics[width=8.cm]{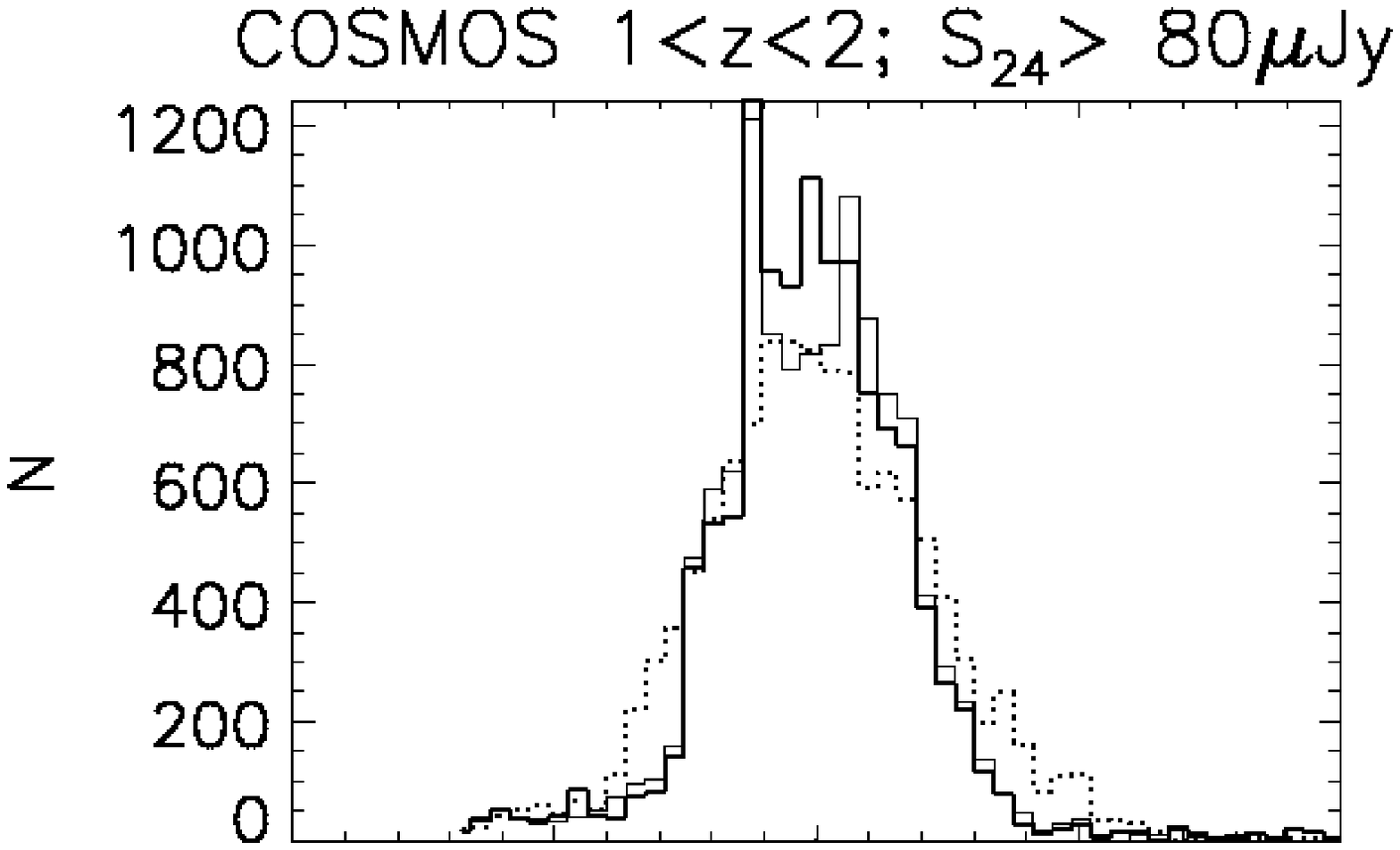}
\includegraphics[width=8.cm]{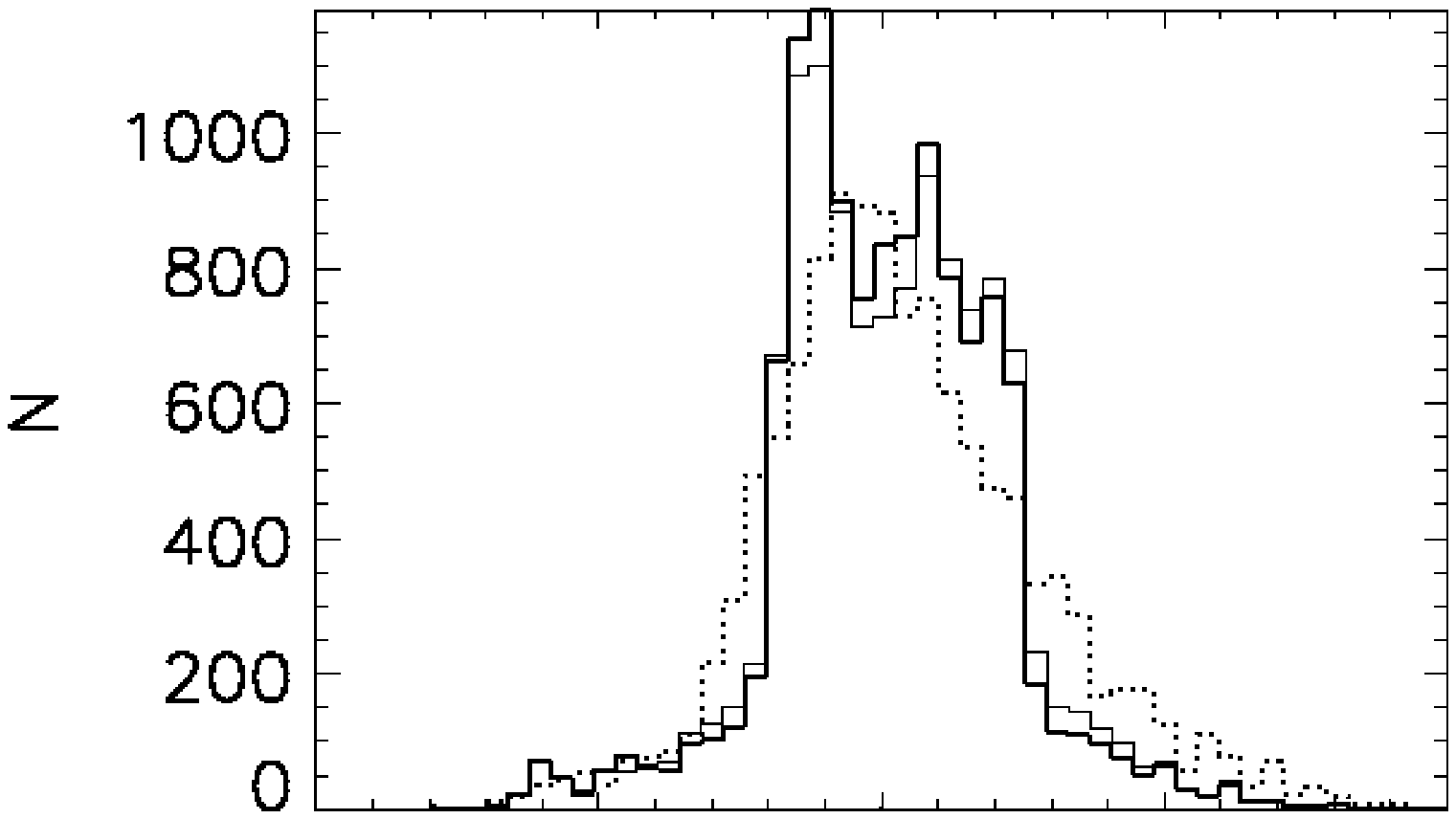}
\includegraphics[width=8.cm]{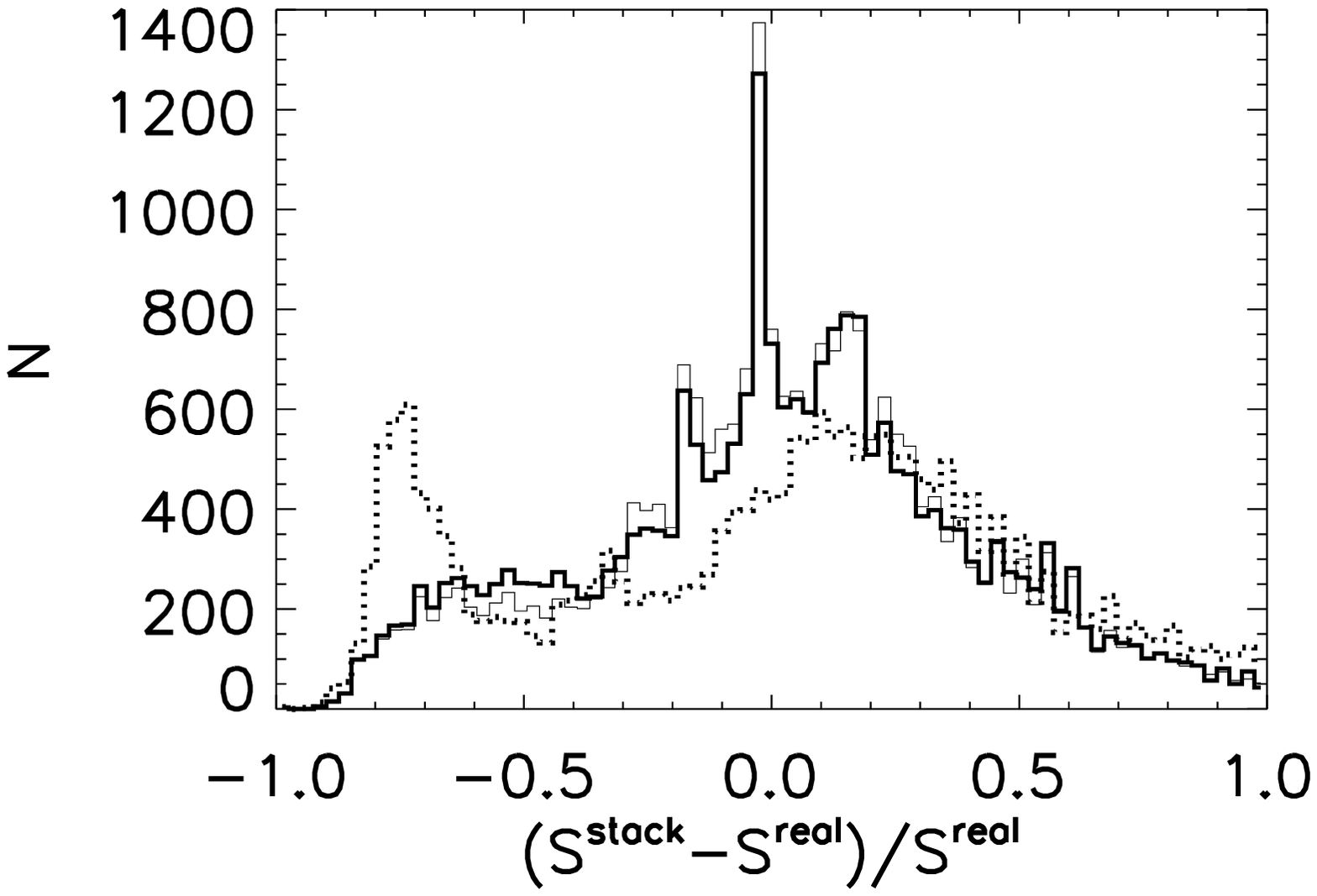}
\end{centering}
 \caption{Relative errors in recovered fluxes for individual sources at 850~$\mu$m in a COSMOS-like observation
for redshift errors $\frac{\triangle z}{z}$ equal to 0 (top), 3\% (middle), and 10\% (bottom),
for $S_{24}>80\,\mu$Jy and $1<z<2$. Three estimates are shown: direct
values obtained with the stacking (dotted line); values obtained with the
stacking and smoothed in $z$ (thin solid line); and values smoothed both in $z$ and in $S_{24}$ (thick
solid line).\label{fig:Cosmos-Scuba-850-3-10}}
\end{figure}

\subsubsection*{SCUBA-2 observation of the COSMOS field at 850~$\mu$m}
We analyze here the stacking of sources in the COSMOS field observed with
SCUBA-2. SCUBA-2 will have a very good sensitivity; we use an  estimate of the noise for these observations
of $\sigma=1$~mJy, close to that specified in the SCUBA-2 webpage\footnote{http://www.jach.hawaii.edu/JCMT/surveys/Cosmology.html}.
Because the signal of the sources at 850~$\mu$m is much fainter relative
to the noise than with Herschel at
350~$\mu$m, we have to increase the size of the redshift bins 
to achieve detections. We take the following boundaries for the redshift slices
$0, 0.1,0.4,0.8,1.,1.2,1.5,1.8, and 2.2$. We use the same detection threshold as that used 
for Herschel at 350~$\mu$m ($D_{thres}=3$).\\

Figure \ref{fig:Cosmos-Scuba-850-3-10} 
shows the errors in the estimate of individual fluxes of
850~$\mu$m sources for $S_{24}>80\,\mu$Jy and redshifts $1<z<2$
with redshift errors of $\frac{\triangle z}{z}$=0\% (top),
3\% (middle), and 10\% (bottom). 
The results are poorer than those at 350~$\mu$m (Fig. \ref{fig:Cosmos-observation-stackedH3}).  
This is because
the signal of the individual sources is weaker relative to the noise
at 850~$\mu$m than at 350~$\mu$m. The results are clearly
dependent on the redshift errors. \\

The sources detected with the stacking technique at $z\sim1$ are
as faint as $S_{850}=0.10\pm0.03$~mJy, which is 10 times smaller than
the noise. At $z\sim2$ we can achieve detections of sources with
$S_{850}=0.17\pm0.05$~mJy, which is 6 times smaller than the noise. This
is equivalent to a gain in the signal-to-noise ratio of a factor
of 30 and 18, respectively, with respect to the 3$\sigma$ detection. 
As for 350~$\mu $m the stacking method  is limited by the Spitzer
detection limit.\\

\begin{figure}

\includegraphics[width=0.4\columnwidth,height=0.5\columnwidth]{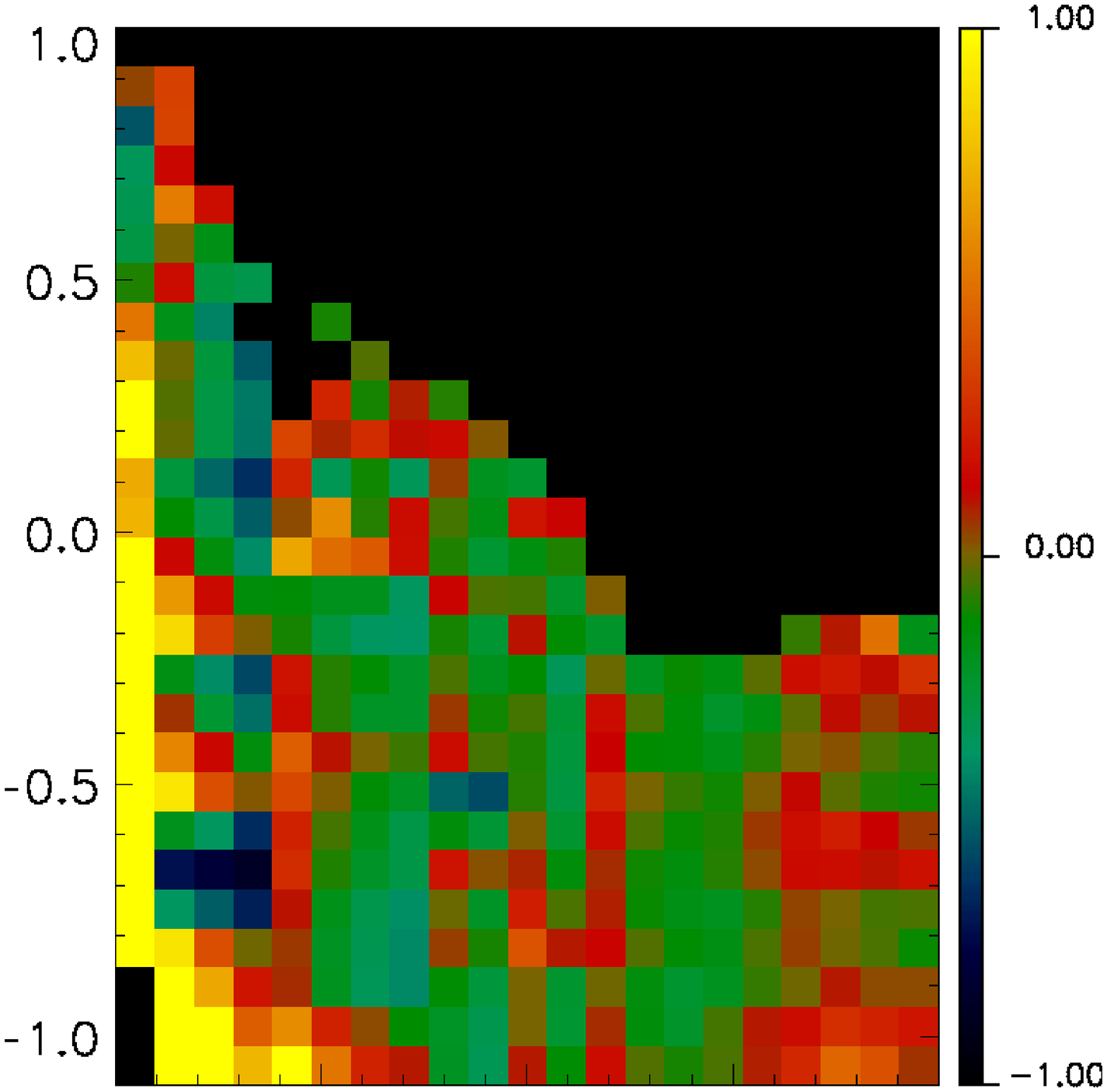}
\includegraphics[width=0.4\columnwidth,height=0.5\columnwidth]{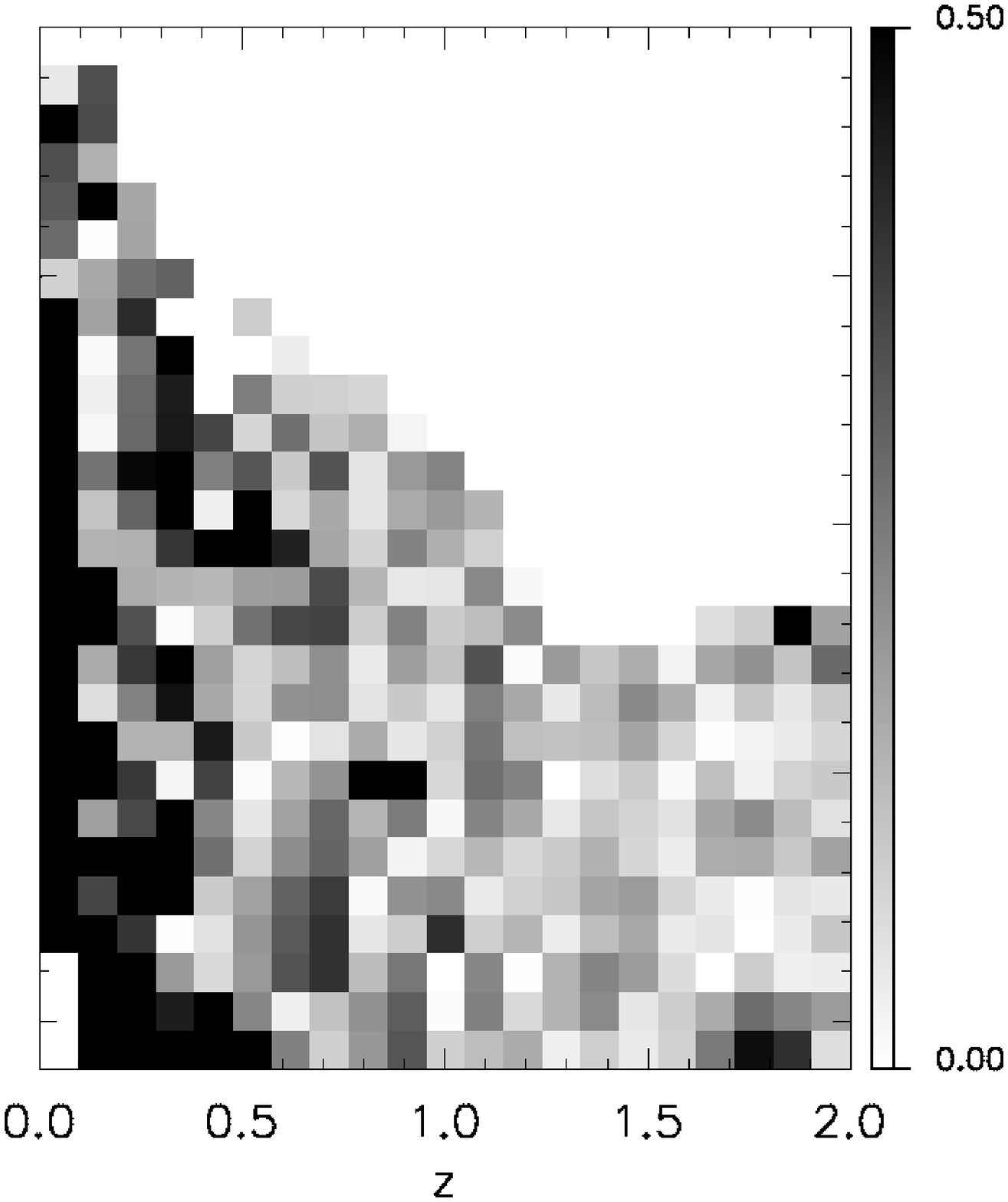}

\includegraphics[width=0.4\columnwidth,height=0.5\columnwidth]{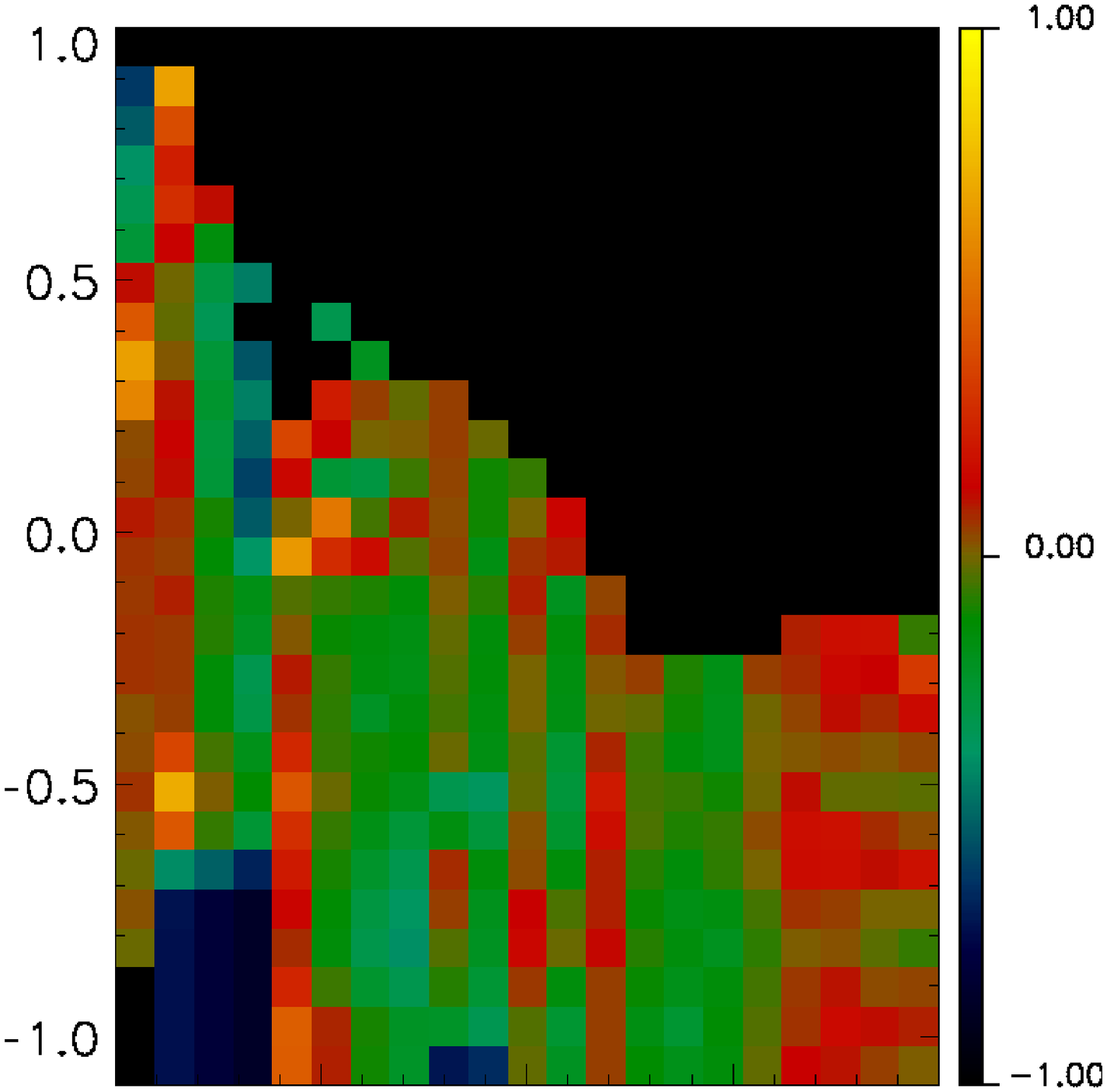}
\includegraphics[width=0.4\columnwidth,height=0.5\columnwidth]{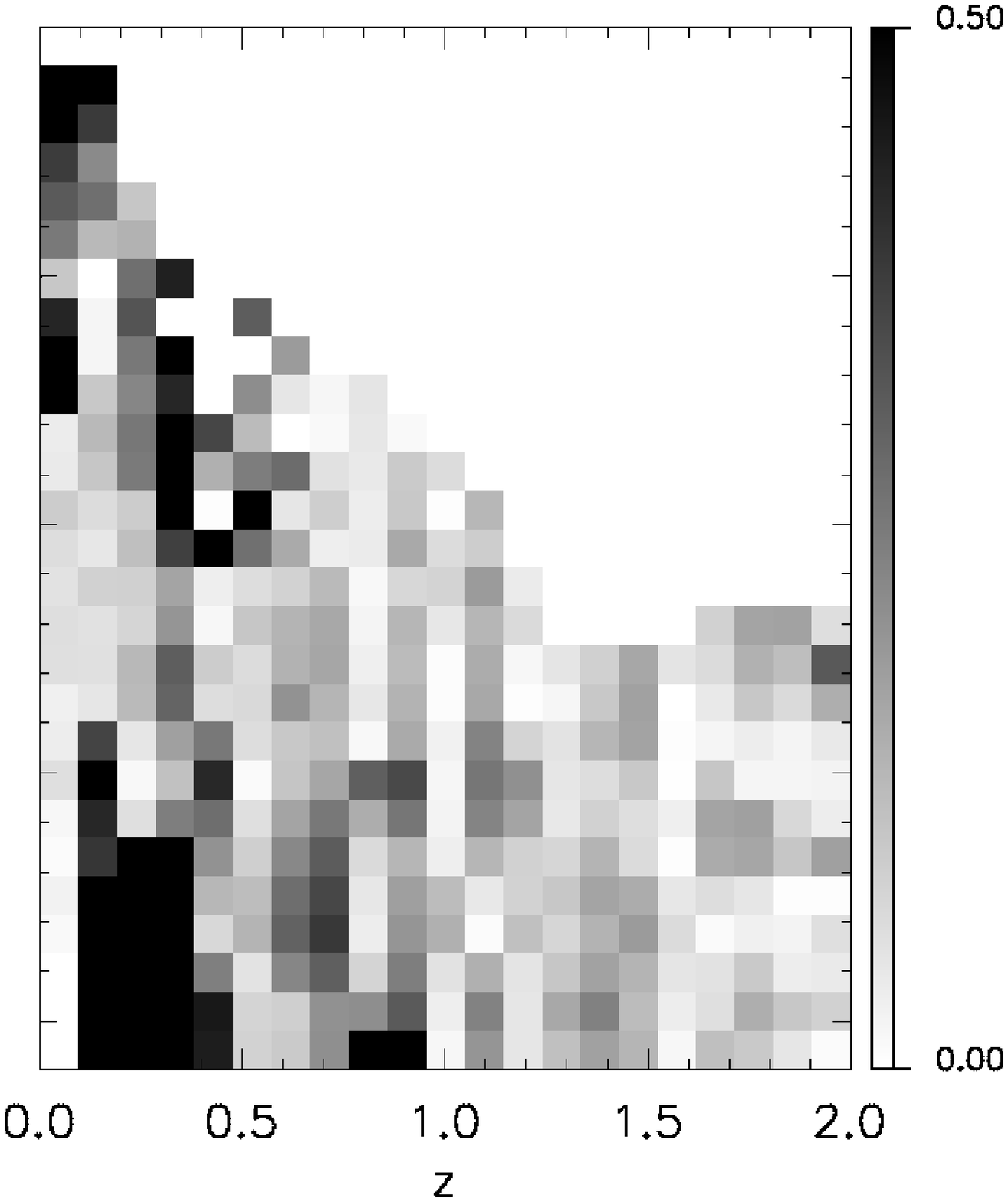}

\includegraphics[width=0.4\columnwidth,height=0.5\columnwidth]{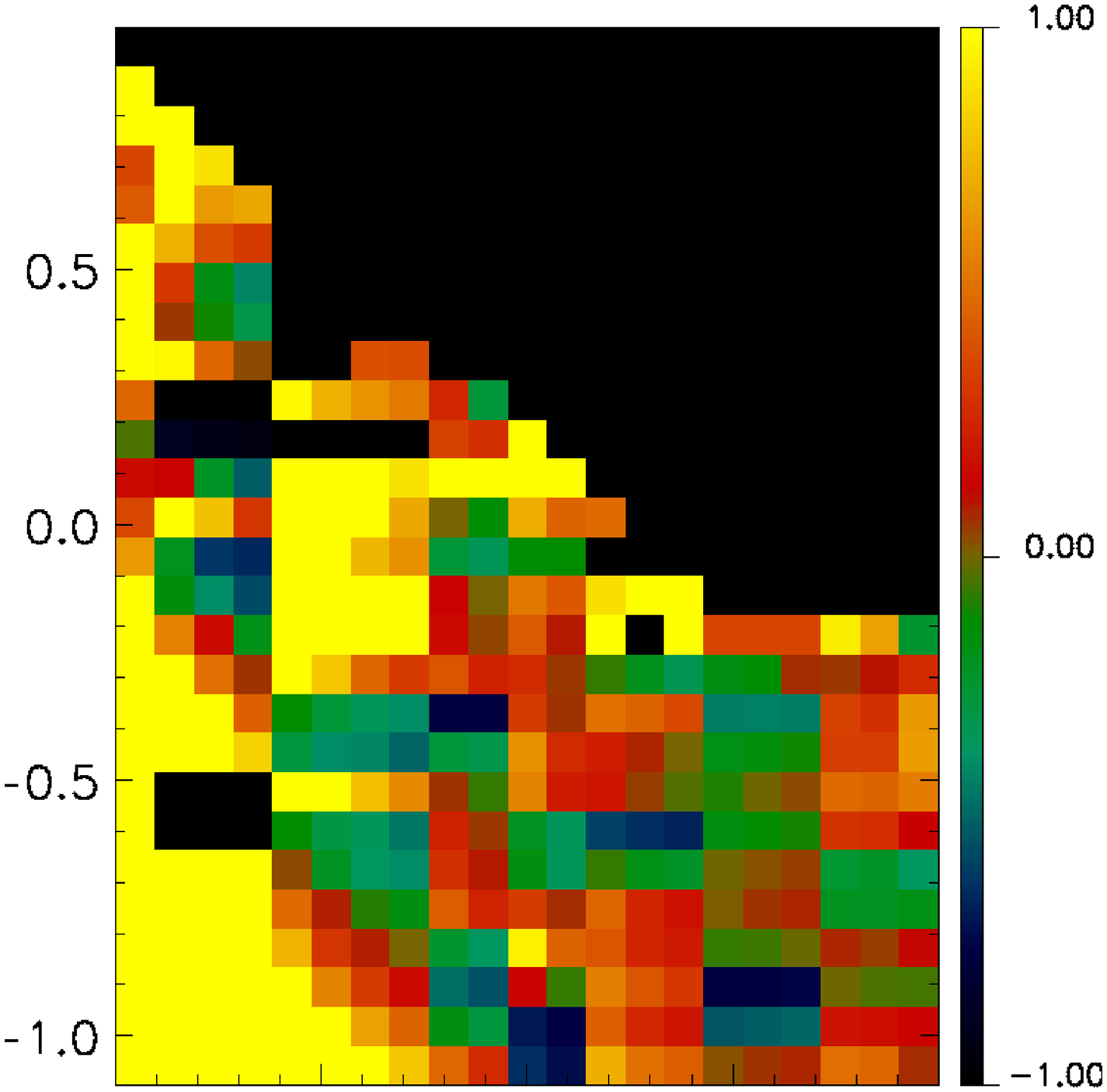}
\includegraphics[width=0.4\columnwidth,height=0.5\columnwidth]{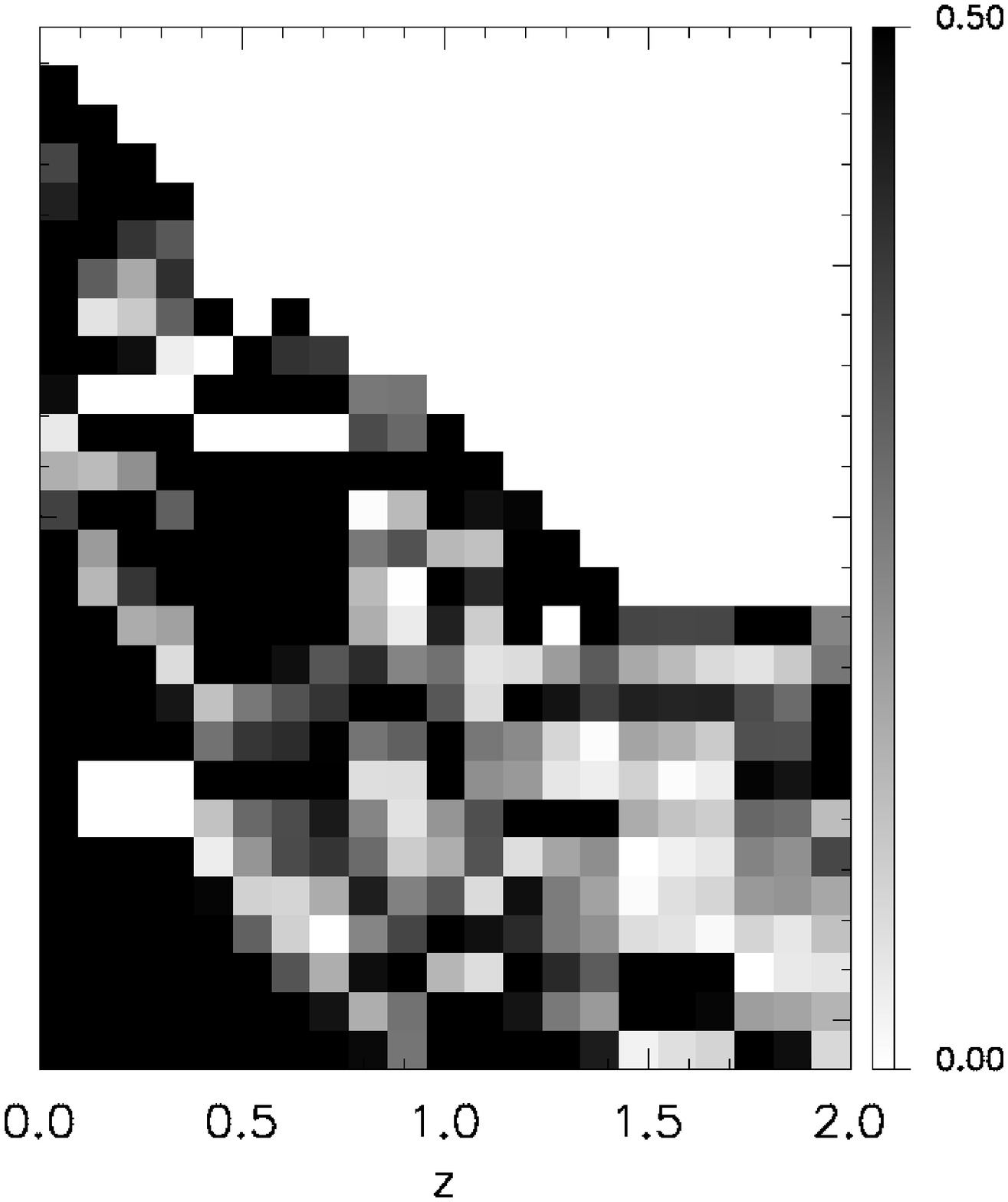} 

\caption{Accuracy of the mean recovered fluxes at 850~$\mu$m in a COSMOS-like observation
when considering 
a redshift error of $\frac{\triangle z}{z}=3\%$ and no smoothing (top),
a redshift error of $\frac{\triangle z}{z}=3\%$ and the smoothing (middle) and 
a redshift error of $\frac{\triangle z}{z}=10\%$ and the smoothing (bottom figures). 
The colors (shading) correspond to different
values of the accuracy, while the vertical axis is $S_{24}$ and the horizontal axis is the redshift bin.
The $S_{24}-z$ space is divided linearly in $z$ and logarithmically
in $S_{24}$. Redshifts are given on the right figures, $log(S_{24})$ (in mJy) 
on the left figures.
Left figures show the relative errors on the mean recovered
fluxes ($\nabla\bar{S}_{850}^{Stack}=(\bar{S}_{850}^{Stack}-\bar{S}_{850}^{Real})/\bar{S}_{850}^{Real}$)
for all the $S_{24}-z$ space. Right figures show the absolute values $\left|\nabla\bar{S}_{850}^{Stack}\right|=\left|(\bar{S}_{850}^{Stack}-\bar{S}_{850}^{Real})/\bar{S}_{850}^{Real}\right|$. 
 \label{fig:Swire-Scuba-Array-StcvsSm}
\label{fig:Cosmos-Scuba-SmoothCompare}}
\end{figure}

Figure \ref{fig:Swire-Scuba-Array-StcvsSm} shows the errors in the estimated
mean fluxes at 850~$\mu$m in the $S_{24}-z$ space for a COSMOS
observation stacked with SCUBA-2 at 850~$\mu$m with redshift error $\frac{\triangle z}{z}=3\%$ 
before and after the ``smoothing'' correction. It shows the
improvement of the accuracy with the ``smoothing'' correction.
Figure \ref{fig:Cosmos-Scuba-SmoothCompare} also shows the errors in the estimate
of the mean fluxes for $\frac{\triangle z}{z}=10\%$ (smoothing applied). As
at 350~$\mu$m, we lose accuracy in our predictions when the redshift
errors are higher. When comparing with 
observations at 350~$\mu$m, we see that our estimates are not as
accurate, the mean errors at 850~$\mu$m being around 15\%
compared to 5-10\% at 350~$\mu$m.
The problems we discussed for 350~$\mu$m observations are yet greater
at 850~$\mu$m. The problem at low redshift is far more important
here because the sources at $z\leq0.9$ 
are in general fainter than at higher $z$. 

\subsubsection*{Planck 850~$\mu$m \label{sub:CorrectionPlanck}}

\begin{figure}
\includegraphics[width=0.5\columnwidth]{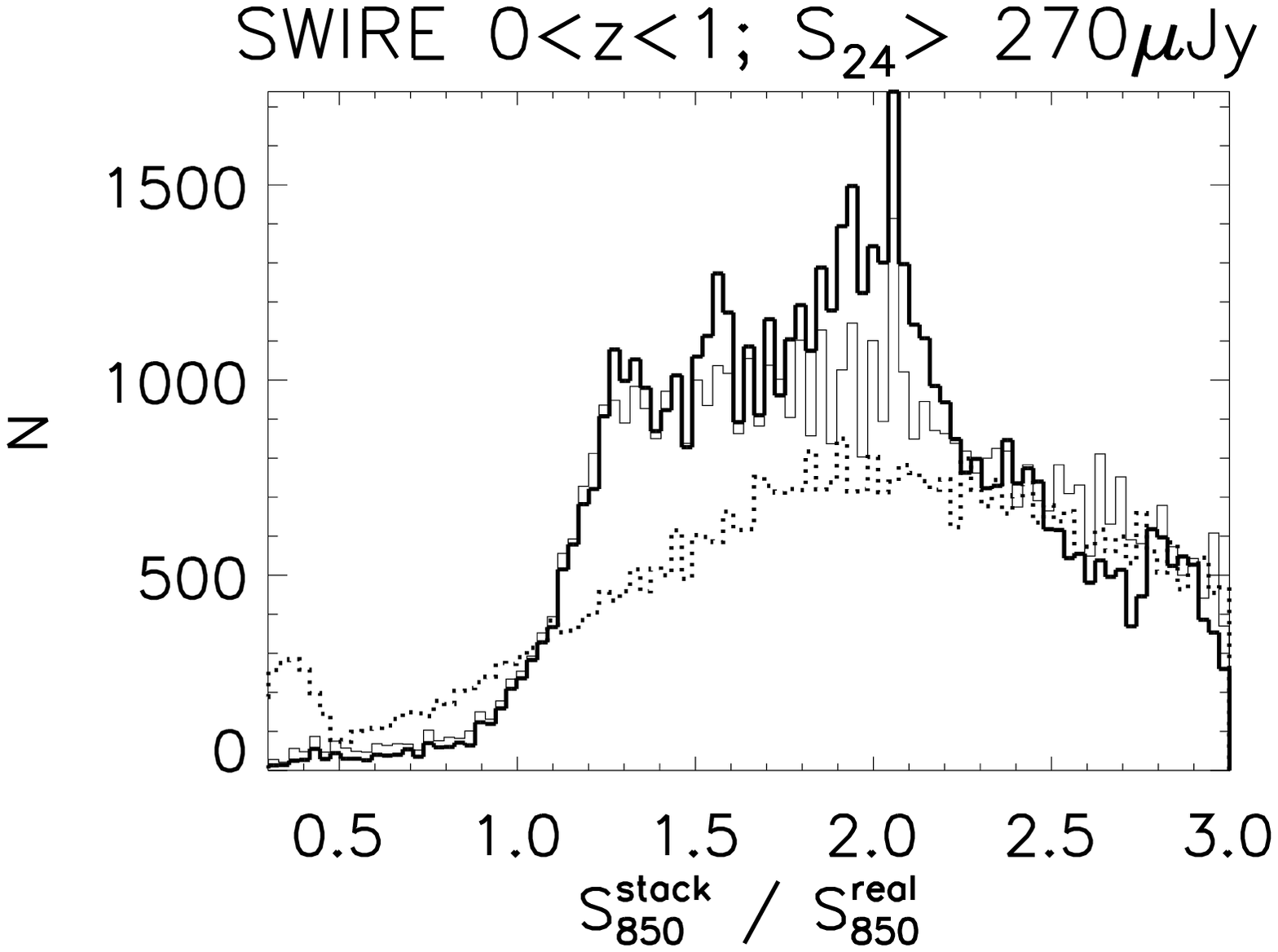}\includegraphics[width=0.5\columnwidth]{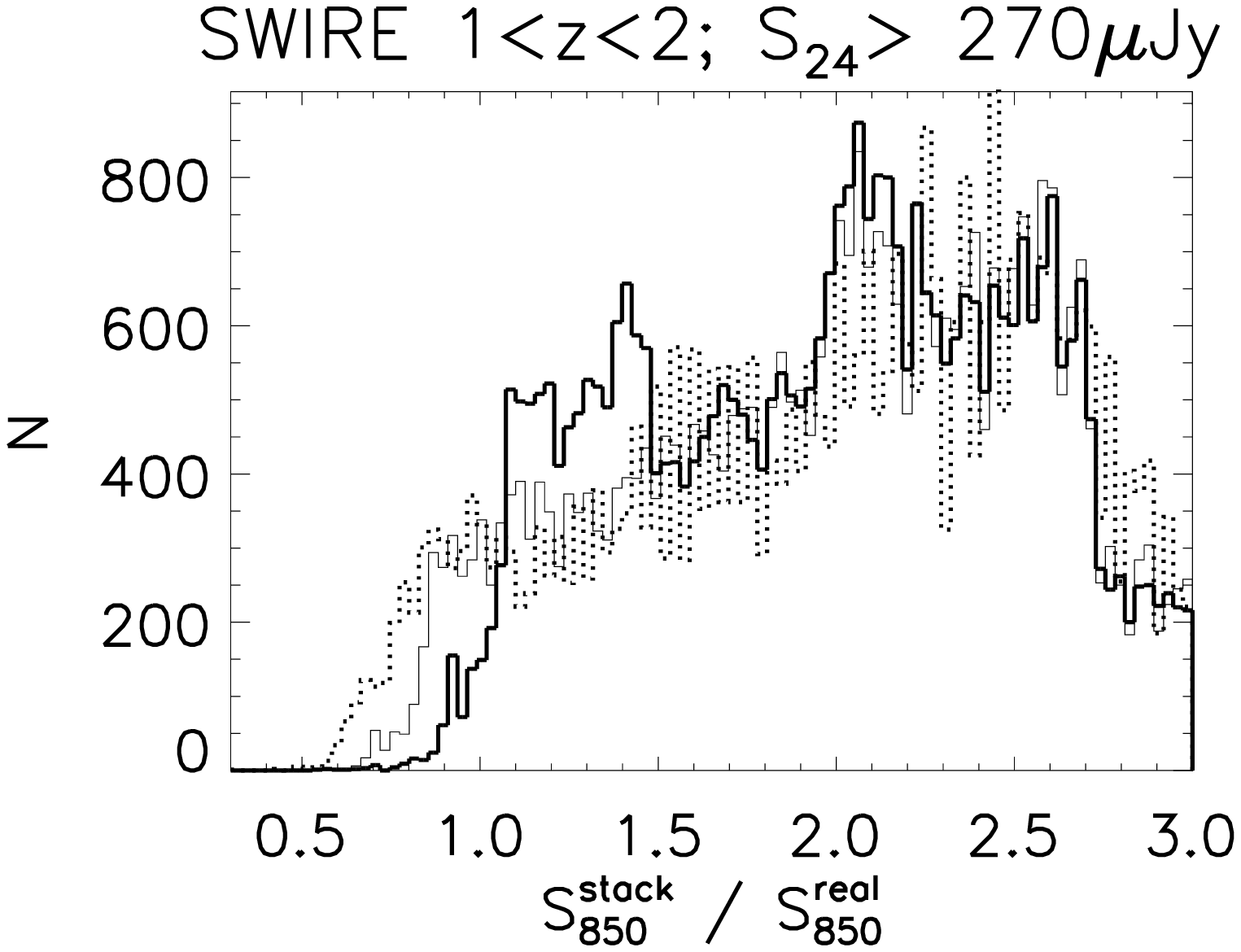}
\caption{Ratio of recovered to input fluxes of individual sources at 850~$\mu$m for an observation of the SWIRE
fields, redshift errors $\frac{\triangle z}{z}$=10\%,
and two redshift bins, $0<z<1$ (left) and $1<z<2$ (right).
Three estimates are shown: direct
values obtained with the stacking (dotted line); values obtained with the
stacking and smoothed in $z$ (thin solid line); and values smoothed both in $z$ and in $S_{24}$ (thick
solid line). 
The value 1 represents a perfect measurement.
The recovered fluxes have not been corrected from the clustering and are thus highly overestimated.
\label{fig:Swire-Planck-Histo}}
\end{figure}

The Planck observation is hindered by the clustering problem caused by its large PSF (5'),
rendering its flux estimates completely useless unless
a correction is applied. The problem is clearly illustrated in Fig. \ref{fig:Swire-Planck-Histo},
where we show the histograms of the ratio of the flux estimates 
to the input fluxes for a Planck observation of the
SWIRE fields for two selected redshift bins. We developed a simple 
method to correct this problem.\\

\begin{figure}
 \includegraphics[width=0.4\columnwidth]{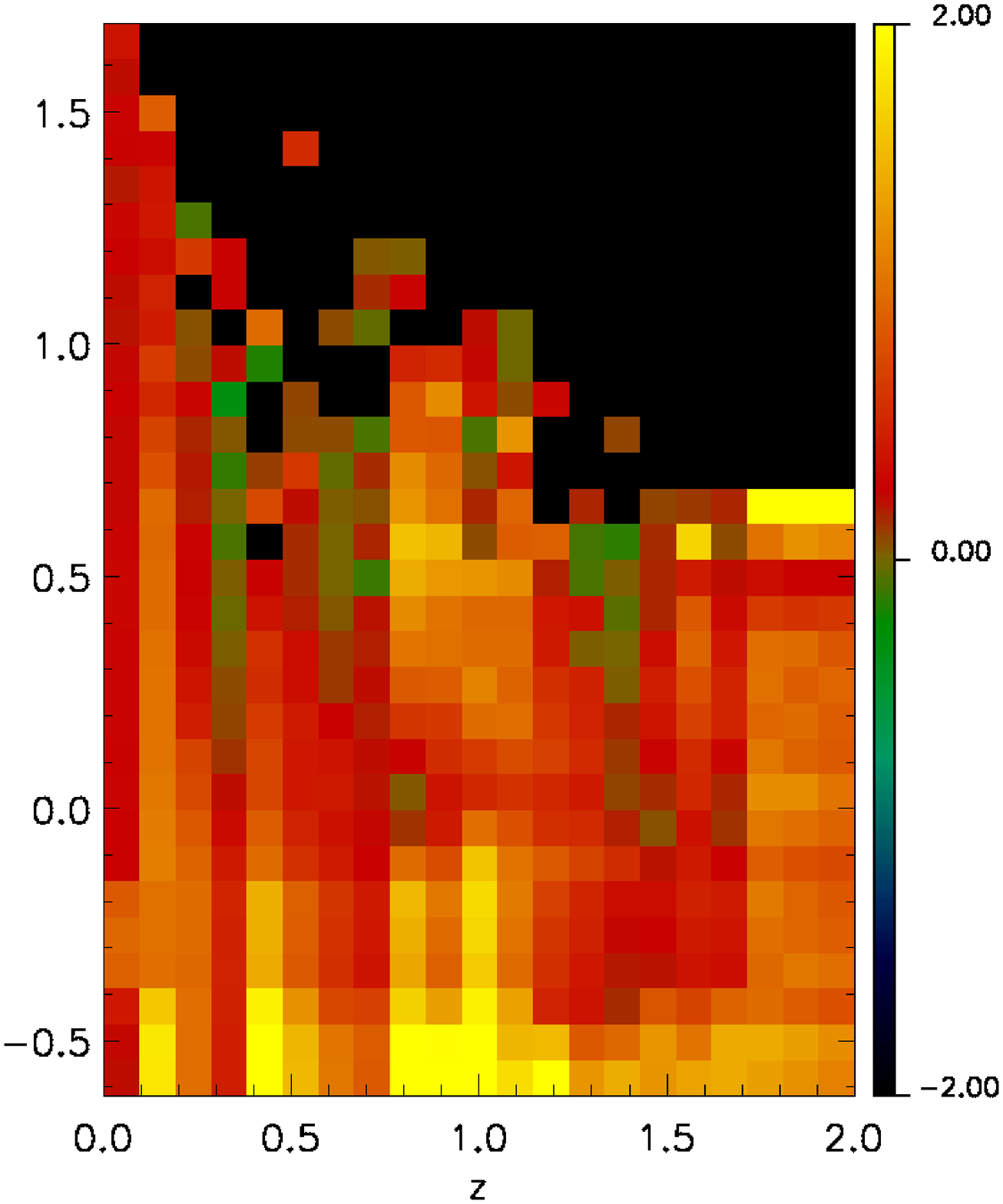}
 \includegraphics[width=0.4\columnwidth]{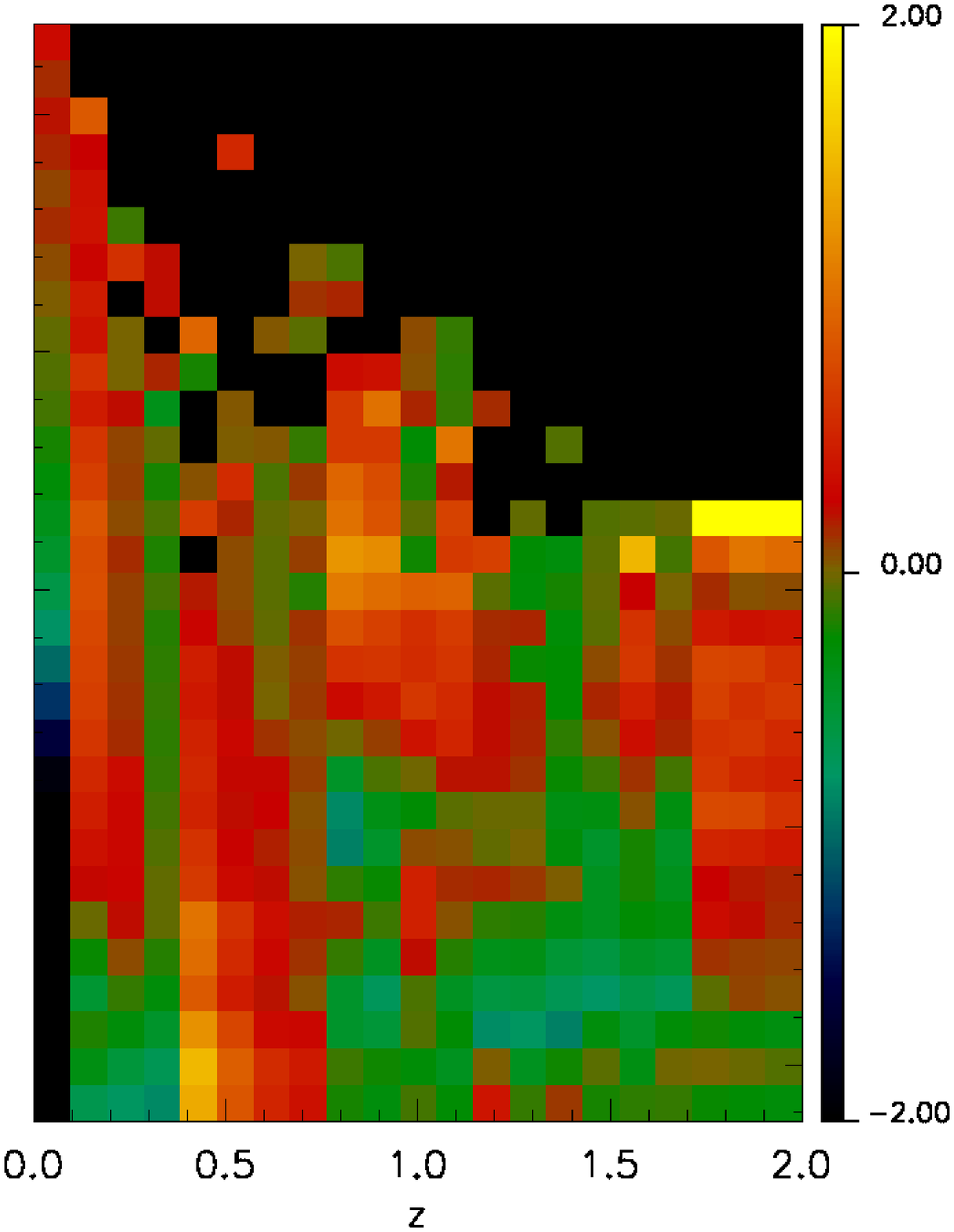}
\caption{Relative error in the mean recovered fluxes at 850~$\mu$m 
($\nabla\bar{S}_{850}^{Stack}=(\bar{S}_{850}^{Stack}-\bar{S}_{850}^{Real})/\bar{S}_{850}^{Real}$)
for a Planck observations of the SWIRE fields
with $\frac{\triangle z}{z}=10\%$ before (left) and after (right) correcting from the
clustering. The colors (shading) correspond to different
values of the relative error, while the vertical axis is $S_{24}$ and the horizontal axis is the redshift bin.}
\label{fig:Planck-Correction-Array}
\end{figure}

When stacking sources in a given redshift bin with Planck, we measure
the added contribution of the sources and the clustering. To correct
the stacked fluxes with Planck for the effects of clustering, we use source fluxes
at 850~$\mu$m obtained by stacking SCUBA-2 data. If we stack
sources detected by Planck for which we have an estimate of their fluxes
inferred from SCUBA-2 data, we can obtain the contribution of the clustering
in the Planck stacking by calculating the difference between the total measured flux
and that measured in the SCUBA-2
stacking. For each redshift bin, we therefore stack Planck data for
all the sources in a SWIRE observation with fluxes $0.27<S_{24}<1$~mJy.
We do not use the brighter sources because their flux estimates
are poorer. Once we have
estimated the effect of the clustering for different redshift
bins, we can correct the fluxes found with Planck. Figure \ref{fig:Planck-Correction-Array}
shows the effect of applying this correction. We can see that the
results are greatly improved. After the correction, the results
for the bright sources $S_{24}>1$~mJy are indeed superior for Planck than
with SCUBA-2, because of its larger sky coverage. We note that the correction is assumed to be the same
inside a redshift bin for all $S_{24}$.
   
\subsection{Combination of different observations \label{sec:Combination-of-different}}

\subsubsection{Observations in the far-infrared (350~$\mu$m)}
\begin{figure}
\includegraphics[width=4.5cm]{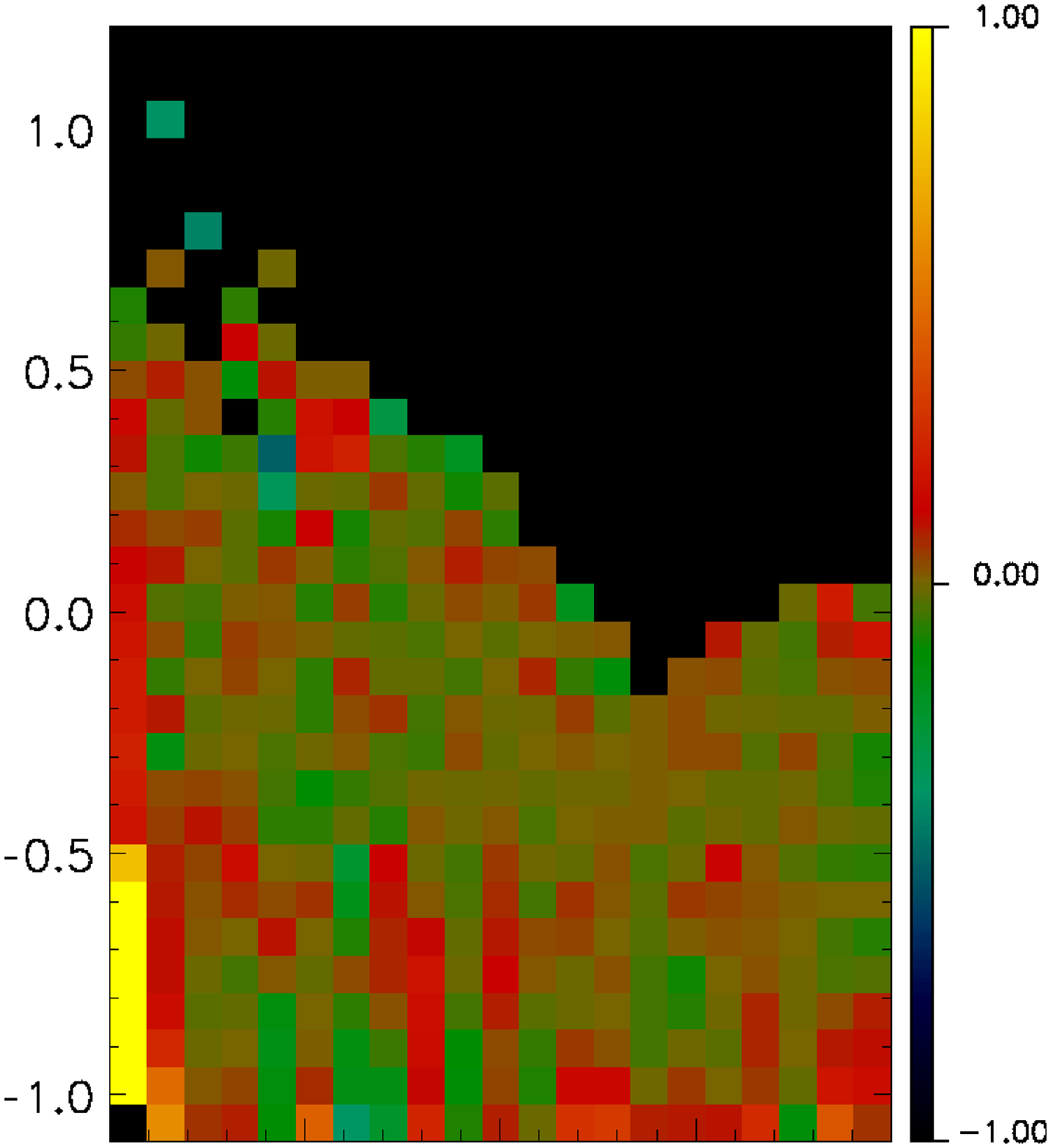}\includegraphics[width=4.5cm]{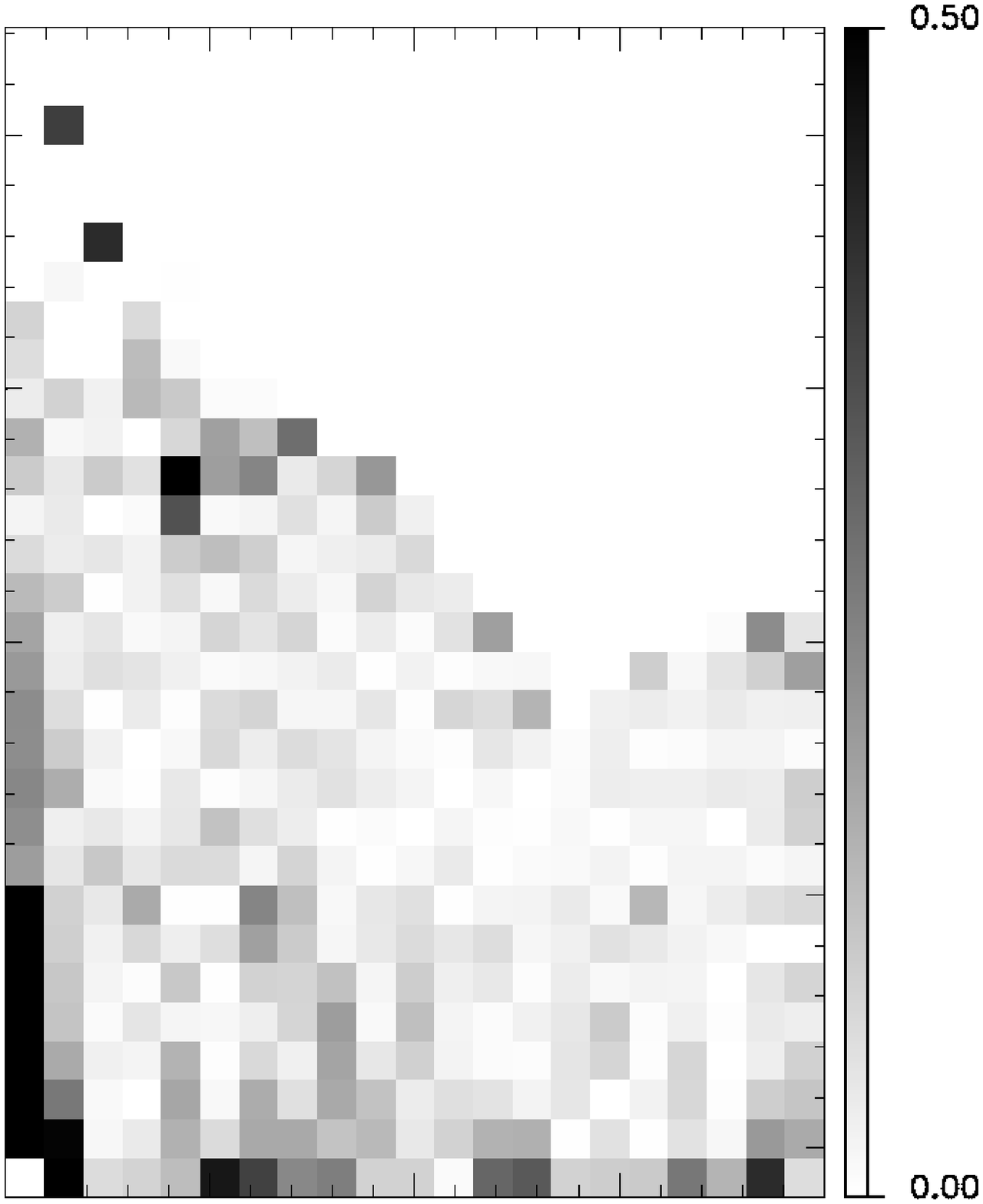}
\includegraphics[width=4.5cm]{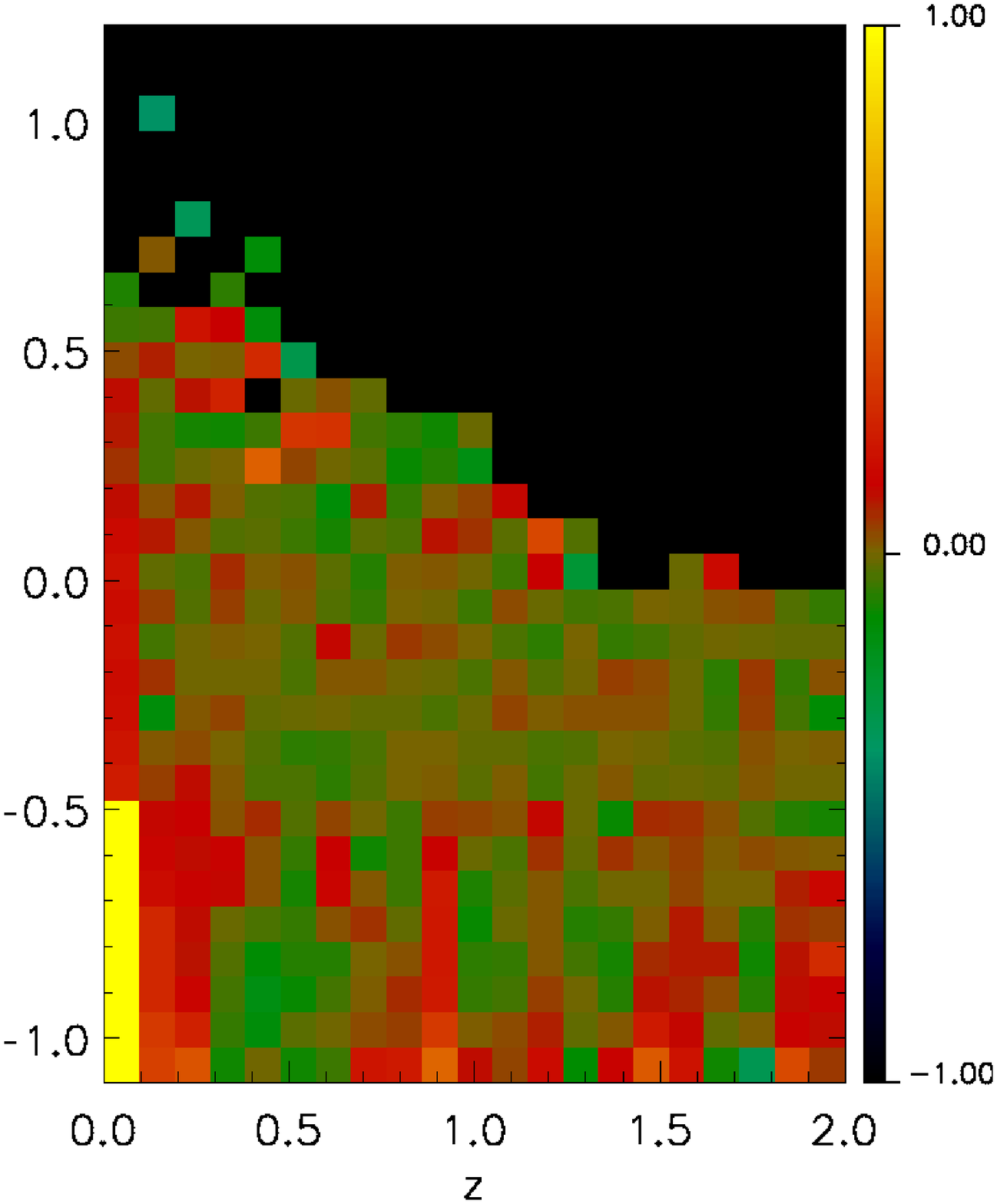}\includegraphics[width=4.5cm]{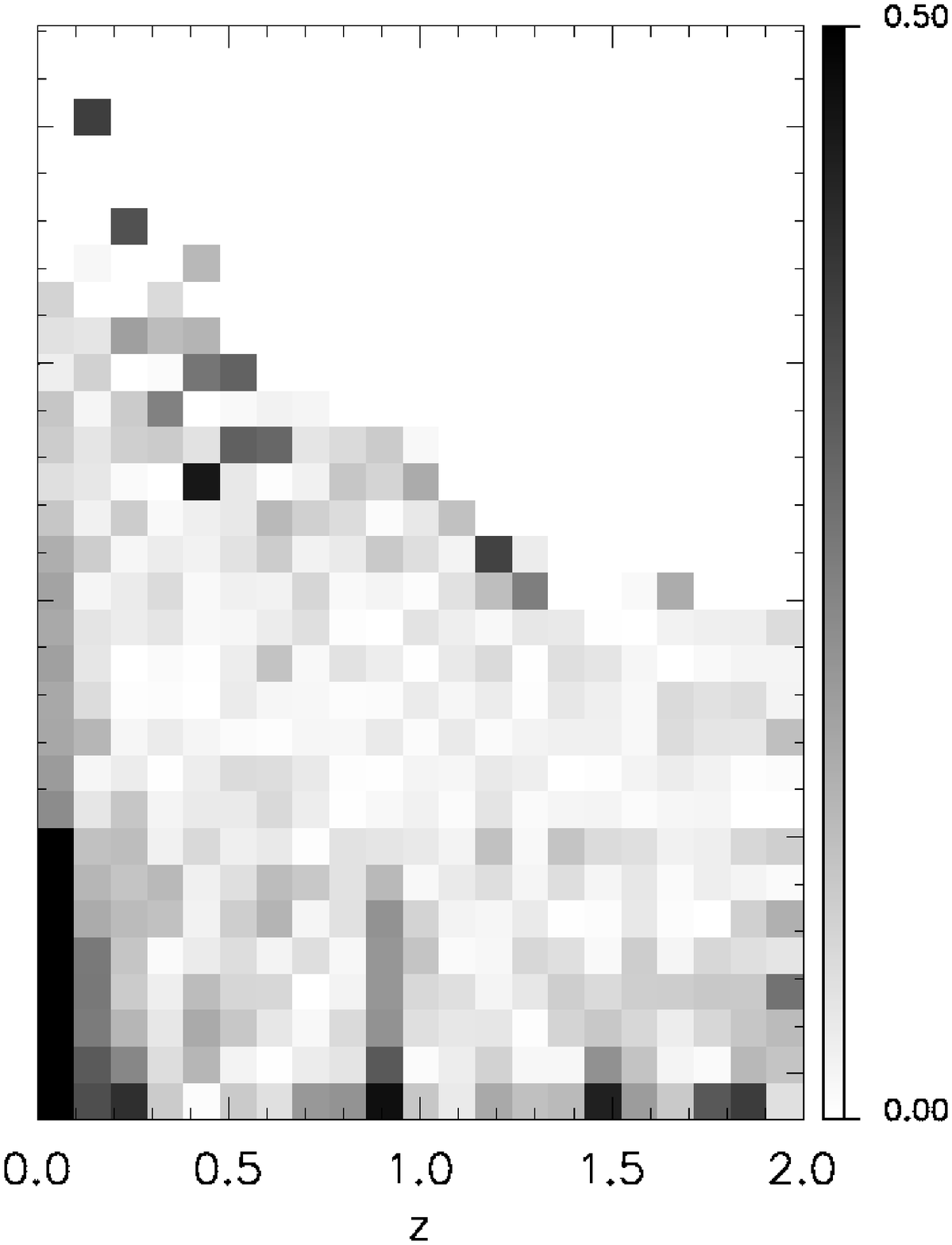}
\caption{Combined observations at 350~$\mu$m with redshift errors $\frac{\triangle z}{z}=3\%$
(top figures) and $\frac{\triangle z}{z}=10\%$ (bottom figures). Left:
relative errors of the estimate of the mean fluxes ($\nabla\bar{S}_{350}^{Stack}=(\bar{S}_{350}^{Stack}-\bar{S}_{350}^{Real})/\bar{S}_{350}^{Real}$) for the $S_{24}-z$ space. Right: absolute values $\left|\nabla\bar{S}_{350}^{Stack}\right|=\left|(\bar{S}_{350}^{Stack}-\bar{S}_{350}^{Real})/\bar{S}_{350}^{Real}\right|$. 
The colors (shading) correspond to different
values of the error, while the vertical axis is $S_{24}$ and the horizontal axis is the redshift bin.\label{fig:Combined-Array-350}}
\end{figure}

We analyzed the Herschel observation of the COSMOS and SWIRE fields. We have seen that
the SWIRE stacking is more accurate when estimating the flux of the
brightest sources. Figure \ref{fig:Combined-Array-350} shows the flux estimates
at 350~$\mu$m when we combine the strengths
of both observations. For sources with $S_{24}<0.27$~mJy, we 
have only COSMOS estimates, which are therefore compelled to use. Since we know
that the SWIRE observations have higher signal-to-noise ratios than COSMOS 
observations at high fluxes, we chose to use these estimates for sources with
$S_{24}>0.34$~mJy. For the fainter sources stacked in SWIRE $0.27<S_{24}<0.34$~mJy data,
we obtained errors larger than those of COSMOS since we assume that the colors of the faintest sources
are as described
in Sect. \ref{sec:Problematic}. For these sources, 
the COSMOS estimates have therefore to be used.

\subsubsection{Observations in the submillimeter (850~$\mu$m)}

\begin{figure}
\includegraphics[width=4.7cm, height=6.cm]{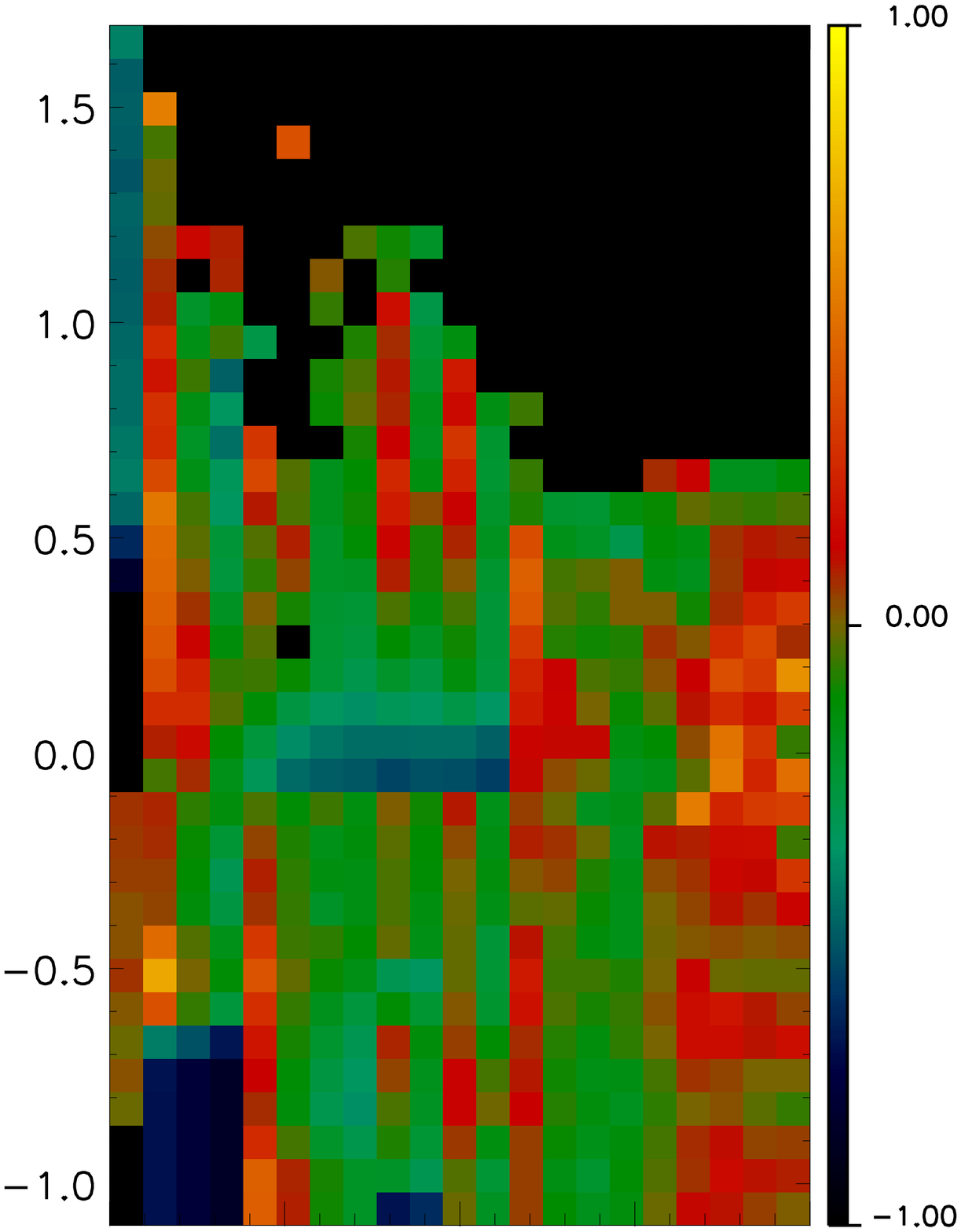}\includegraphics[width=4.7cm, height=6cm]{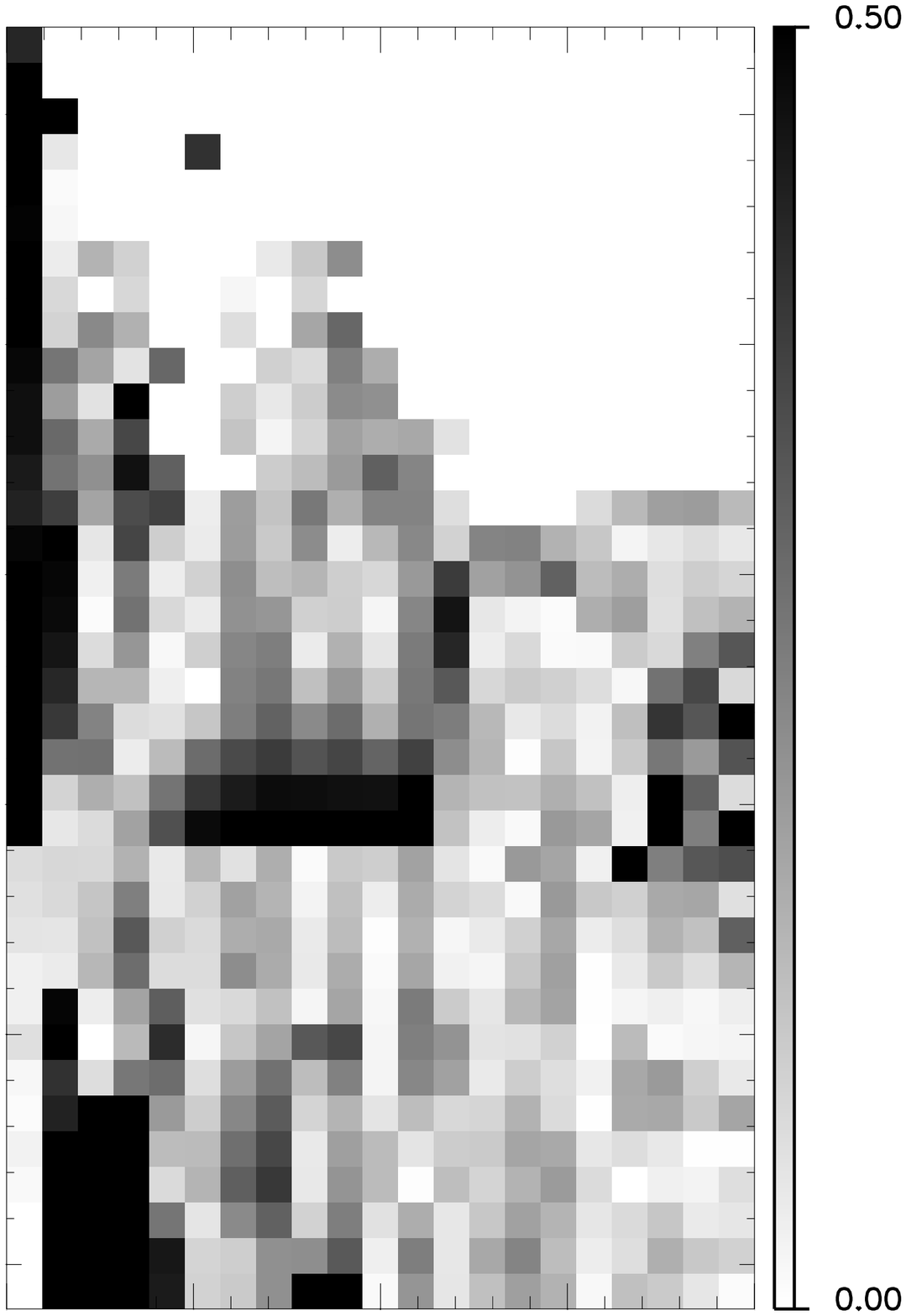}
\includegraphics[width=4.7cm, height=6cm]{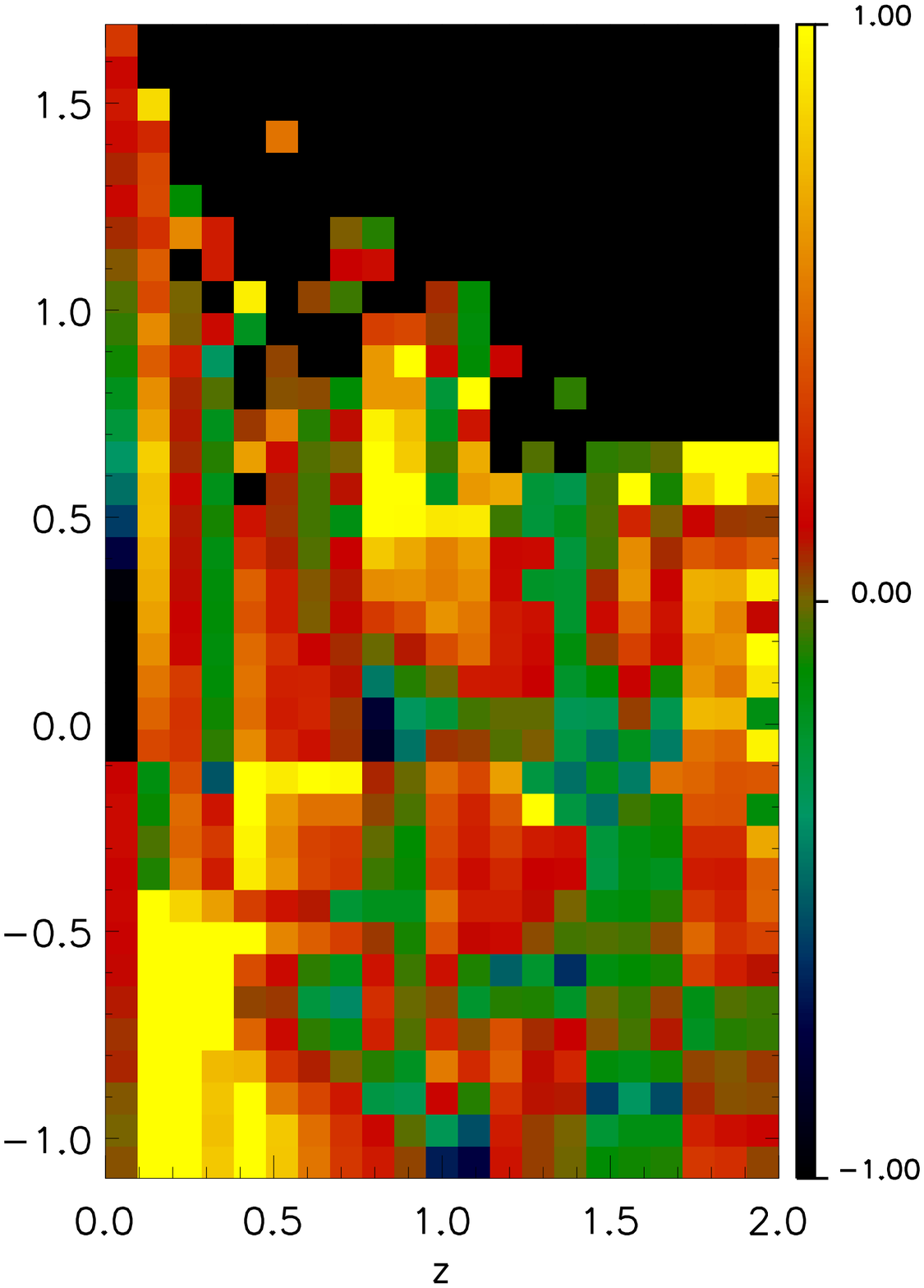}\includegraphics[width=4.7cm, height=6cm]{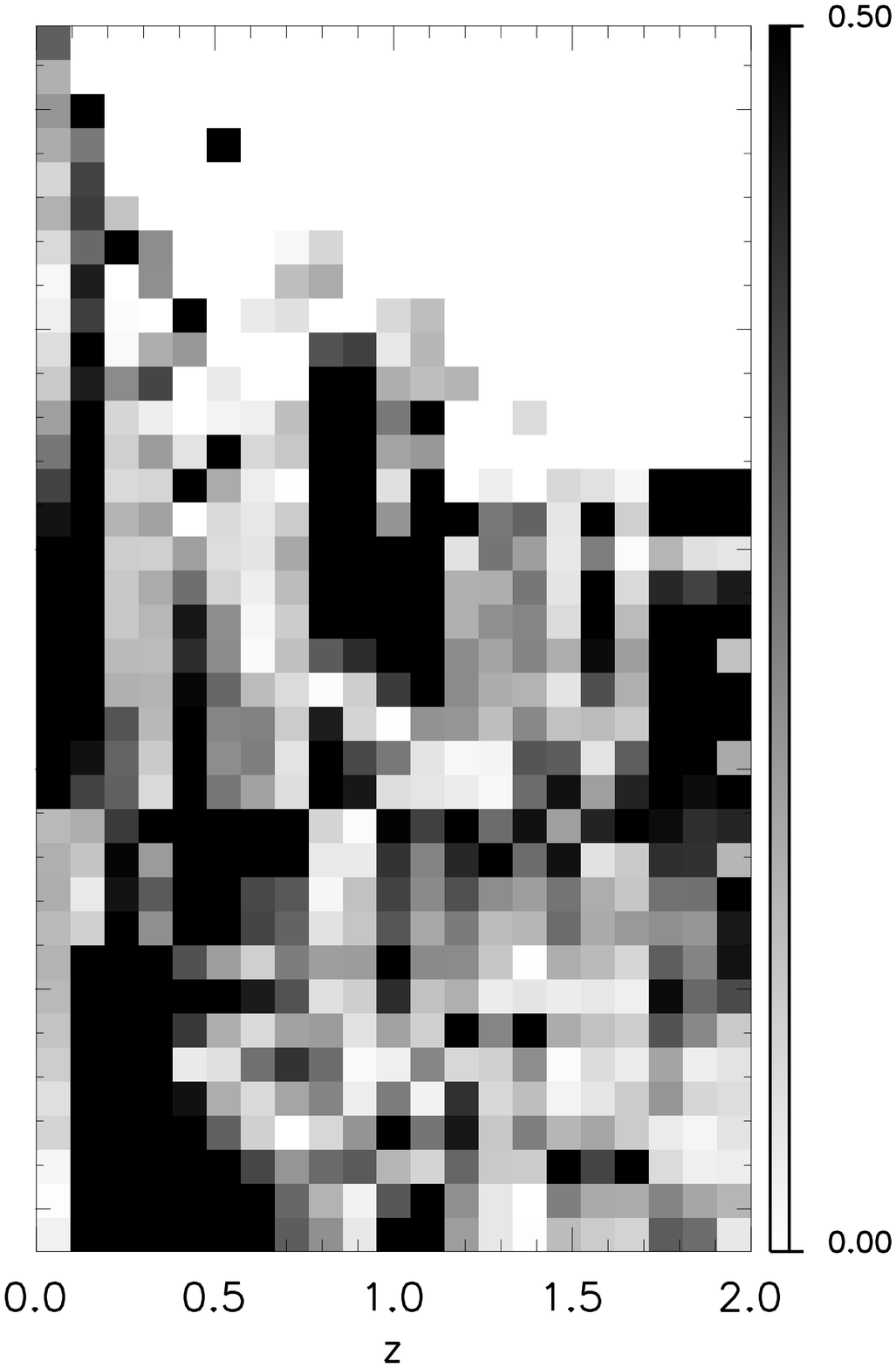}
\caption{Same as Fig. \ref{fig:Combined-Array-350} for combined observations at 850~$\mu$m.\label{fig:Combined-Array-850}}
\end{figure}

As performed at 350~$\mu$m, 
we analyzed the COSMOS/SCUBA-2 and SWIRE/Planck observations
separately and we now combine their respective strengths. Figure
\ref{fig:Combined-Array-850} shows the error estimates for
these combined observations. 
For faint sources with $S_{24}<0.27$~mJy, we use
COSMOS/SCUBA-2. For the faintest sources stacked
in SWIRE ($0.27<S_{24}<1$~mJy), it is more accurate to
use COSMOS/SCUBA-2 than Planck
measurements due to the errors induced by the uncertainty
in the clustering contribution. For brighter sources
($S_{24}>1$~mJy), the corrected Planck estimations are more accurate
than those of SCUBA-2 and we prefer to use them. Figure 
\ref{fig:Combined-Array-850} shows the relative errors in the mean recovered fluxes 
with respect to the input fluxes at
850~$\mu$m, when combining both observations.
They are typically of the order of 15\% for $\frac{\triangle z}{z}=3\%$.

\subsubsection{Observations at other wavelengths}
For observations in the far-infrared and because of the issues discussed
in Sect. \ref{sub:Planck-850} and lower typical noise level, the stacking
technique produces more accurate estimates of the fluxes with Herschel
than with Planck, although the latter has the advantage of covering
the entire sky. We did not present separately the Herschel observations
at 250~$\mu$m or 500~$\mu$m since the analysis of the results
at these two wavelengths are similar to those for 350~$\mu$m
observations. At 550~$\mu$m, a wavelength
where there is a Planck but not a Herschel channel, it is more advisable
to use the values found by Herschel at 500~$\mu$m after applying
a small correction than to use the Planck values. At 850~$\mu$m,
we combined the Planck observations with those of SCUBA-2 although other submillimeter
data (e.g., LABOCA) could have been used. At 1380~$\mu$m (Planck/HFI 217 GHz), we tested the same
approach using MAMBO/IRAM simulated observations
to complement the Planck observations, obtaining similar results as for 850~$\mu$m. \\ 

\begin{figure}
\includegraphics[width=0.5\columnwidth]{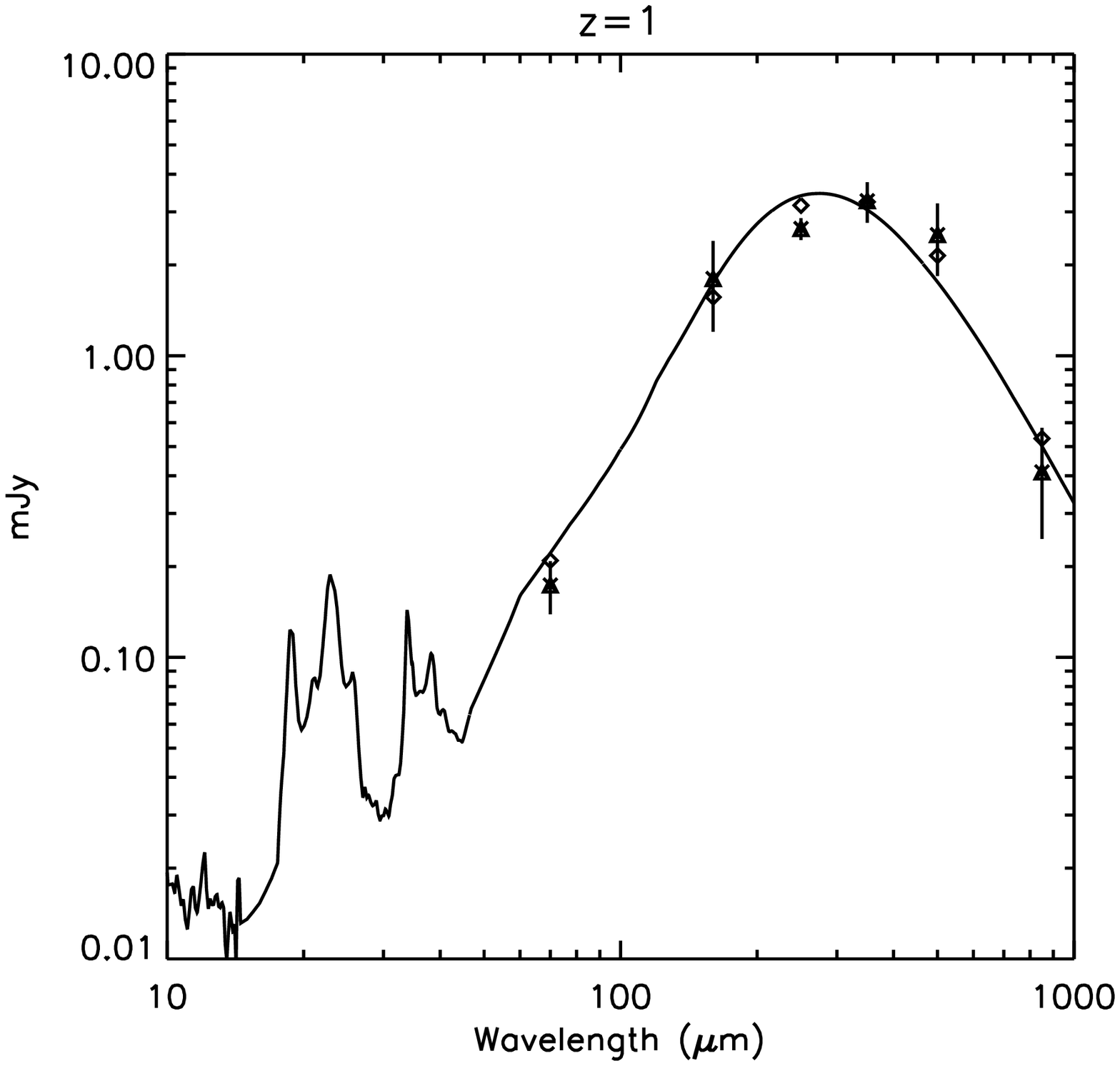}\includegraphics[width=0.5\columnwidth]{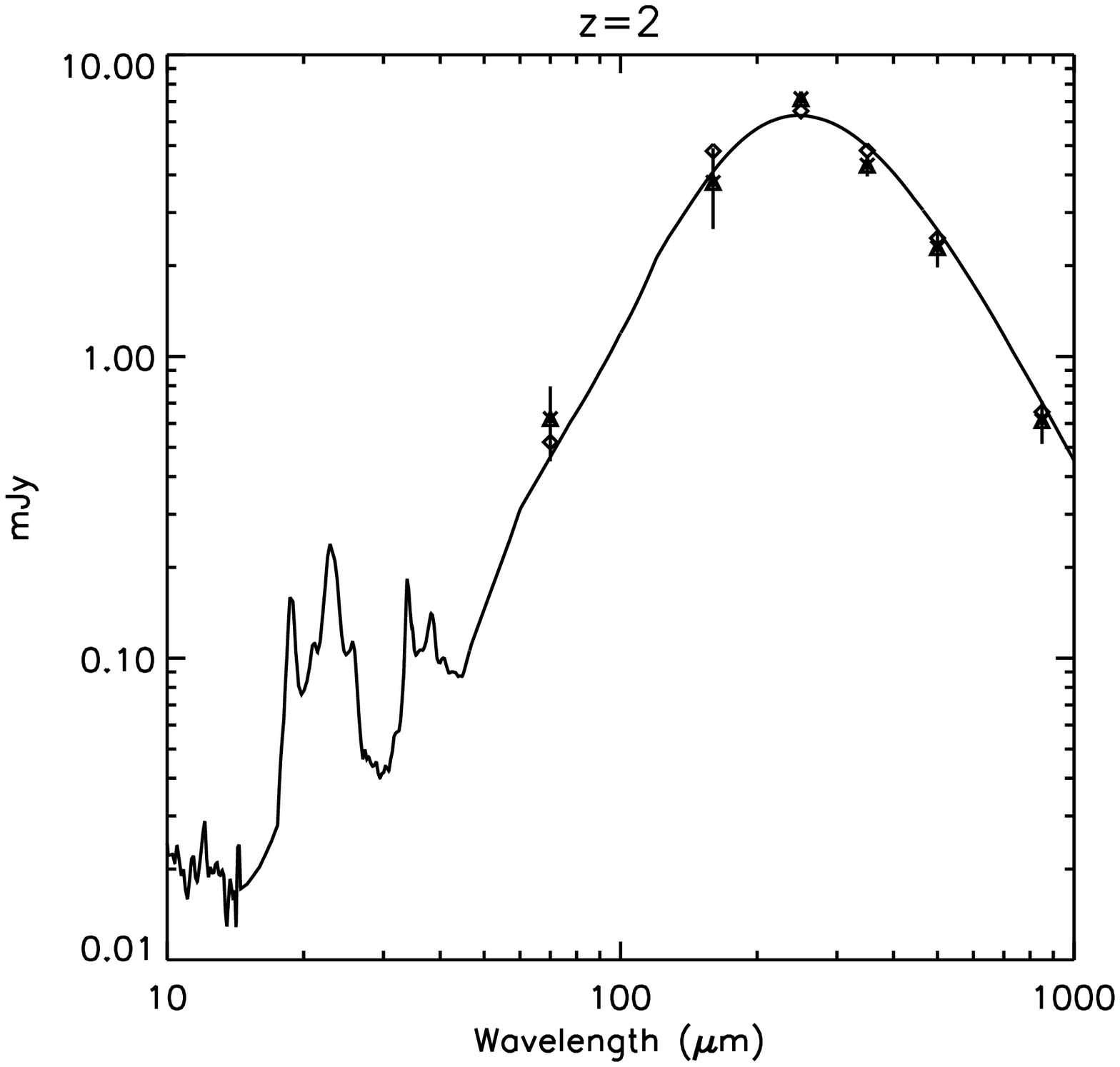}
\caption{True mean fluxes (diamonds) compared to the mean
fluxes found by stacking (triangles) with the estimated errors at
70, 160, 250, 350, 500, and 850~$\mu$m of the 800 faintest sources
above $80\,\mu$Jy (mean fluxes $S_{24}=110\:\mu$Jy and $S_{24}=135\,\mu$Jy
at redshifts $1<z<1.1$ and $2<z<2.1$, respectively) in our simulated
COSMOS observation. The points at 70 and 160~$\mu$m are from a Spitzer
simulation with instrumental noise taken from \citet{2007ApJS..172...86S}. The
solid lines show the Starburst SEDs from the library of \citet{2004ApJS..154..112L} that has the same 
mean $S_{24}$ at those two redshifts.\label{fig:SED-Stacked-Real}}
\end{figure}

The complete mean SEDs for the different populations can provide
information about the mean galaxy properties, such as 
star-formation rate and dust content. Figure \ref{fig:SED-Stacked-Real}
shows our measurements at 70, 160, 250, 350, 500, and 850~$\mu$m
of the flux of the 800 faintest sources detected in our simulated
COSMOS survey at $1<z<1.1$ and at $2<z<2.1$ relative to both their true fluxes
and the SED of a typical source at these fluxes and redshifts. The largest errors are
found at 70~$\mu$m, 160~$\mu$m, and 850~$\mu$m. For both redshifts,
the errors in our estimates are smaller than $10\,\%$. The same
method could be applied to fainter populations, if they were detected
individually with Spitzer. As mentioned before, the limitation of the method is
the detection limit of the Spitzer observations at 24~$\mu$m.

\section{Cleaning maps of undetected source populations\label{sec:Cleaning}}

\subsection{Contribution to the CIB \label{sec:Contribution-to-CIB}}
An obvious application of the results
provided by the stacking technique is the measurement of the total energy emitted by
different galaxy populations at wavelengths where they can not be seen
directly. This would give us the CIB fraction at those wavelengths
coming from the chosen population. We compare the total contribution from 
sources brighter than $S_{24}=80\,\mu$Jy at redshifts $z < 2$ in
our simulations with that determined using the stacking technique, and obtain very similar results.
At 350~$\mu$m, we find (using our stacking estimates) that these 
sources account for $35.4\%$ and $35.8\%$ of the CIB  when the redshift errors are 
$3\%$ and $10\%$, respectively. This is a $0.4\%$ and $0.8\%$ overestimate 
of their contribution ($35\%$) to the CIB of the underlying model.
At 850~$\mu$m, we estimate that these sources account for $19\%$ and
$20\%$  of the CIB when the redshift errors are $3\%$ and $10\%$, respectively, which
is a slight $2-3\%$ overestimate of their contribution ($17\%$) to the
CIB in the model. 

\subsection{Removing anisotropies due to low-z infrared galaxies}
A more sophisticated use of the present results is the statistical
removal of the contribution of these populations at long wavelengths.
If we accurately extract a sufficiently large fraction of the
background anisotropies at low $z$, this will allow us to study the CIB anisotropies at high
$z$. For the first time, we could then separate the contributions to
the CIB anisotropies at different redshifts. 
This would allow us to study large-scale
structures at high redshift.\\

\begin{figure}
\includegraphics[width=0.48\columnwidth]{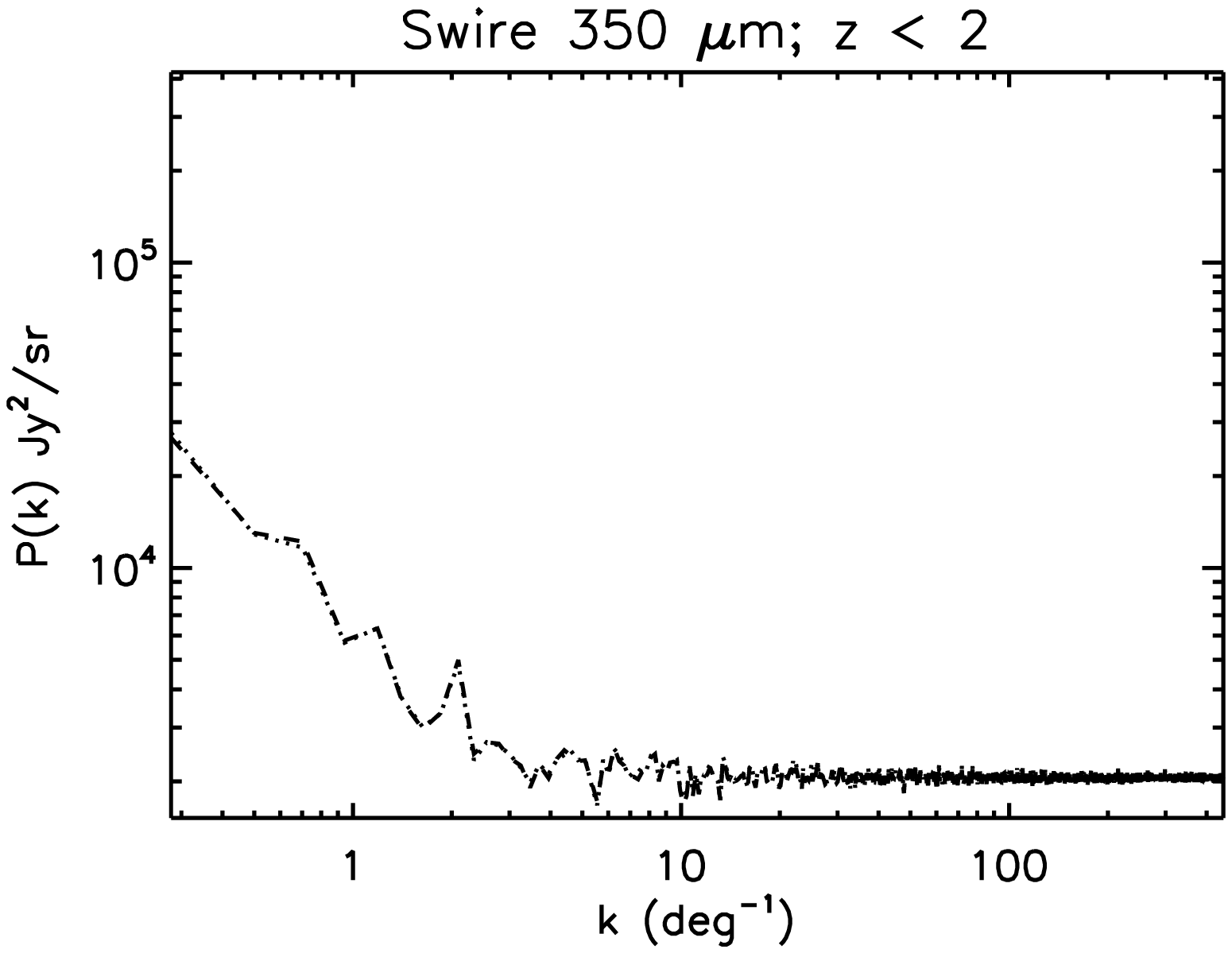}\includegraphics[width=0.505\columnwidth]{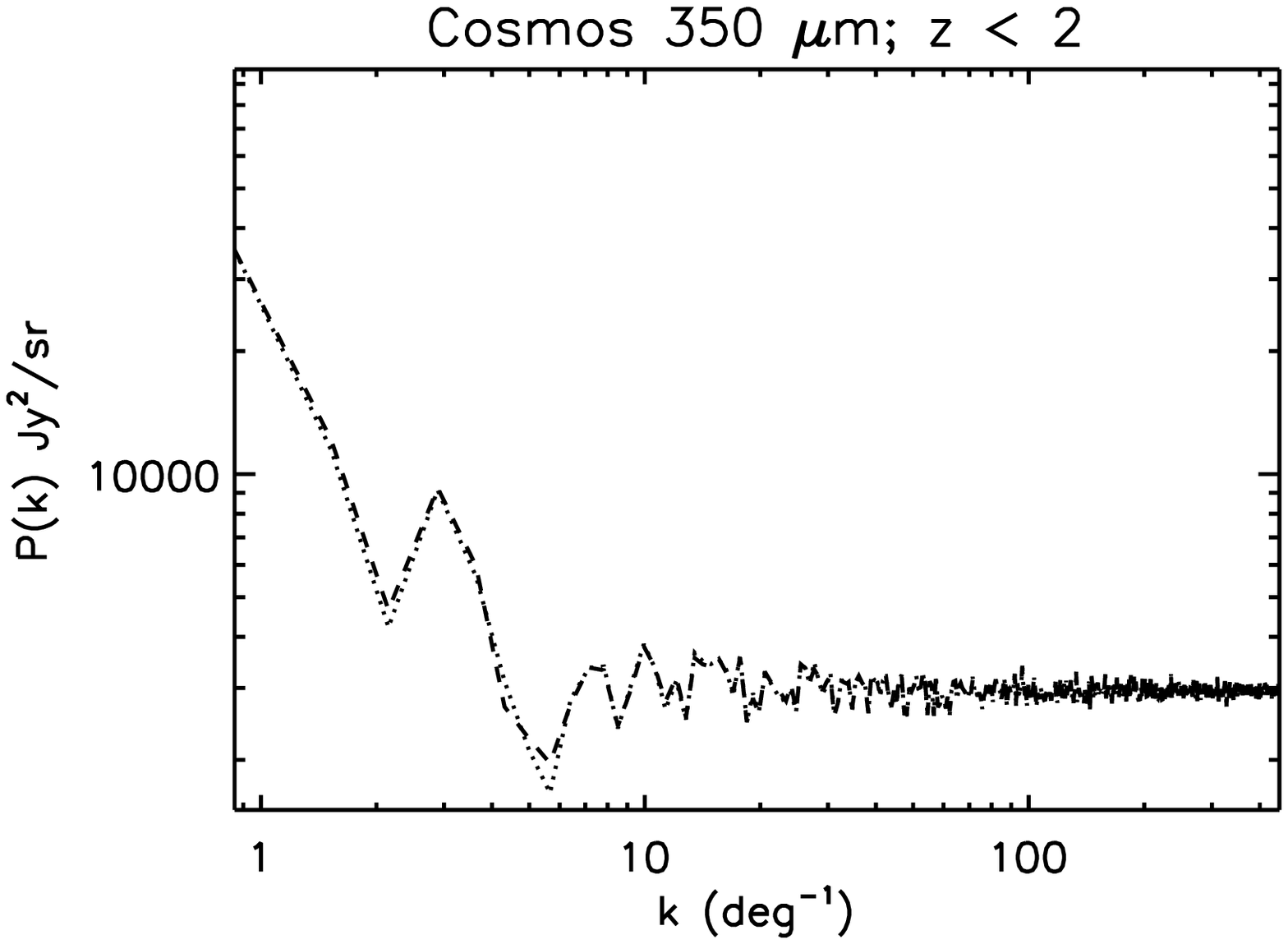}
\includegraphics[width=0.48\columnwidth]{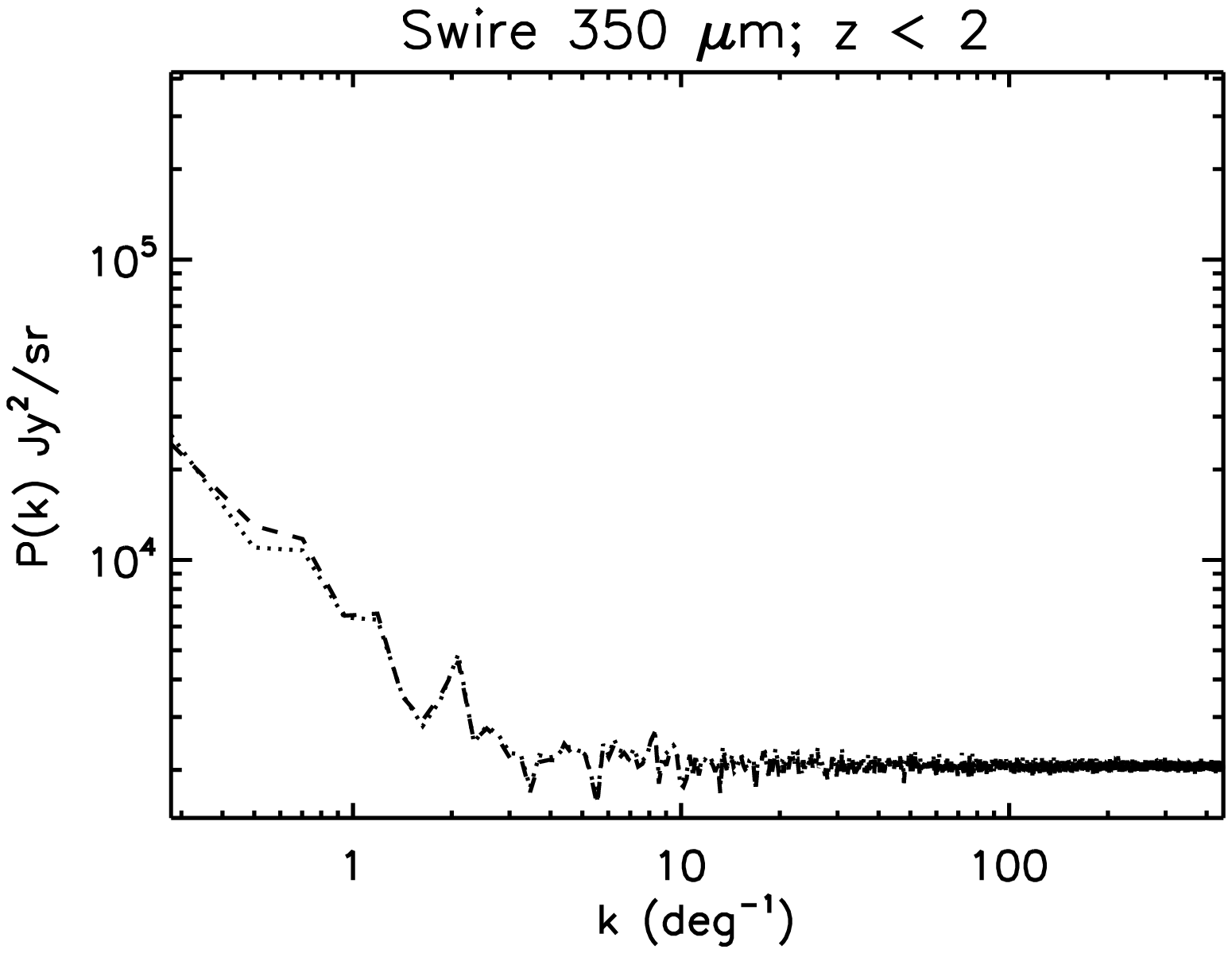}\includegraphics[width=0.505\columnwidth]{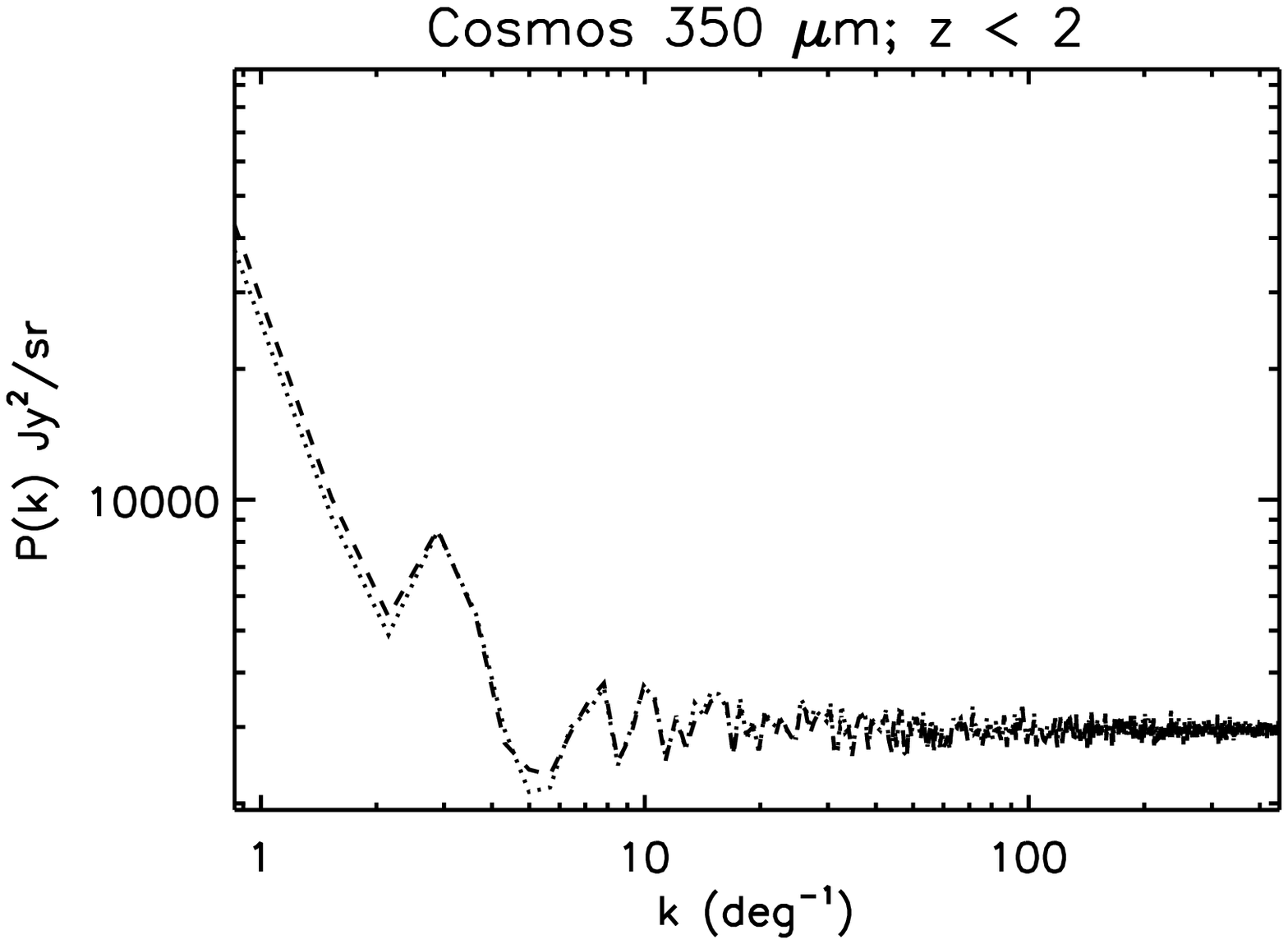}
\caption{Power spectra of two maps in which we placed the sources
with either their input fluxes from the simulations (dotted line) or their stacked fluxes (dashed
line). The results are shown for a SWIRE observation (left figures)
and COSMOS observation (right figures) at 350~$\mu$m for stacked
sources up to $z=2$ with redshift errors $\frac{\triangle z}{z}=3\%$ (top)
and $\frac{\triangle z}{z}=10\%$ (bottom).\label{fig:P0-350-3&10}}
\end{figure}

\begin{figure}
\includegraphics[width=0.51\columnwidth]{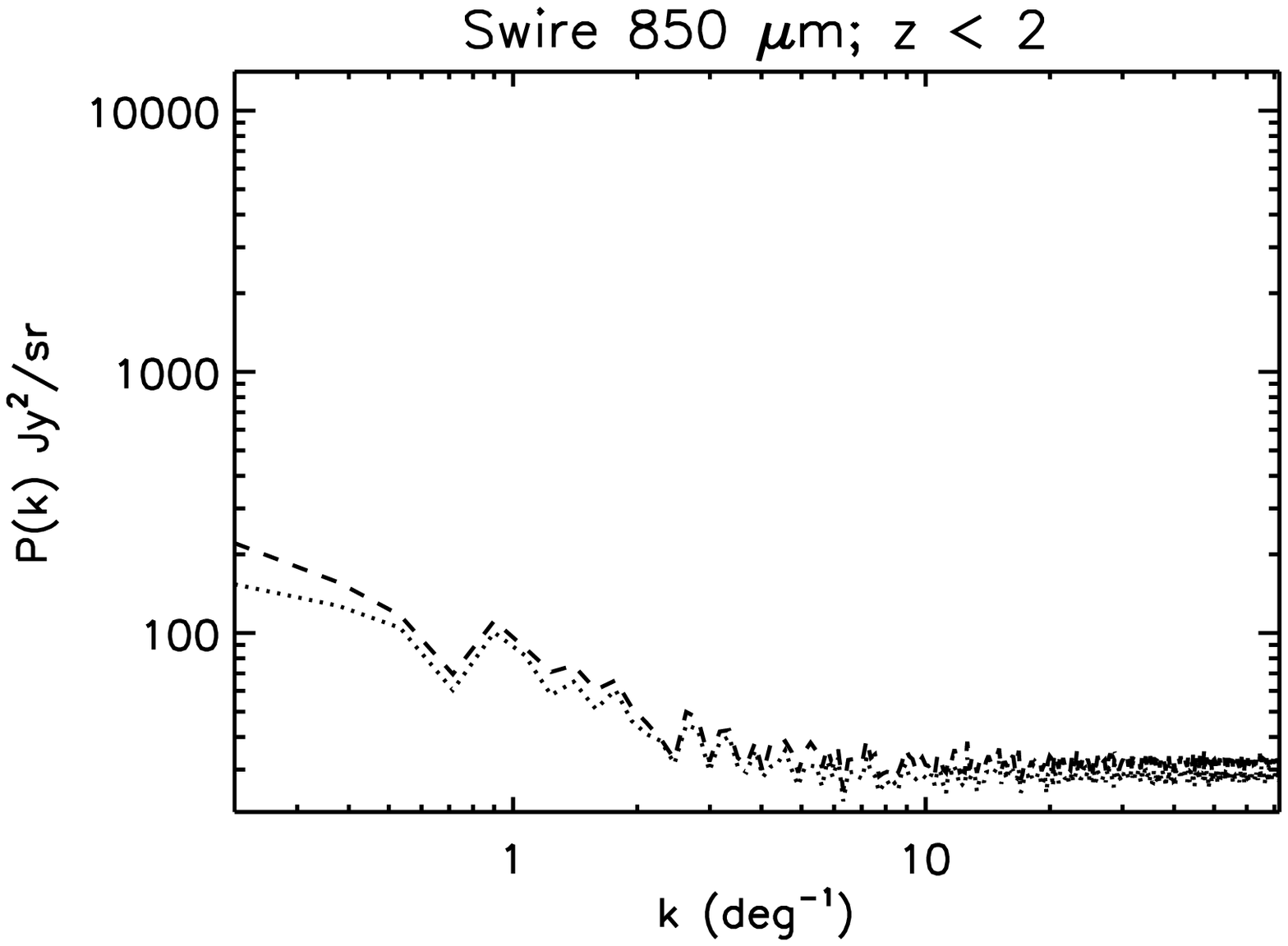}\includegraphics[width=0.500\columnwidth]{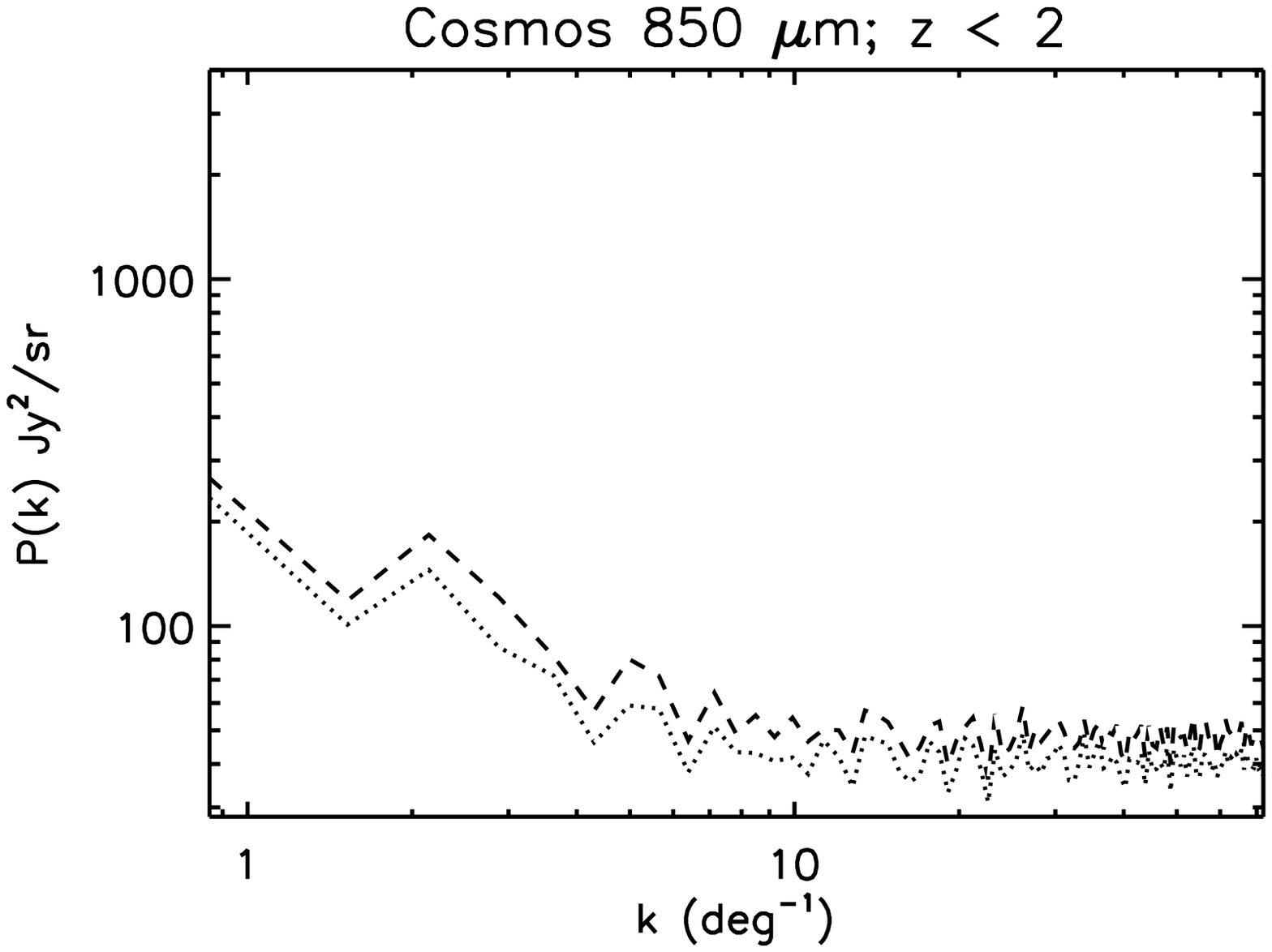}
\includegraphics[width=0.51\columnwidth]{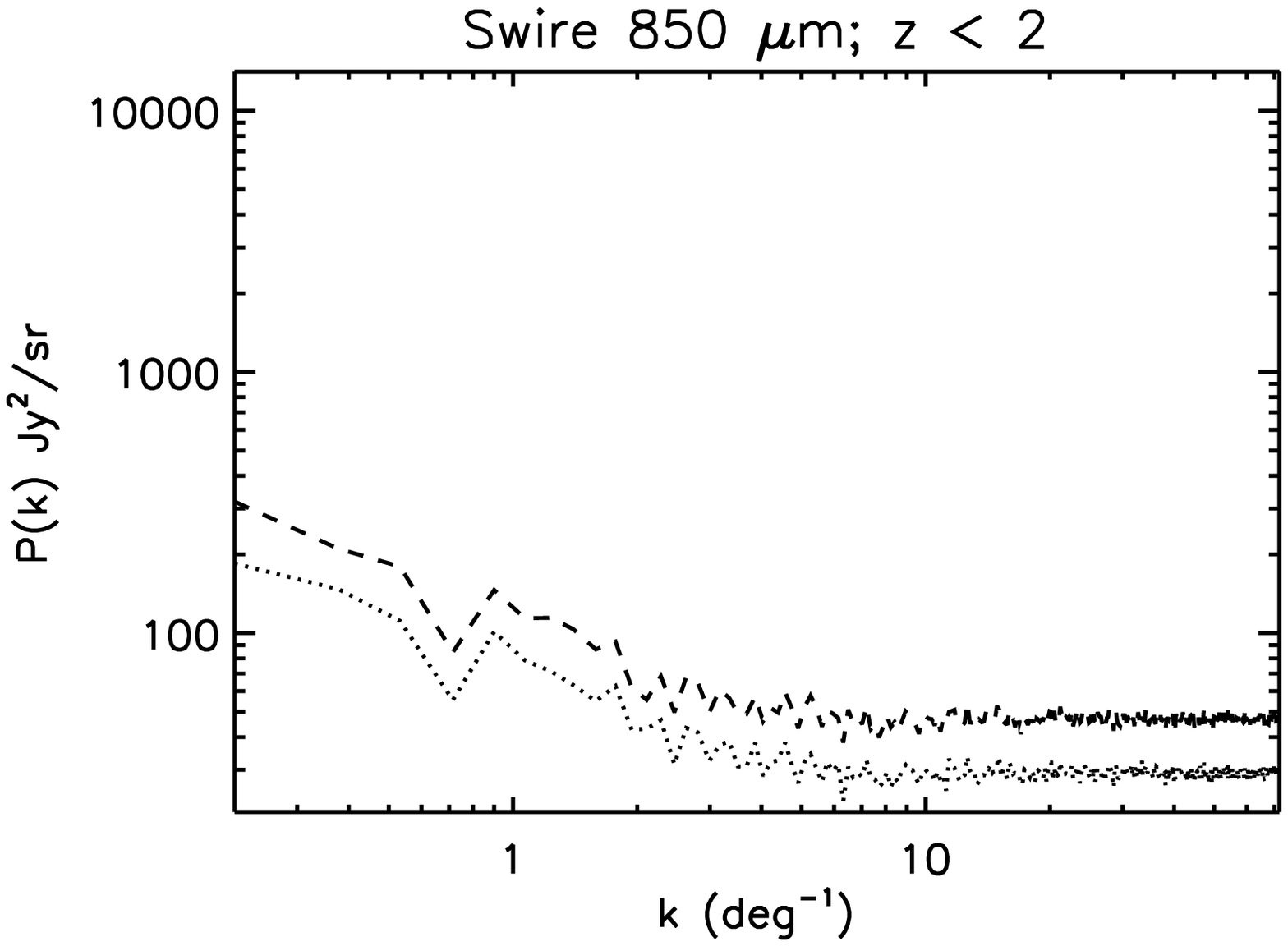}\includegraphics[width=0.500\columnwidth]{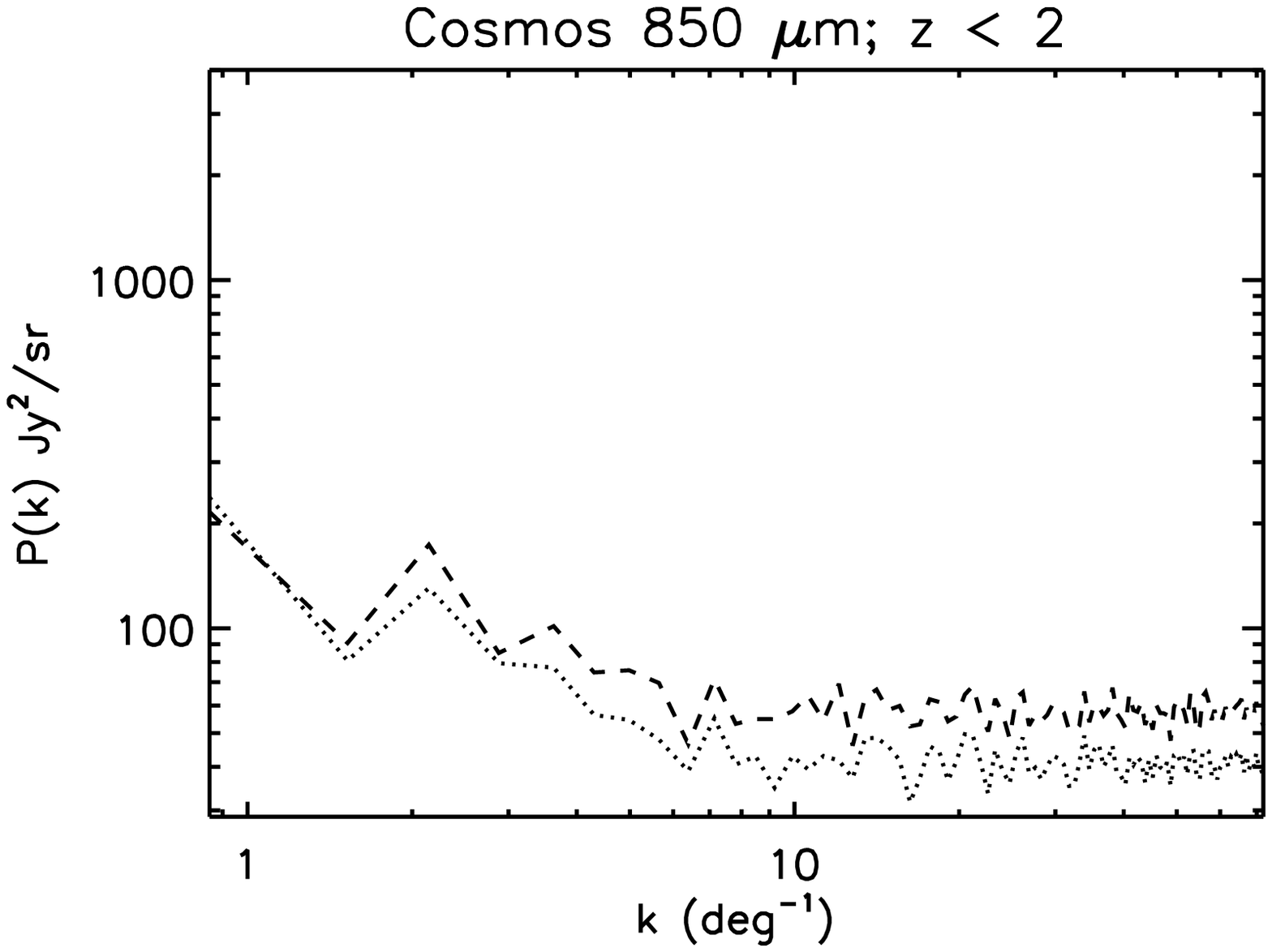}
\caption{Same as Fig. \ref{fig:P0-350-3&10} at 850~$\mu$m \label{fig:P0-850-3&10}}
\end{figure}

To remove from the observed maps the contribution of sources up to a certain redshift, 
we create a map of sources for whose fluxes were estimated using
the stacking technique. We subtract this map from the
observed maps, which is equivalent to individually subtracting all the
stacked sources. We estimate the source fluxes from the colors 
obtained by combining the different observations, as described
in Sect. \ref{sec:Combination-of-different}. However, we know that
the flux estimates have significant errors for very
bright sources and sources at redshifts $z<0.1$
at 350~$\mu$m and $z<0.8$ at 850~$\mu$m. These errors will affect 
the accuracy of our removal of the low-$z$ background
anisotropies. We also studied the effect of a Gaussian dispersion
in the fluxes of the sources (as described in Sect. \ref{sec:Smoothing}) 
on the power spectra. For dispersions as high as $25 \%$, the results are equivalent with and
without dispersion. This is because of the large number of sources contributing to each bin.\\

To assess the importance of these errors, we compare
the map compiled using the flux estimates by stacking with a second
map where these sources have their true input fluxes. Comparing
the power spectrum of both maps gives the accuracy of 
the anisotropy estimates for the first map. Figure \ref{fig:P0-350-3&10}
shows the two power spectra at 350~$\mu$m for sources at $z<2$ for
both a SWIRE observation (with $S_{24}>270\,\mu$Jy) and a COSMOS
observation (with $S_{24}>80\,\mu$Jy) and for two redshift errors
$\frac{\triangle z}{z}=3\%$ and $\frac{\triangle z}{z}=10\%$. 
At $350\,\mu m$, the accuracy of our estimation is superior to
$0.5\%$ for both the correlated and Poissonian part of the
spectrum in both the SWIRE and COSMOS observations in the case of a
small redshift error ($\frac{\triangle z}{z}=3\%$). When the redshift error is greater, our
estimate of the Poissonian noise increases moderately with mean errors of
$3\%$. Figure \ref{fig:P0-850-3&10} shows the same result at 850~$\mu$m. Because of the small
redshift error in the COSMOS survey, we overestimate the correlated 
part by $40\%$ and the Poissonian part by $24\%$. For larger redshift errors,
our overestimates increase to $60\%$ and $50\%$ of the correlated and
Poissonian part, respectively. In this case, this shows the importance of 
accurate redshifts. The differences in the
overestimates of the Poissonian and correlated part
are caused by the populations contributing to these two
regimes not being exactly the same, bright sources contributing more
in relative terms to the Poissonian fluctuations than to the
correlated part.

\subsection{High-redshift power spectra of CIB anisotropies}

\subsubsection{Observations at 350~$\mu$m}

\begin{figure*}
\begin{centering}
\includegraphics[width=0.6\columnwidth]{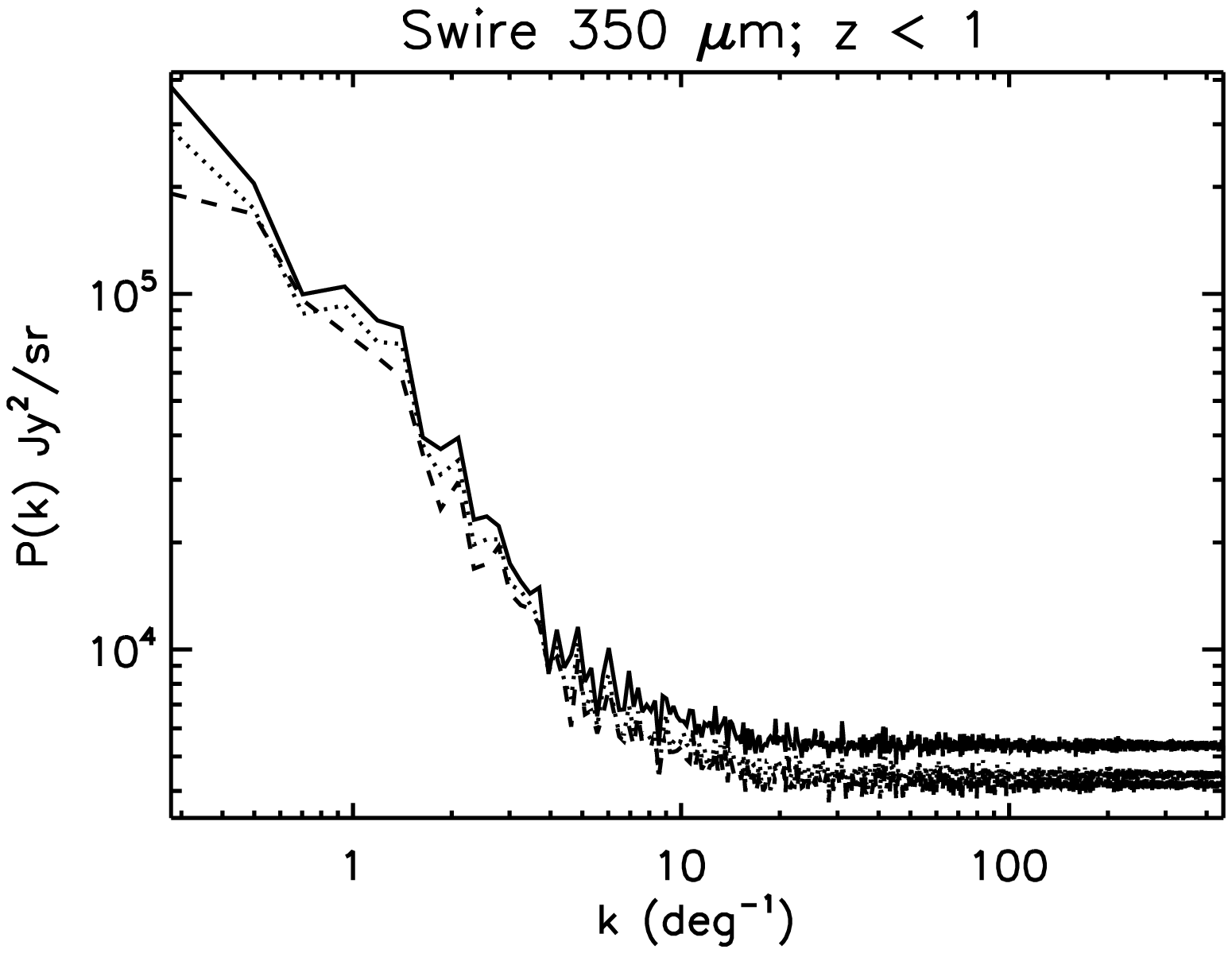}\includegraphics[width=0.67\columnwidth]{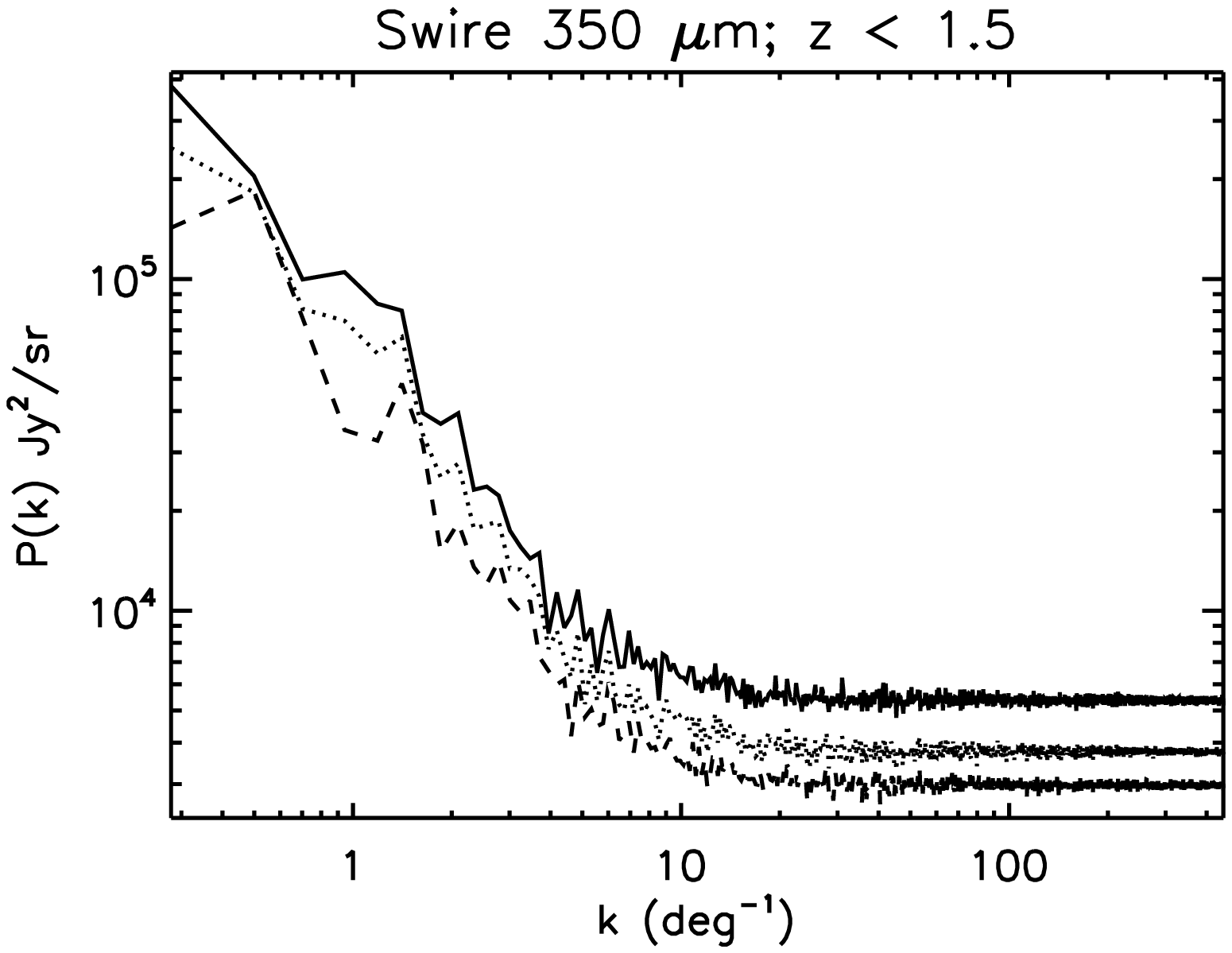}\includegraphics[width=0.6\columnwidth]{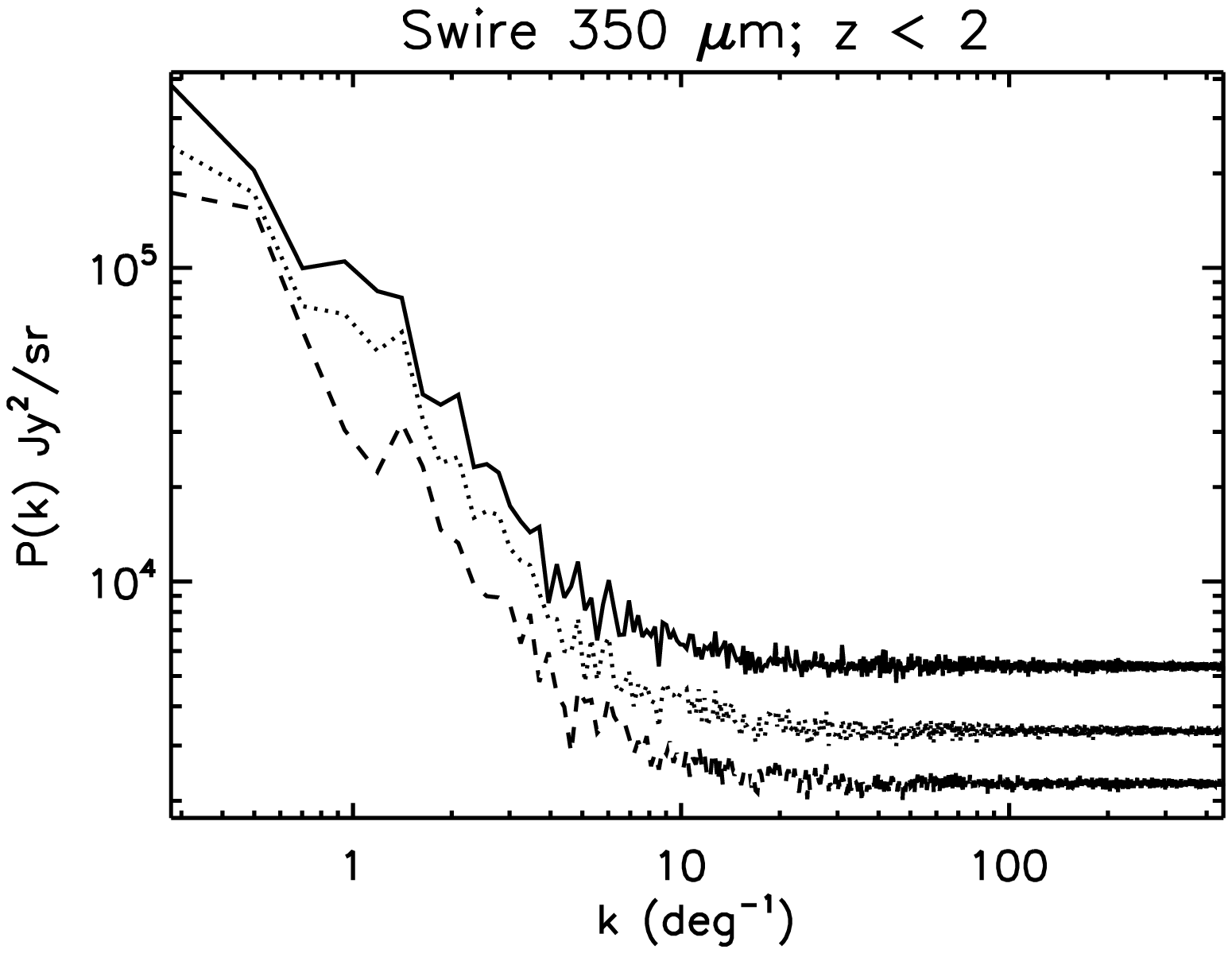}
\par\end{centering}
\begin{centering}
\includegraphics[width=0.6\columnwidth]{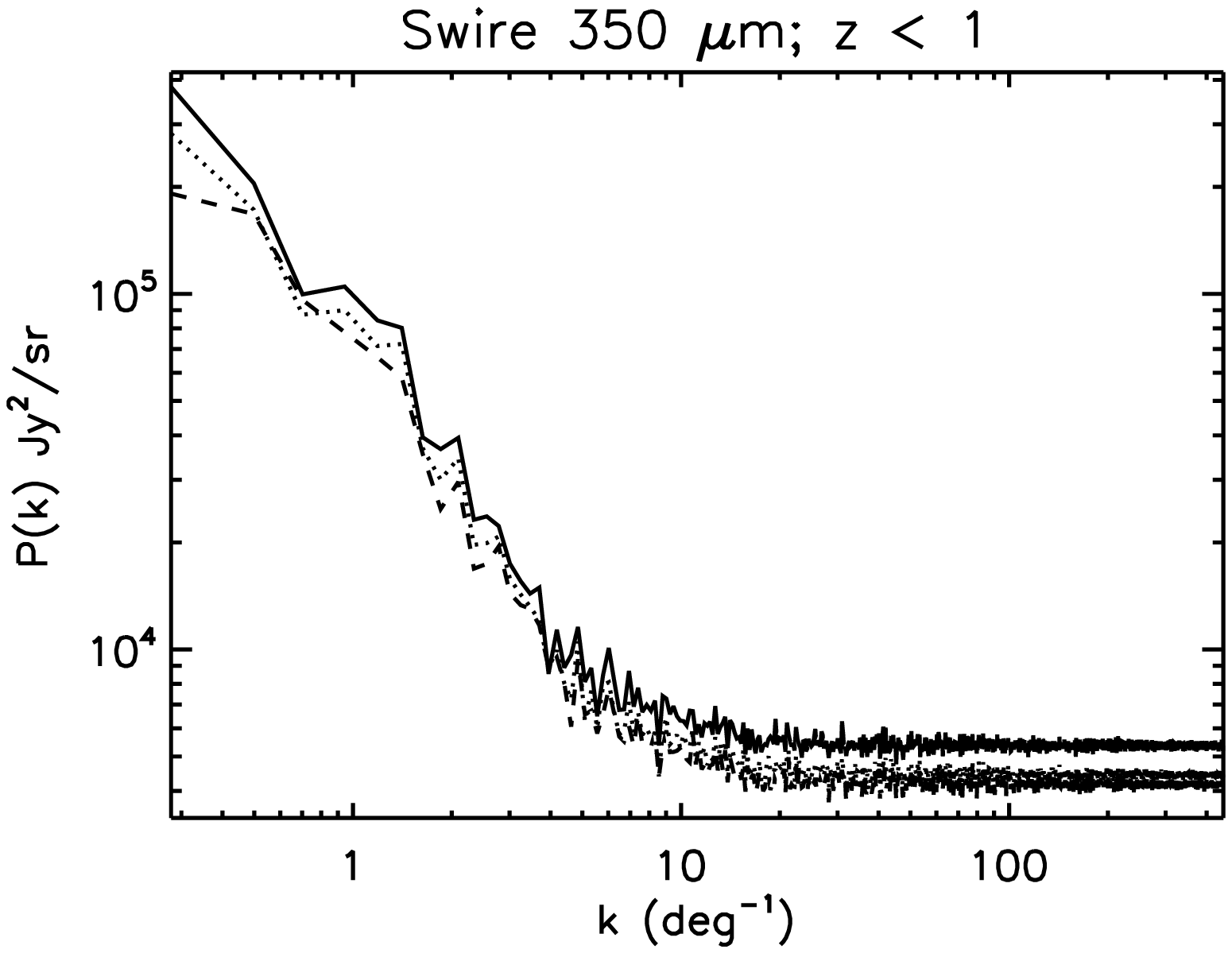}\includegraphics[width=0.67\columnwidth]{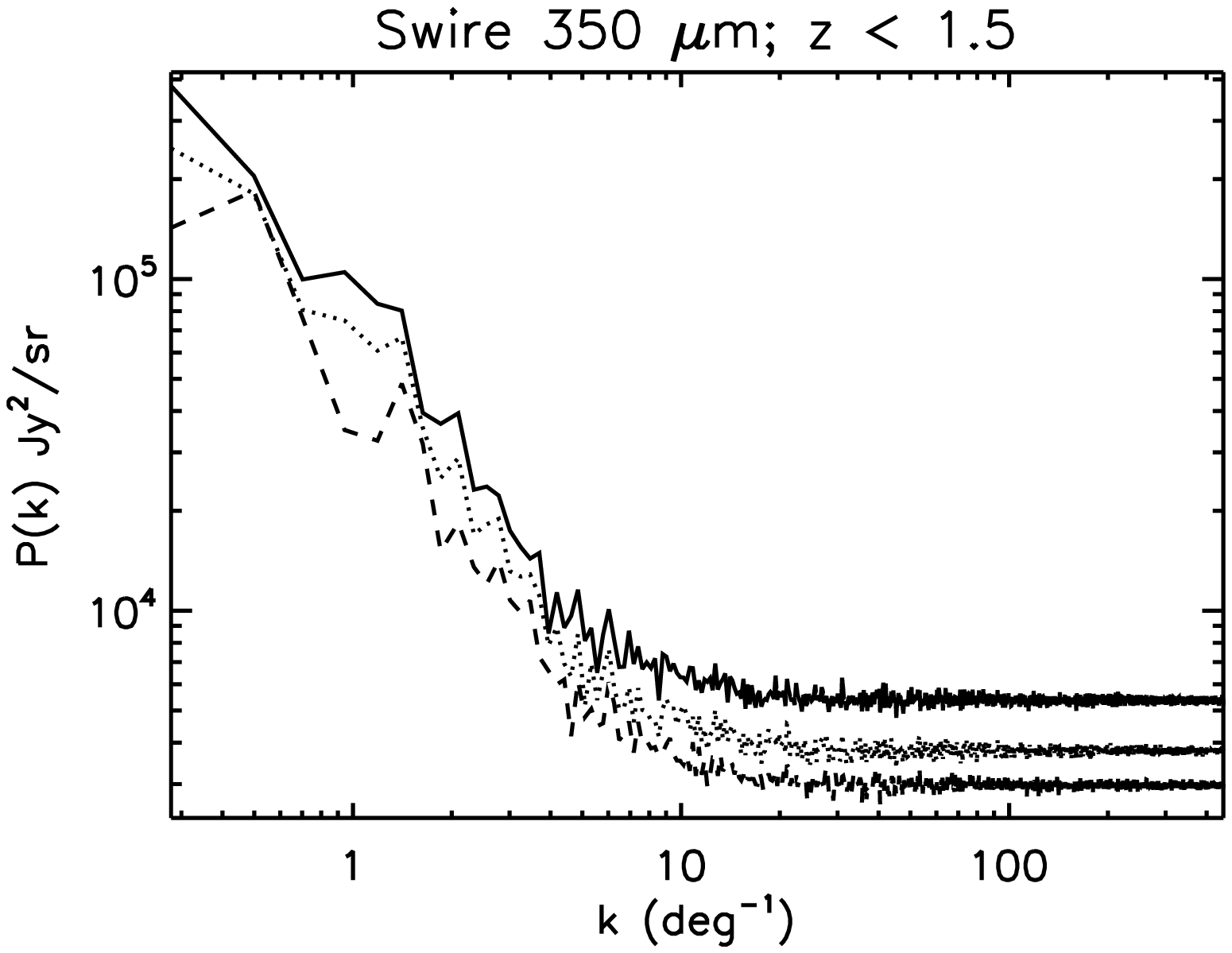}\includegraphics[width=0.6\columnwidth]{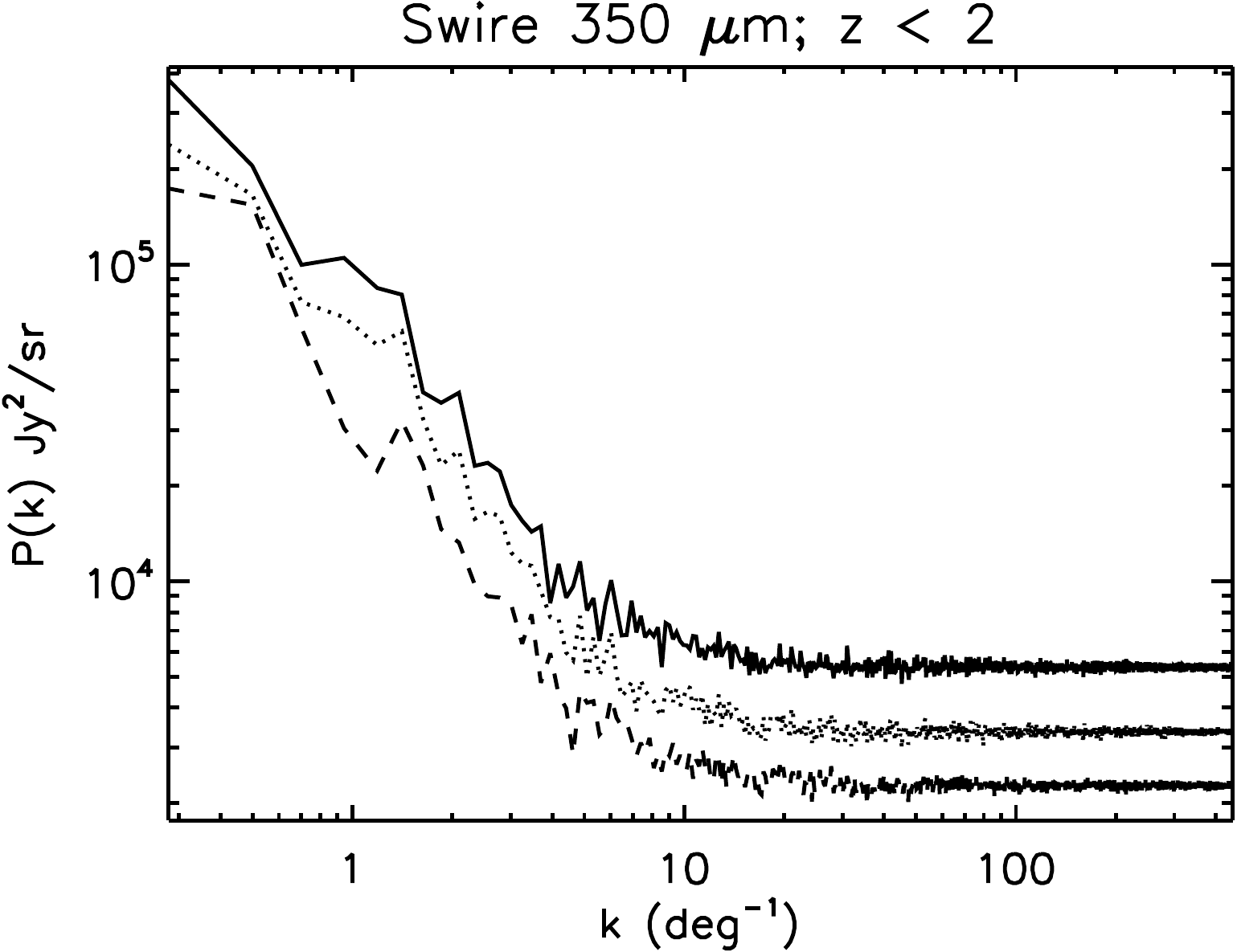}
\par\end{centering}
\caption{Power spectra of the map for a SWIRE observation at 350~$\mu$m. The
solid line is the total power spectrum of the background, the
dashed line is the power spectrum of the background for $z>z_{lim}$
(where $z_{lim}$ is a redshift limit), and the dotted line is the
power spectrum of the total background from which we have subtracted
the stacked sources at $z<z_{lim}$. The redshift limit $z_{lim}$
is $z_{lim}=1$ (left figures), $z_{lim}=1.5$ (middle figures), and
$z_{lim}=2$ (right figures). The redshift errors are $\frac{\triangle z}{z}=3\%$
(top) and $\frac{\triangle z}{z}=10\%$ (bottom). \label{fig:P2Swire-350-3-10=000025}}
\end{figure*}

After analyzing the accuracy of the map that we intend
to subtract, we investigate
our capabilities to subtract a significant part of the background anisotropies
for different redshift limits.
Figure \ref{fig:P2Swire-350-3-10=000025} compares the power spectra
of the total background anisotropies to those at $z>1$, $z>1.5$, and
$z>2$  in a SWIRE observation. It also shows the power spectra of the map of CIB anisotropies
from which we have subtracted the $z<1$, $z<1.5$,
and $z<2$ contribution, which were estimated by stacking. Since
our subtraction is rather accurate,
the very small difference between these last two sets of power spectra is caused by us
 not subtracting all the sources but only
those above $S_{24}>270\, \mu$Jy. 
We subtract
approximately half the correlated part ($k < 8$~deg$^{-1}$) and two thirds of the
Poissonian part ($k > 8$~deg$^{-1}$) independently of redshift errors.\\

\begin{figure*}
\begin{centering}
\includegraphics[width=0.65\columnwidth]{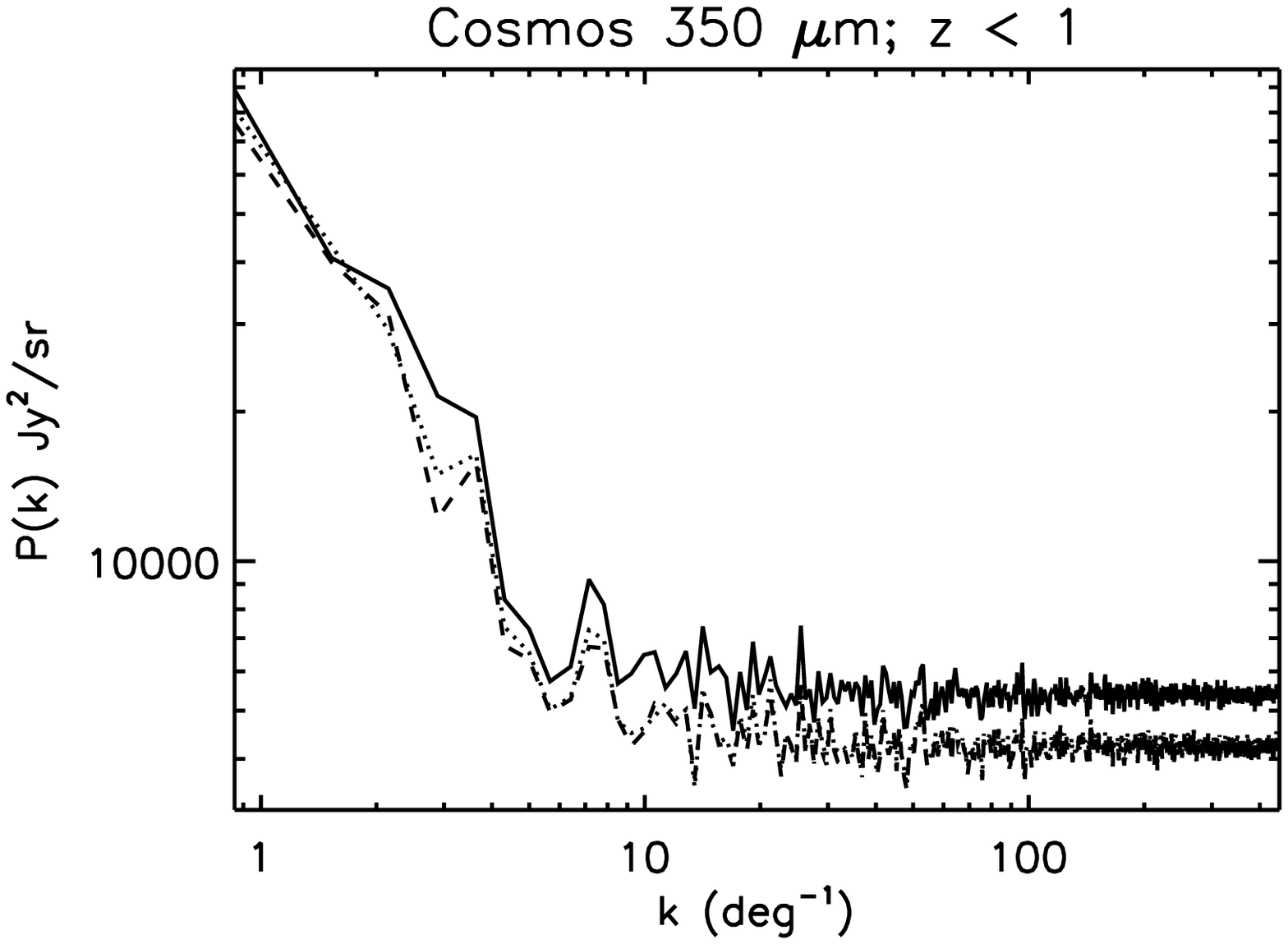}\includegraphics[width=0.65\columnwidth]{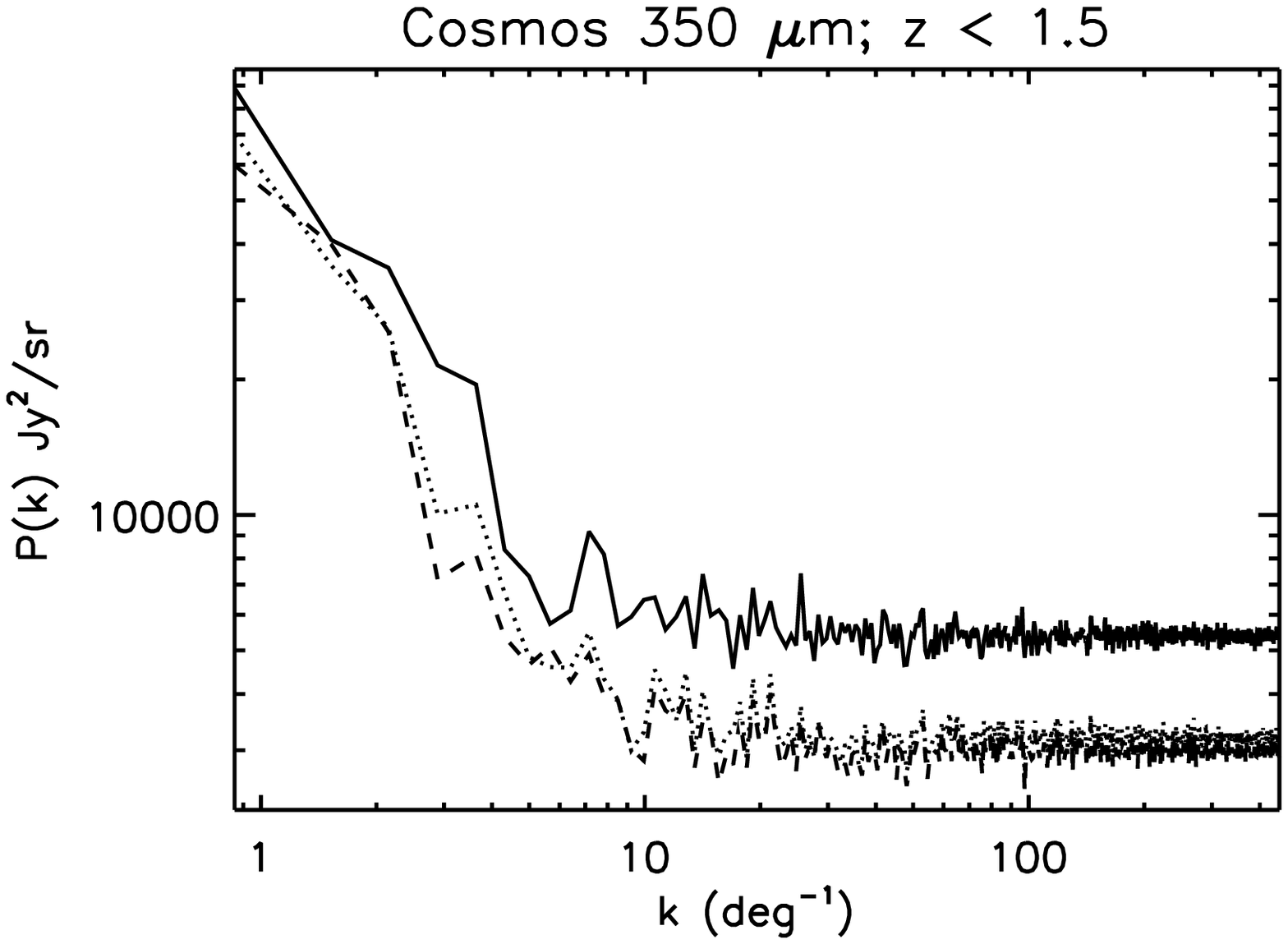}\includegraphics[width=0.65\columnwidth]{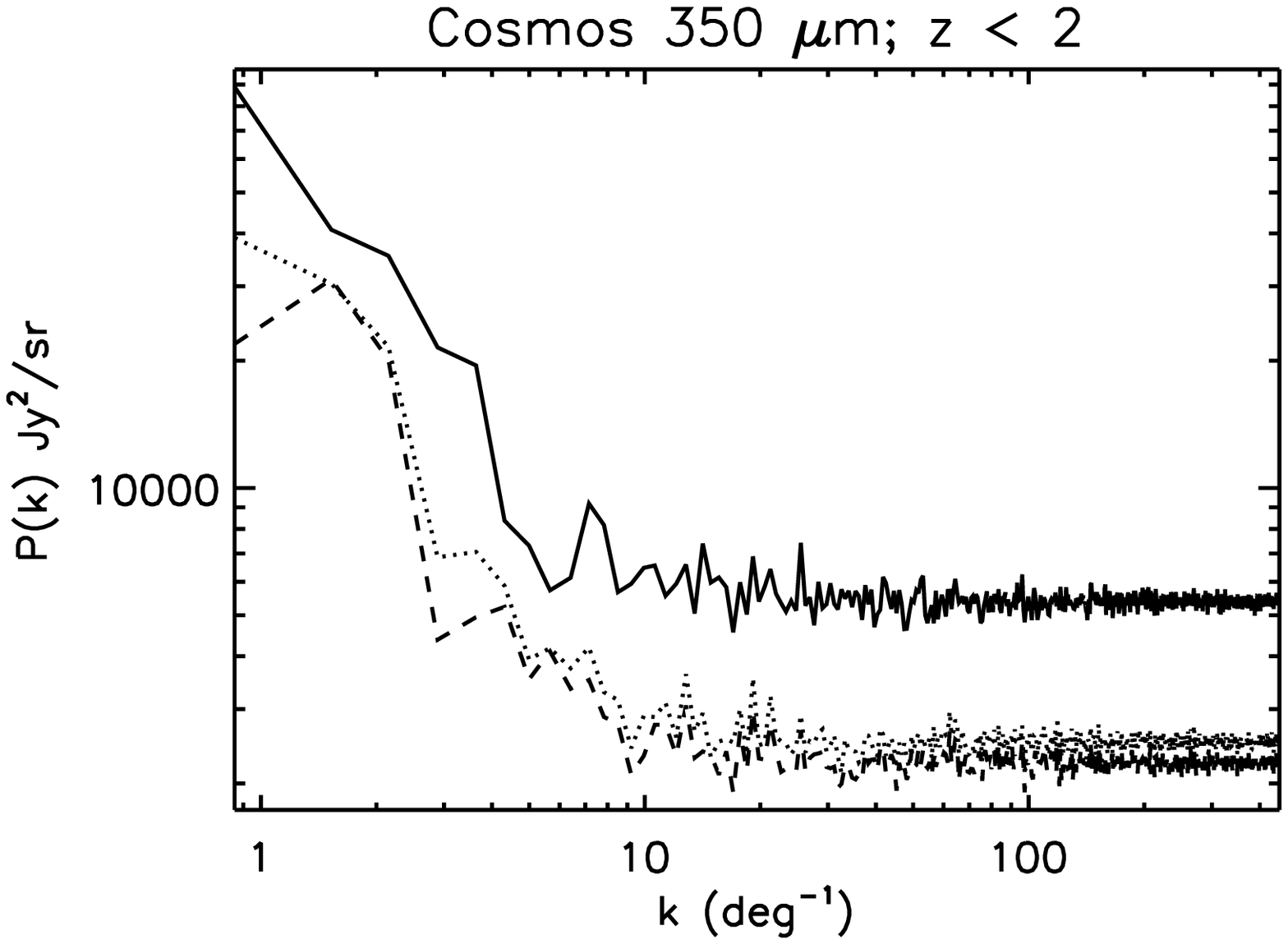}
\par\end{centering}
\begin{centering}
\includegraphics[width=0.65\columnwidth]{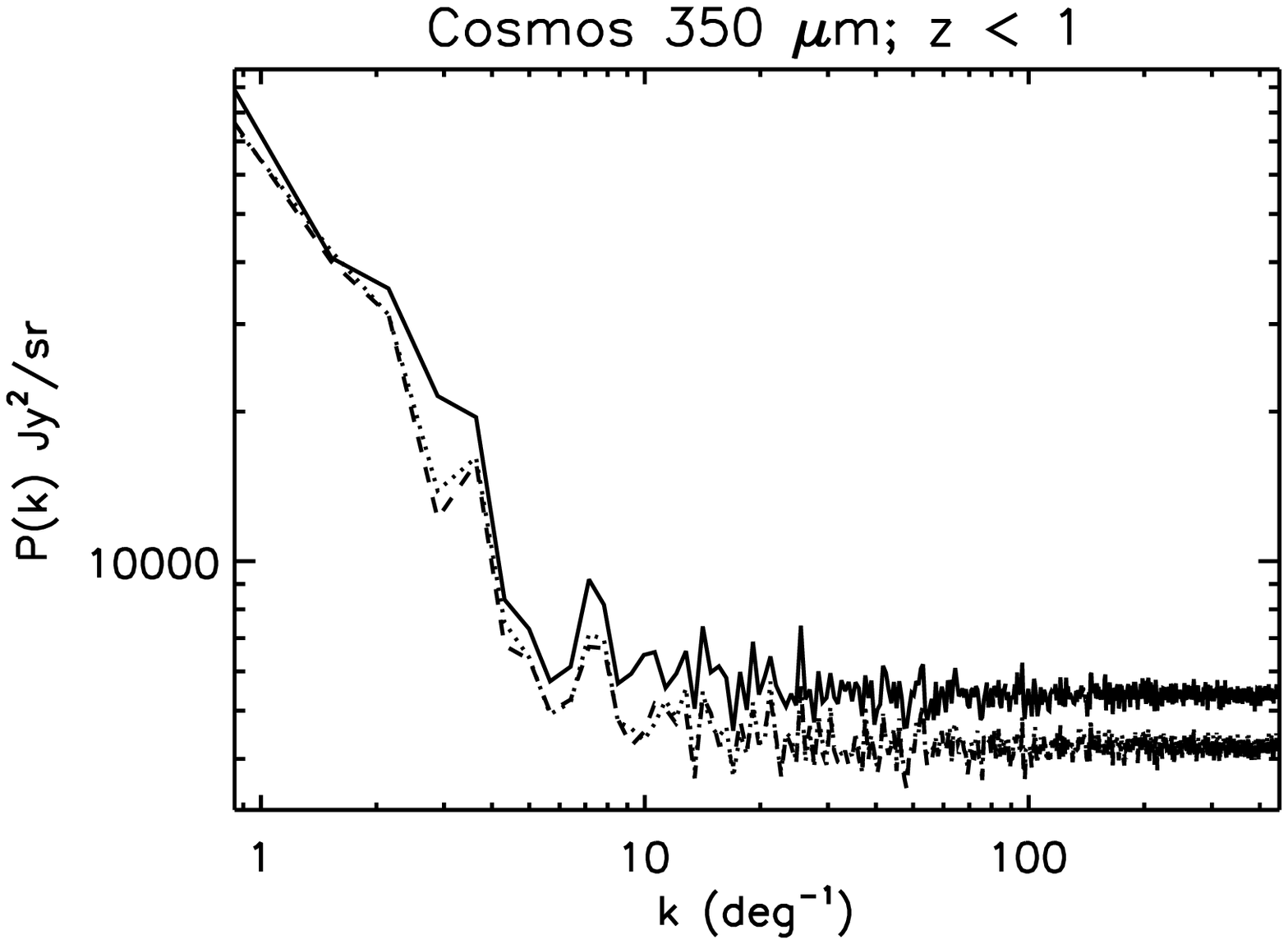}\includegraphics[width=0.65\columnwidth]{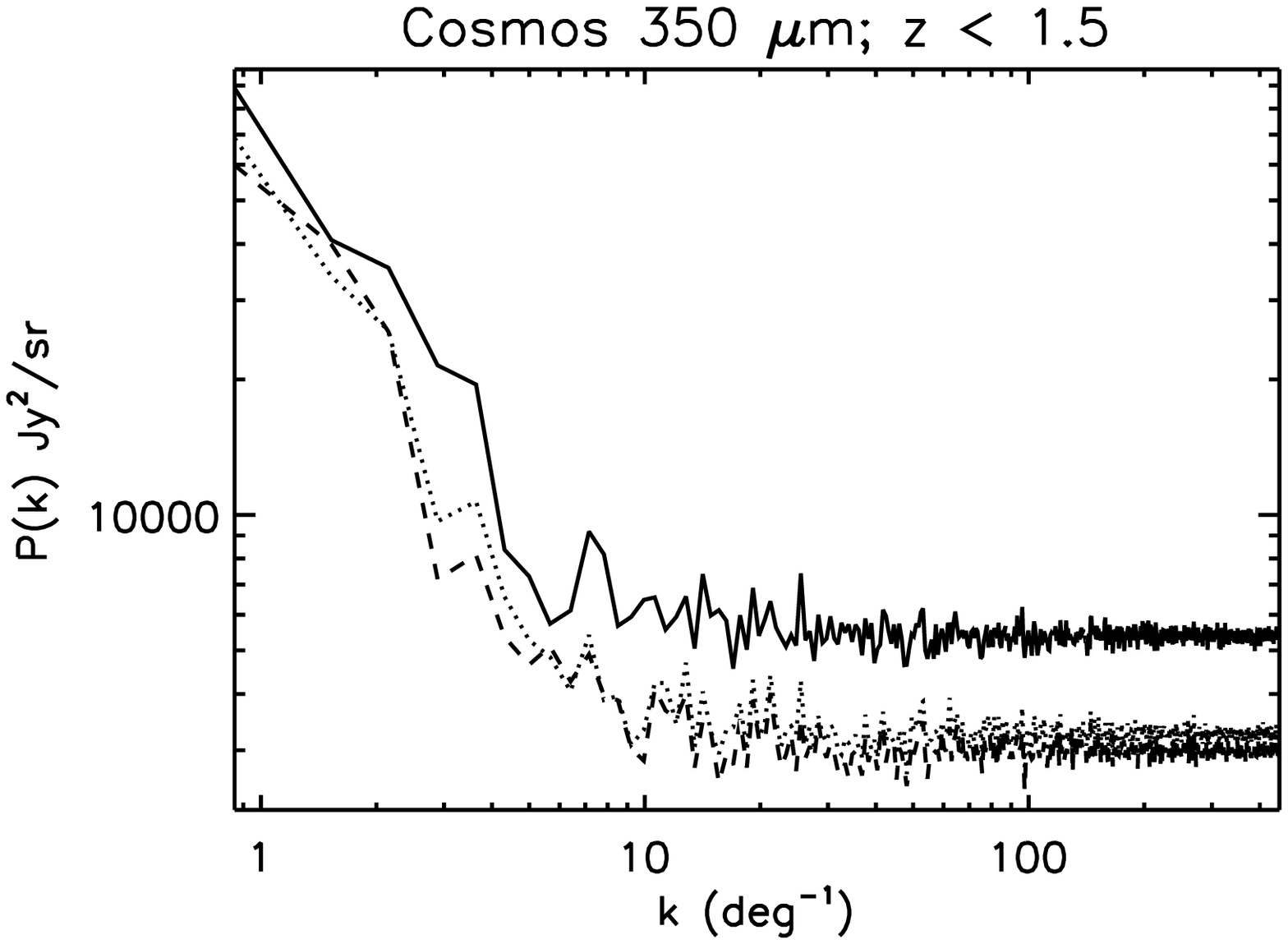}\includegraphics[width=0.65\columnwidth]{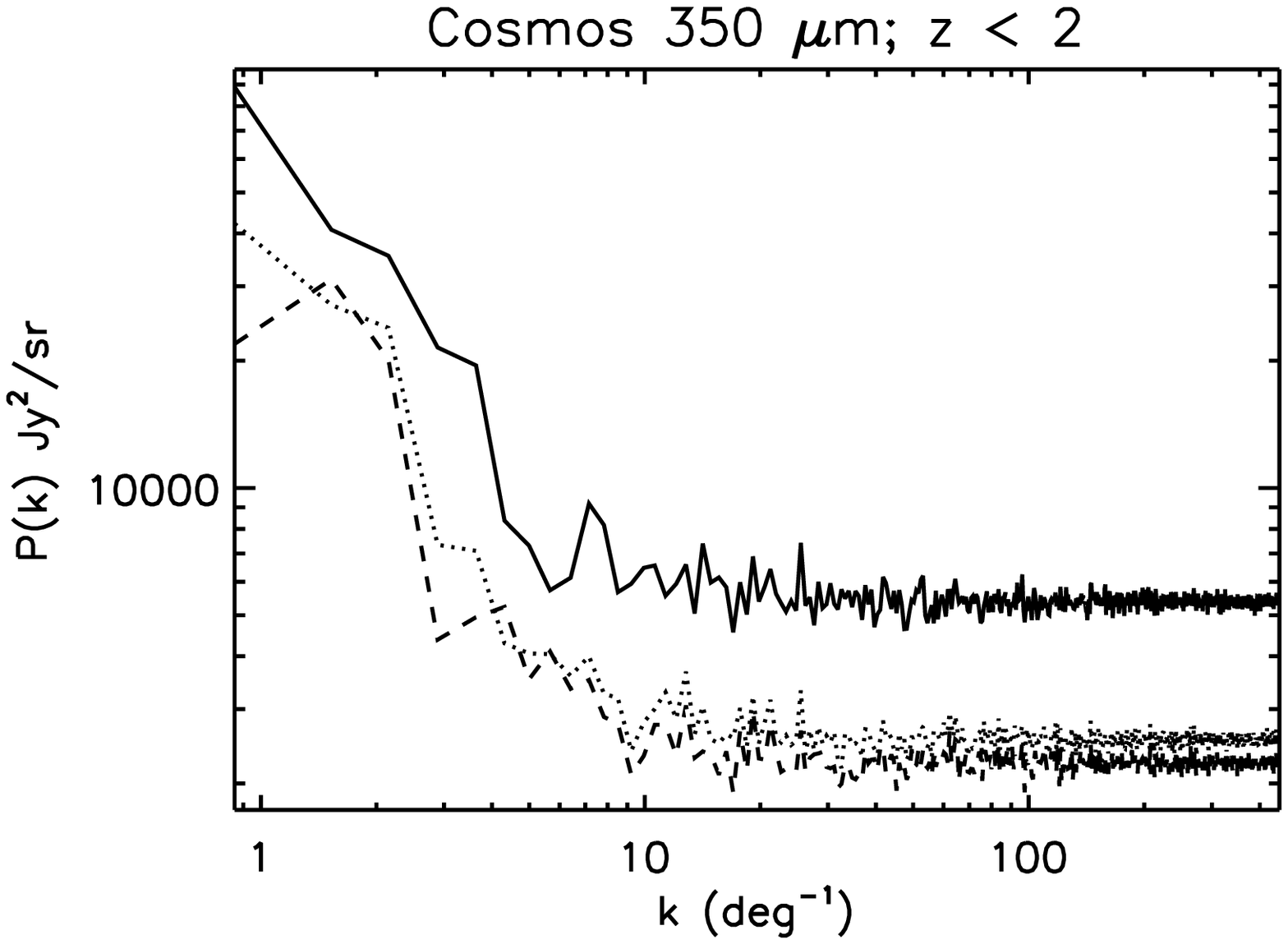}
\par\end{centering}
\caption{Same as Fig. \ref{fig:P2Swire-350-3-10=000025} but for the COSMOS field.\label{fig:P2Cosmos-350-3&10=000025}}
\end{figure*}

Figure \ref{fig:P2Cosmos-350-3&10=000025} shows the same results
for a COSMOS observation. We have
the positions of sources with $S_{24}>80\,\mu$Jy which allows
us to subtract a larger fraction of the background than in the SWIRE
survey. Unfortunately because of the smaller size of the field, we do not have access 
to the largest scales that we were able to analyze with SWIRE. 
We subtract approximately $\sim 99\%$ of the correlated part and $\sim 90\%$ of the Poissonian
for the small redshift error. For the large redshift error, these
fractions become $\sim 85\%$ and $\sim 90\%$ of the correlated and Poissonian
parts, respectively. For each of the considered redshift limits, the power spectrum of the residual left after our
subtraction of the $z<z_{lim}$ stacked source is in close agreement with the power spectrum at high redshifts
($z > z_{lim}$). This remains true when we consider a large redshift error.

\subsubsection{Observations at 850~$\mu$m}

\begin{figure}
\begin{centering}
\includegraphics[width=0.48\columnwidth]{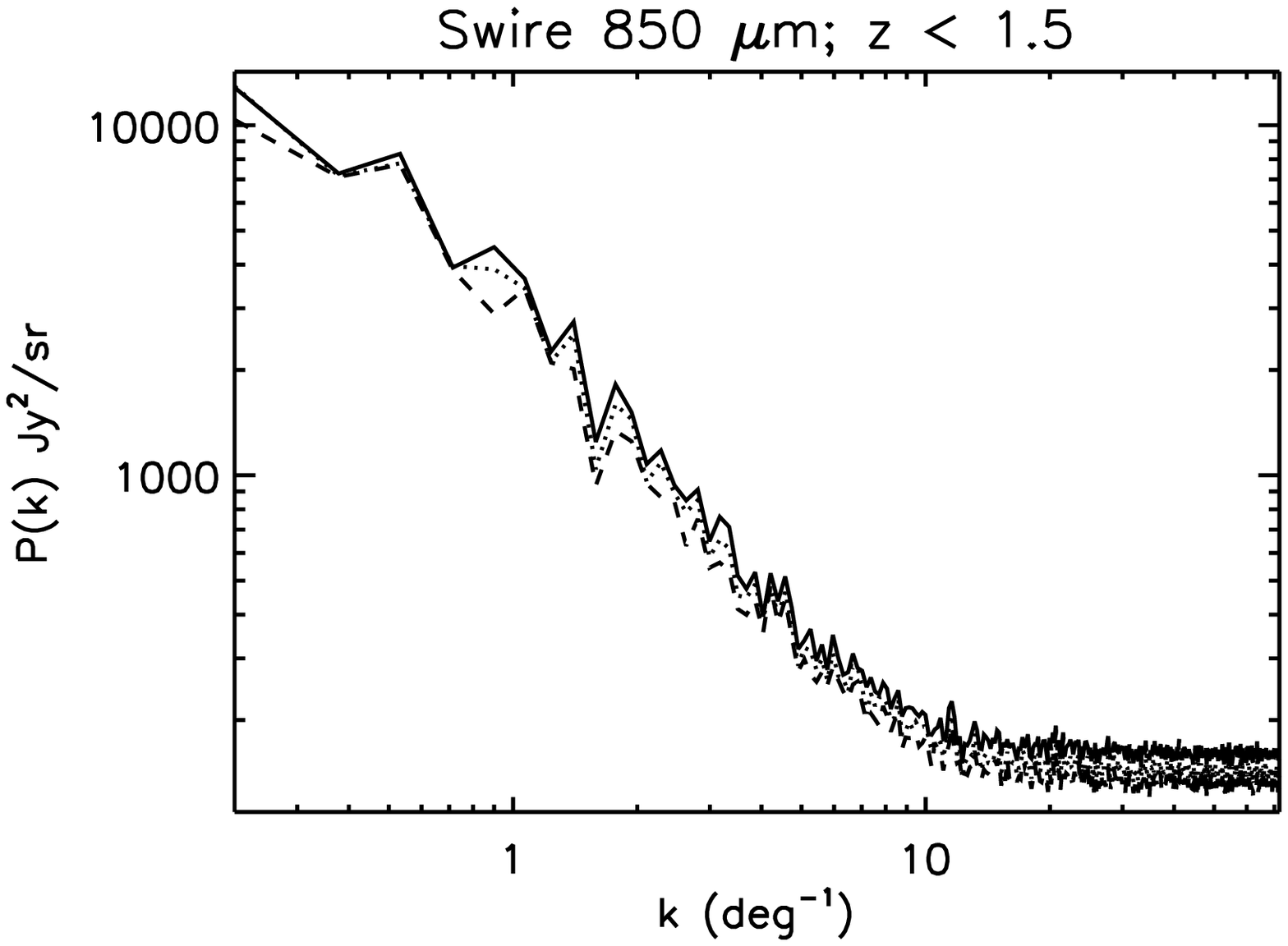}\includegraphics[width=0.48\columnwidth]{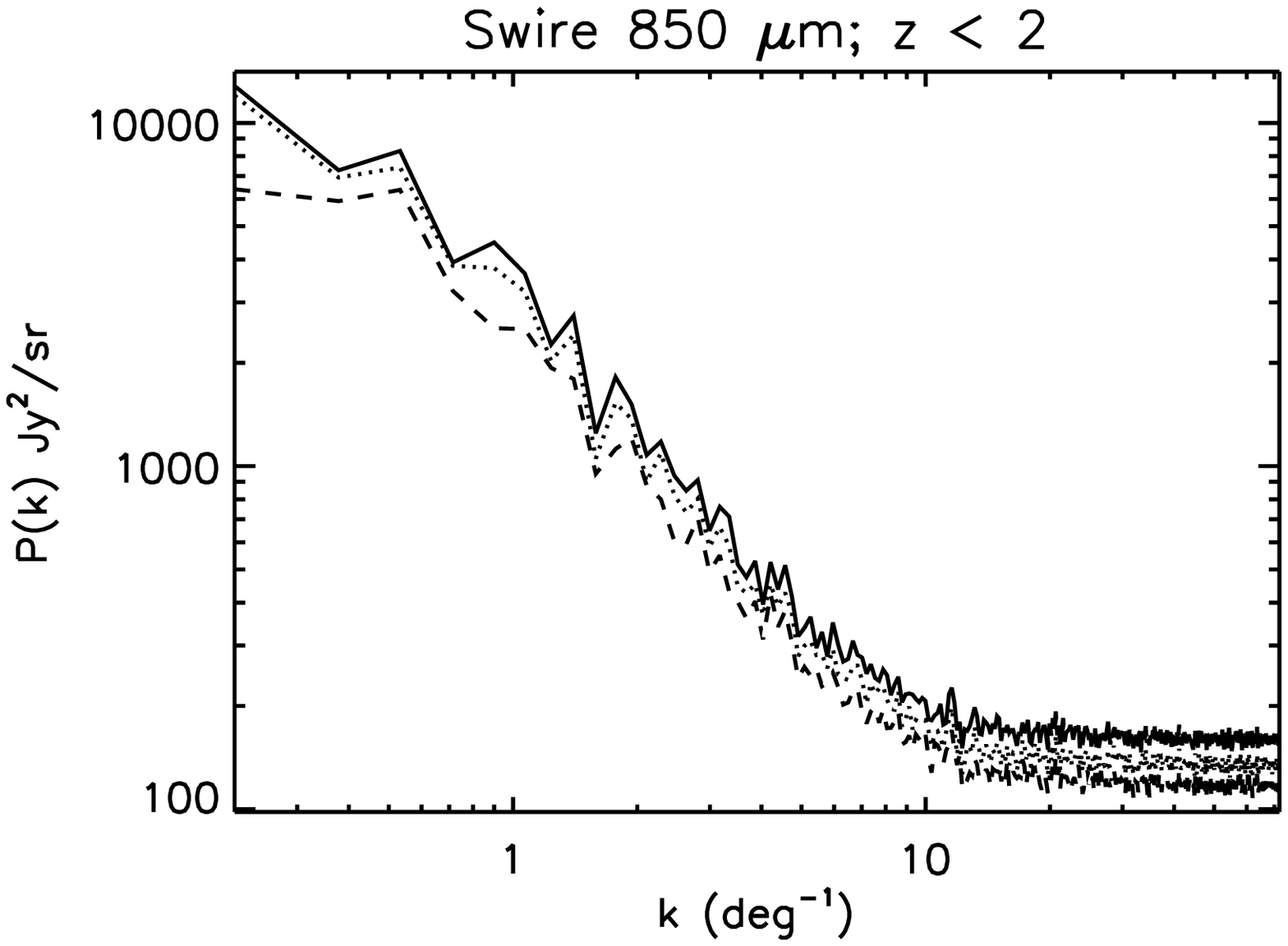}
\par\end{centering}
\begin{centering}
\includegraphics[width=0.48\columnwidth]{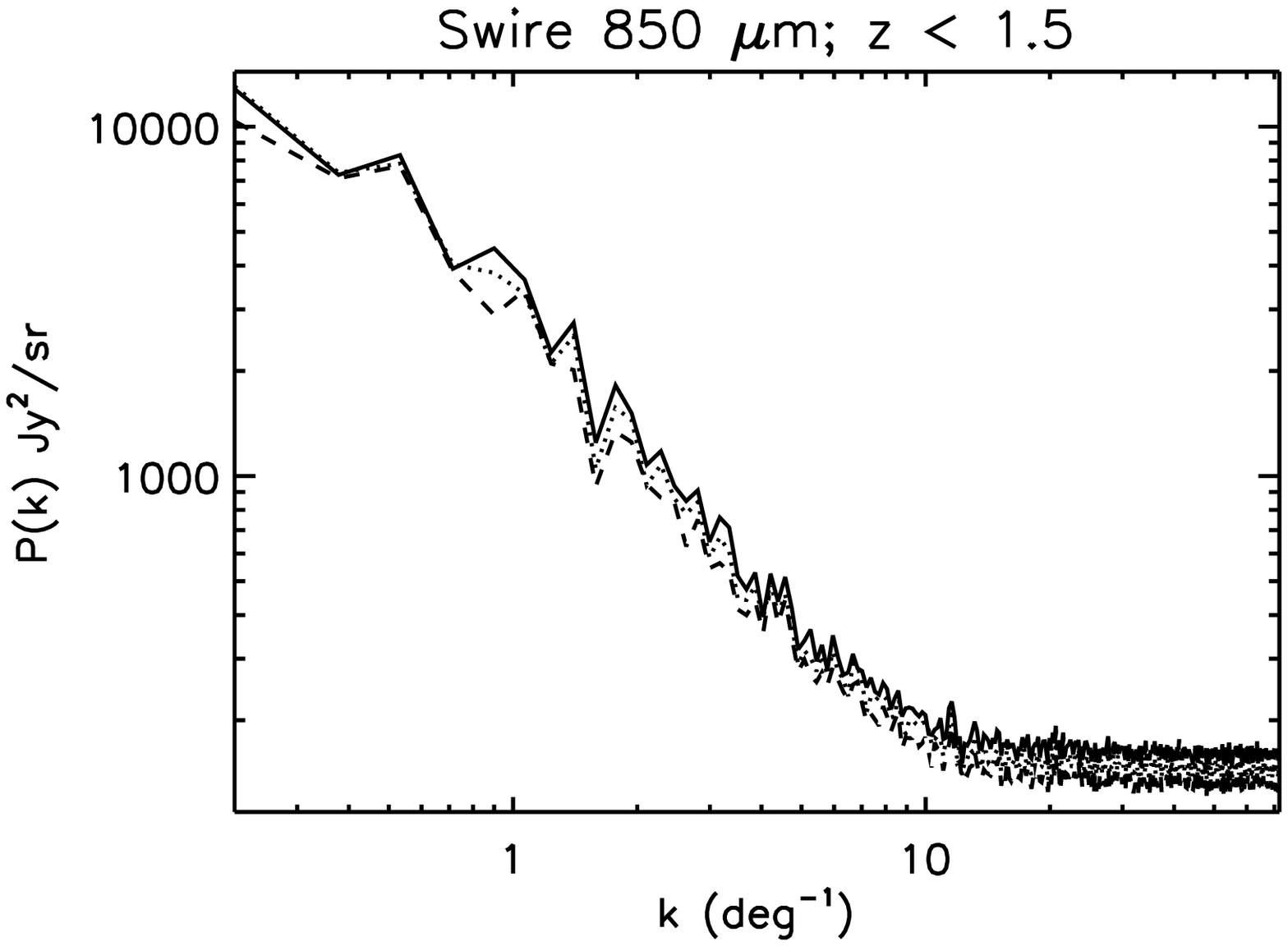}\includegraphics[width=0.48\columnwidth]{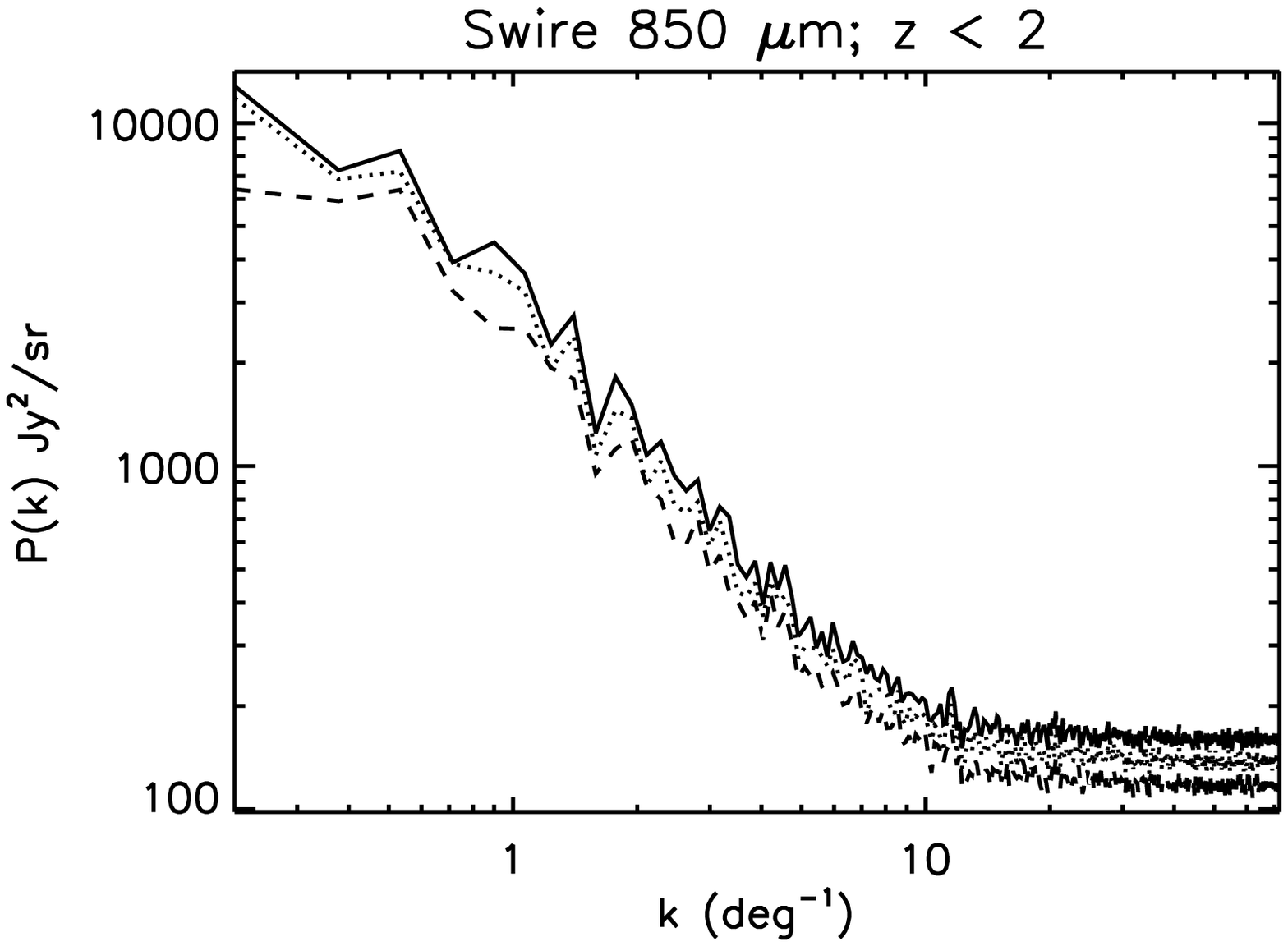}
\par\end{centering}
\caption{Power spectra of the 850~$\mu$m map of the SWIRE fields. The
solid line is the total CIB power spectrum, the
dashed line is the CIB power spectrum for $z>z_{lim}$
(where $z_{lim}$ is a redshift limit), and the dotted line is the
power spectrum of the total CIB from which we have subtracted
the stacked sources at $z<z_{lim}$. The redshift limit $z_{lim}$
is $z_{lim}=1.5$ (left) and $z_{lim}=2$ (right).
The redshift errors are $\frac{\triangle z}{z}=3\%$ (top) and $\frac{\triangle z}{z}=10\%$
(bottom).\label{fig:P2Swire-850-3-10=000025}}
\end{figure}

\begin{figure}
\begin{centering}
\includegraphics[width=0.48\columnwidth]{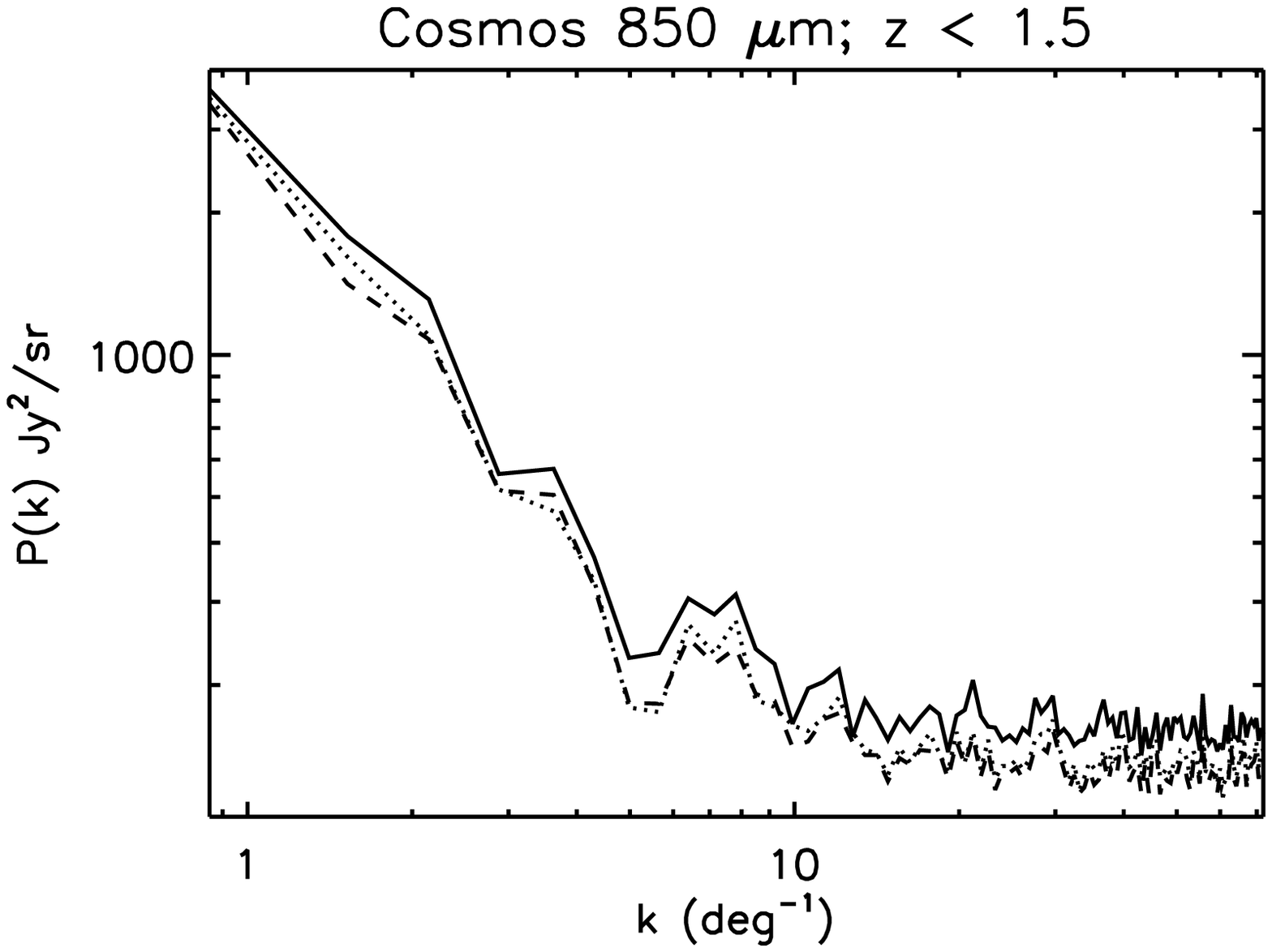}\includegraphics[width=0.48\columnwidth]{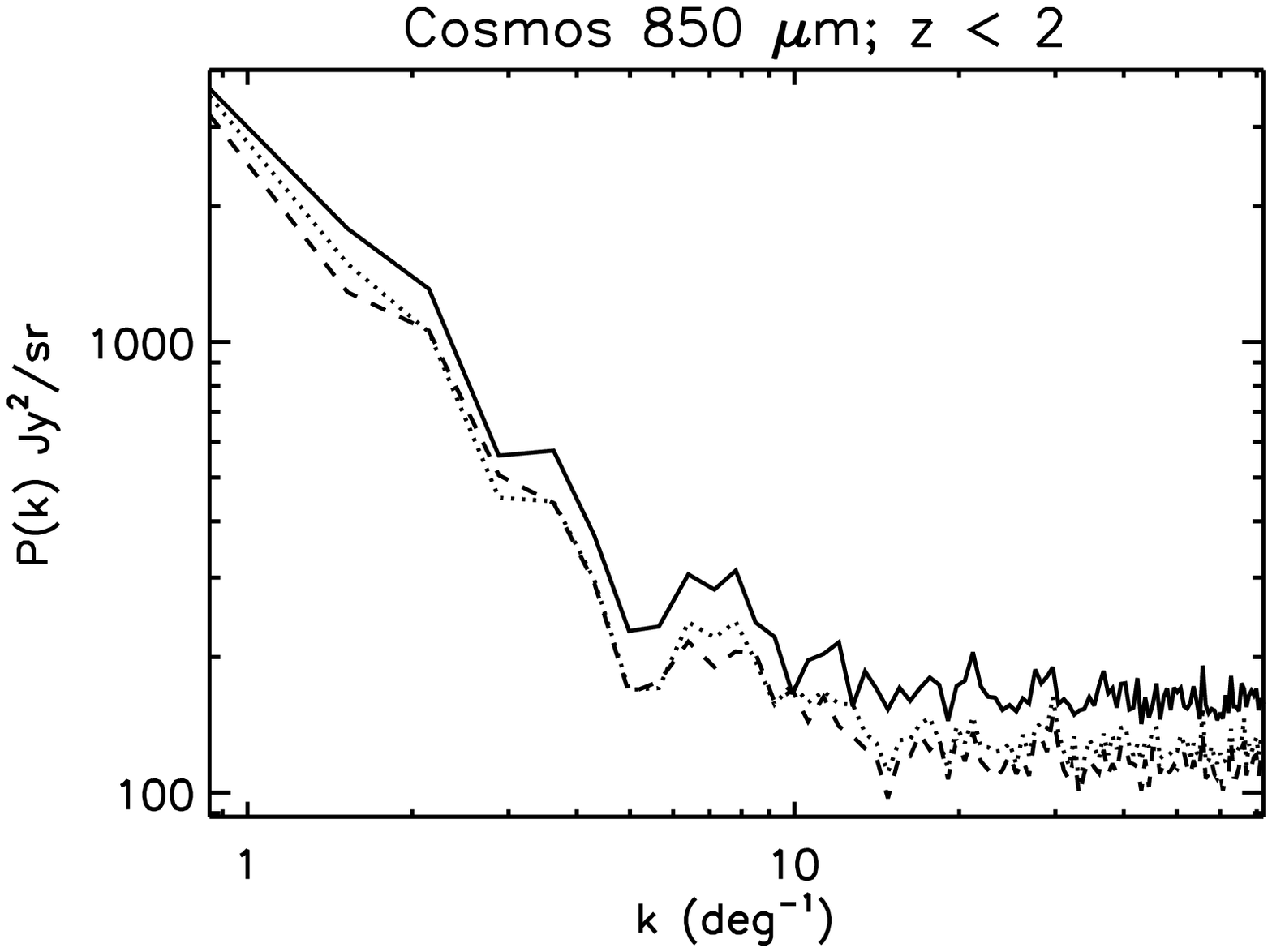}
\par\end{centering}
\begin{centering}
\includegraphics[width=0.48\columnwidth]{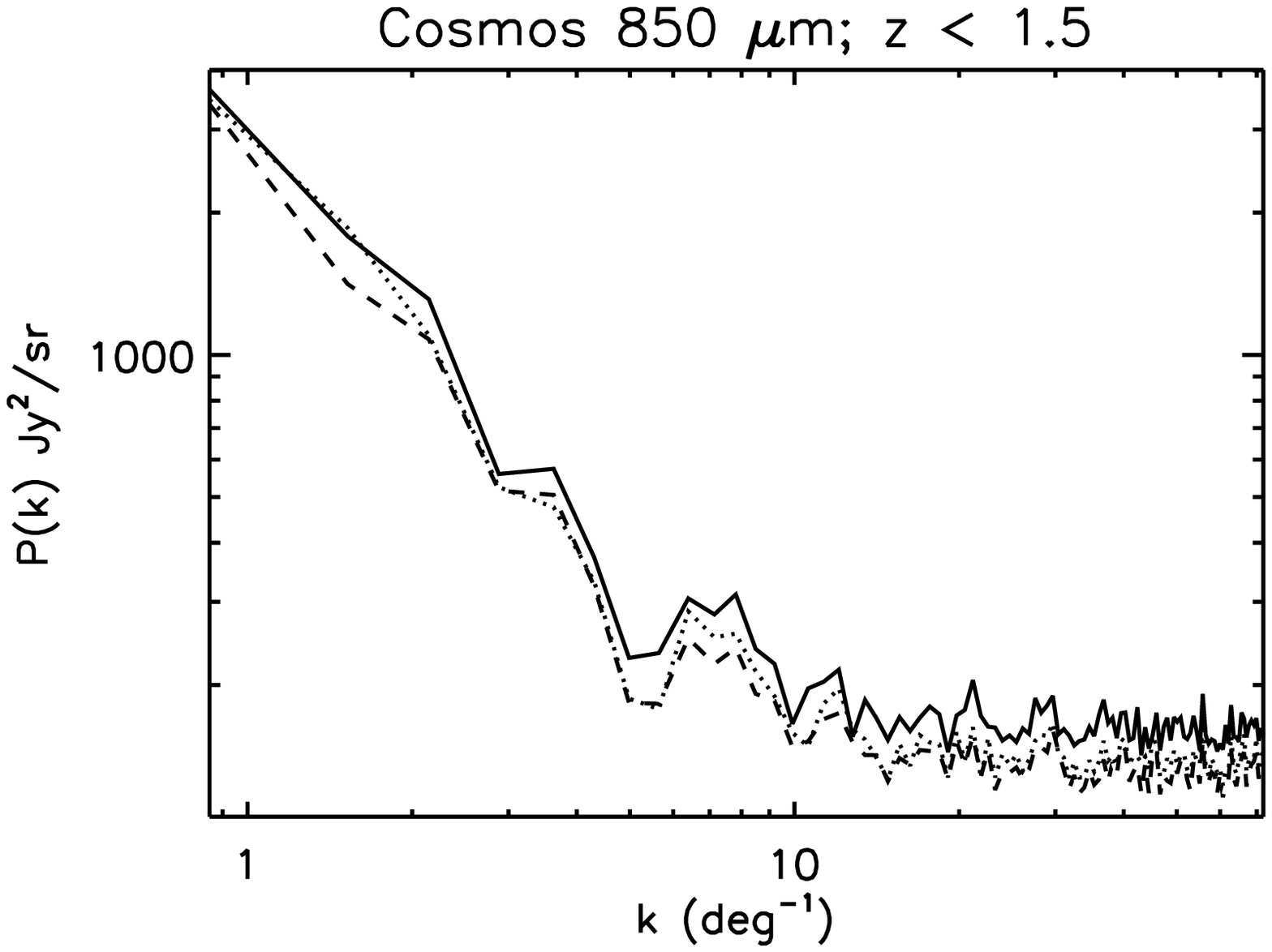}\includegraphics[width=0.48\columnwidth]{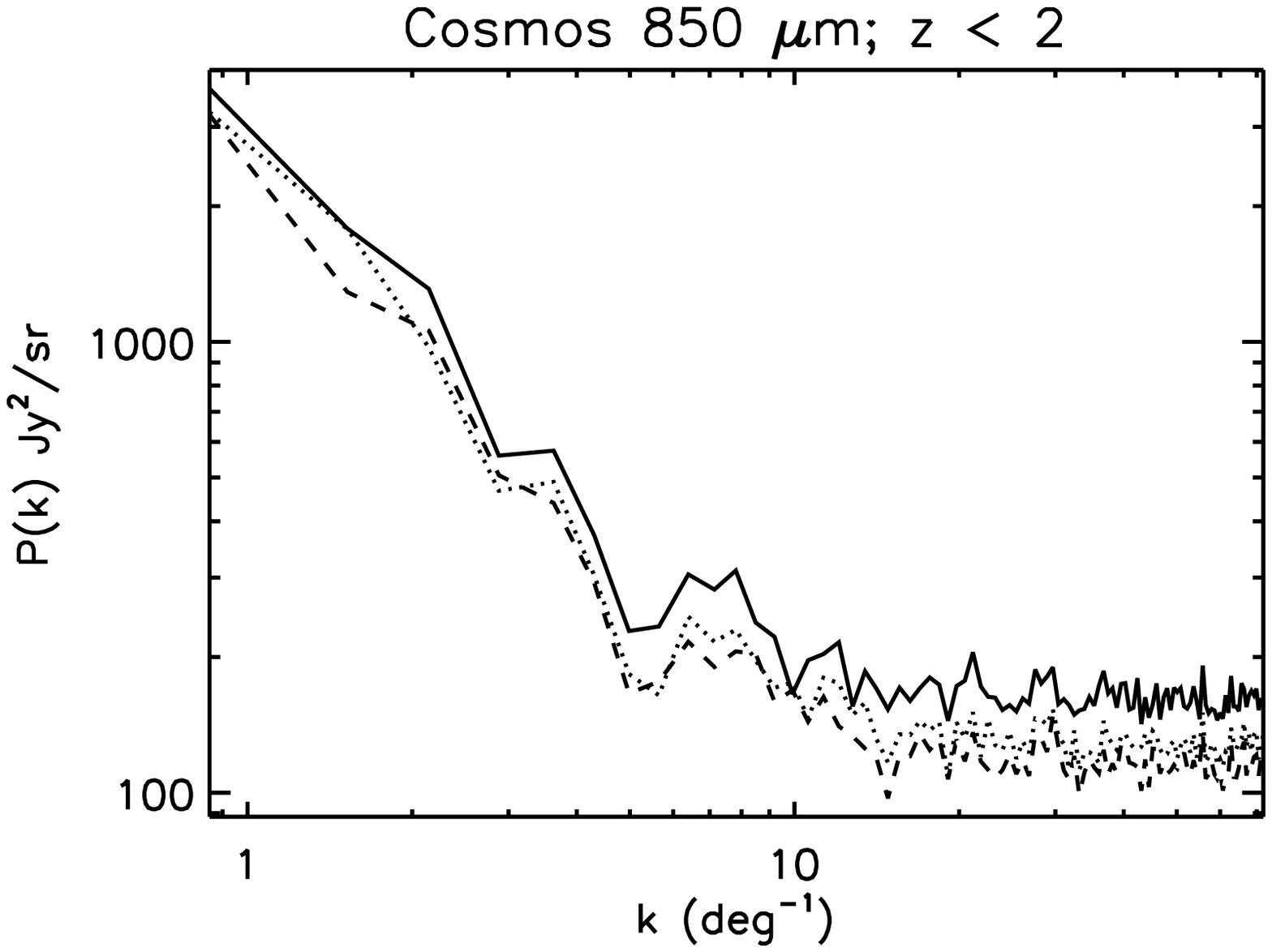}
\par\end{centering}
\caption{Same as Fig. \ref{fig:P2Swire-850-3-10=000025} but for the COSMOS field.\label{fig:P2Cosmos-850-3-10=000025}}
\end{figure}

Figures \ref{fig:P2Swire-850-3-10=000025} and \ref{fig:P2Cosmos-850-3-10=000025} 
show the similar results but at 850~$\mu$m. For these observations,
we needed to use COSMOS data because for SWIRE data we do not subtract a
significant fraction of the CIB anisotropies. In terms of power
spectra, we are able with SWIRE to subtract only $\sim 30\%$ of the
correlated part and $\sim 50\%$ of the Poissonian part. 
In the case of COSMOS, we subtract approximately $\sim 75\%$ of both the correlated and 
Poissonian part of the power spectra. Figure \ref{fig:P2Cosmos-850-3-10=000025} (top-right) shows 
that, for errors of $\frac{\triangle z}{z}=3\%$, our method is very efficient in subtracting $z<2$ anisotropies.

\section{Summary \label{sec:Summary}}

We have described a stacking algorithm and illustrated its
capabilities using Spitzer observations. We
have studied the accuracy of the stacking method as a means of determining the
average fluxes of classes of undetectable sources at long wavelengths. The results
show that the technique will be capable of measuring accurate fluxes
at both far-infrared and submillimeter wavelgnths for sources
as faint as 80~$\mu$Jy at 24~$\mu$m using average colors.\\

With the successful commissioning of the Planck and Herschel missions, large maps
(even all-sky for Planck) from 250~$\mu$m to the millimeter
wavelength range are now available. SCUBA-2 and other submillimeter cameras 
(e.g., LABOCA) will provide data of higher angular resolution in the submillimeter. 
We have applied the stacking
method to the Herschel, Planck, and SCUBA-2 simulated data and measured
the full average SED of populations of sources detected at 24~$\mu$m. 
The strong variation in the $S_{24}/S_{\lambda}$ color 
 with redshift requires us to define the
populations to which the method will be applied not only in ranges of $S_{24}$
but also in terms of (photometric) redshift. 
We show we are able to measure the mean flux of populations 4
to 6 times fainter  than the total noise at 350~$\mu$m at redshifts
$z=1$ and $z=2$, respectively, 
and 6 to 10 times fainter than the total noise at 850~$\mu$m, at the same redshifts. We have been
able to reproduce the SED at wavelengths 70, 160, 250, 350, 500, 
and 850~$\mu$m of a population of sources with mean flux $S_{24}=0.11$~mJy
and $S_{24}=0.135$~mJy at redshifts $z=1$ and $z=2$, respectively. \\

In the deep Spitzer fields, the detected 24~$\mu$m sources constitute a
large fraction of the anisotropies. We have shown that the method presented in this paper 
enables an excellent (350-850 COSMOS) to good (350-850 SWIRE)
removal of both the Poissonian and correlated low-z anisotropies. 
The relative contribution of sources to the background anisotropies 
up to $z=2$ decreases with wavelength in the model. This property
is expected to remain valid independently of the details of the model 
from 250~$\mu$m to
the millimeter range. Although the accuracy of the subtracted map is
lower at  850~$\mu$m, 
the cleaning of the power spectrum is quite effective (because the contribution of the 
low-redshift sources is small at these submillimeter wavelengths). \\

The same technique could also be used to remove from the observations
all the contributions from sources for which we have estimated a 
flux, to decrease the confusion noise
caused by infrared galaxies. This would be interesting for the
detection of other types of sources (for example, SZ sources in Planck data).\\

The method allows us to build $z\gtrsim1-2$ CIB maps
from the submillimeter to the millimeter. We have found that the method can also 
be successfully applied at the other Herschel and Planck wavelengths 
than those tested in this paper. The longer wavelengths at which
this can be achieve will depend on the success of the component separation and
not on the removal of the $z<2$ sources. We can then hope to have a set of large CIB maps
dominated by high-redshift galaxies. 
This set of CIB maps at different wavelengths dominated by $z>2$ sources will
be a powerful tool for studying the evolution of the large-scale structure of infrared galaxies.
The effect of the K-correction ensures that each of these maps (at different wavelengths) are
dominated by particular high-redshift ranges. Methods of independent
component separation based on the correlation matrix between these maps \citep [e.g.,][] {2003MNRAS.346.1089D} should allow us to extract maps and power spectra for a number of redshift ranges equal to the number of maps.
This last step will fulfill the main objective of this work. It will allow the study of the evolution of the
IR galaxy clustering at high redshifts by means of the power spectrum analysis of
CIB anisotropies. These maps may also be used to help us understand the contribution of high-z IR galaxies both to 
the CIB and the star-formation history.

\bibliographystyle{natbib}
\bibliography{ArticleReferences}

\begin{appendix} \section{Alternative correction for the clustering contribution to the stacked fluxes in
  Planck maps}

We developed an alternative method for correcting the photometry of
a group of stacked sources for the effects of the clustering. If we consider
that the signal measured for a population of stacked sources at a given wavelength is the 
combination of the signal originating from the sources and from
the clustering, we can write the measured flux as:
\begin{equation}
S_{\lambda}^{measured}=S_{\lambda}^{sources}+S_{\lambda}^{clustering}+{\sigma}
\end{equation}
where $S_{\lambda}^{measured}$ is the total measured signal, $S_{\lambda}^{sources}$ is the
part of the signal coming from the sources, and $S_{\lambda}^{clustering}$ is
the part of the signal coming from the sources correlated with the
detected sources that we are stacking, and ${\sigma}$ is the noise.\\

If two populations of sources have very similar fluxes at the
wavelength of detection (24~$\mu$m) and are situated at similar
redshifts, we can assume that their sources have very
similar physical characteristics and hence their colors  
$S_{\lambda}/S_{24}$ are very similar. In this case, we can write:
\begin{equation}
\left(\frac{S_{24}^{sources}}{S_{\lambda}^{sources}}\right)_{A} \simeq \left(\frac{S_{24}^{sources}}{S_{\lambda}^{sources}}\right)_{B}
\end{equation}
where the A and B subscripts represent the first and second population
of sources. We can measure the total flux (from the sources and the
clustering) for the stacking of both source populations and
express them as: 
\begin{equation}
(S_{\lambda}^{total}=S_{\lambda}^{sources}+S_{\lambda}^{clustering}+{\sigma})_{A}\\
\end{equation}
\begin{equation}
(S_{\lambda}^{total}=S_{\lambda}^{sources}+S_{\lambda}^{clustering}+{\sigma})_{B}.
\end{equation}
If we were to assume that the contribution of the correlated sources to the
flux is the same for both populations
$(S_{\lambda}^{clustering})_{A}=(S_{\lambda}^{clustering})_{B}$, as
expected for sources with similar spatial distributions, and that the
noise is negligible, we would have a
system of three equations with three unknowns that we can solve.\\

The main problem for the applicability of this method is that we need
to stack many sources to ensure that the noise becomes negligible compared to
the signal. Because of this, it is preferable to combine an observation whose photometry is affected by the 
clustering with another observation for which this problem does not
exist, as illustrated by our present analysis. If the photometry of this second observation is affected by
smaller errors (as it is the case of SCUBA-2 data relative to 
Planck data at 850~$\mu$m), the results will be improved by combining the two
observations. However, the method discussed in this appendix is applicable to cases where we do not have an alternative observation with which we can correct from the clustering problem.

\end{appendix}

\end{document}